\PassOptionsToPackage{unicode}{hyperref}
\PassOptionsToPackage{hyphens}{url}
\PassOptionsToPackage{dvipsnames,svgnames,x11names}{xcolor}
\documentclass[
  12pt]{article}

\usepackage{amsmath,amssymb}

\usepackage{setspace}

\usepackage{iftex}
\ifPDFTeX
  \usepackage[T1]{fontenc}
  \usepackage[utf8]{inputenc}
  \usepackage{textcomp} 
\else 
  \usepackage{unicode-math}
  \defaultfontfeatures{Scale=MatchLowercase}
  \defaultfontfeatures[\rmfamily]{Ligatures=TeX,Scale=1}
\fi
\usepackage{lmodern}
\ifPDFTeX\else  
\fi
\IfFileExists{upquote.sty}{\usepackage{upquote}}{}
\IfFileExists{microtype.sty}{
  \usepackage[]{microtype}
  \UseMicrotypeSet[protrusion]{basicmath} 
}{}
\makeatletter
\@ifundefined{KOMAClassName}{
  \IfFileExists{parskip.sty}{%
    \usepackage{parskip}
  }{
    \setlength{\parindent}{0pt}
    \setlength{\parskip}{6pt plus 2pt minus 1pt}}
}{
  \KOMAoptions{parskip=half}}
\makeatother
\usepackage{xcolor}
\makeatletter
\ifx\paragraph\undefined\else
  \let\oldparagraph\paragraph
  \renewcommand{\paragraph}{
    \@ifstar
      \xxxParagraphStar
      \xxxParagraphNoStar
  }
  \newcommand{\xxxParagraphStar}[1]{\oldparagraph*{#1}\mbox{}}
  \newcommand{\xxxParagraphNoStar}[1]{\oldparagraph{#1}\mbox{}}
\fi
\ifx\subparagraph\undefined\else
  \let\oldsubparagraph\subparagraph
  \renewcommand{\subparagraph}{
    \@ifstar
      \xxxSubParagraphStar
      \xxxSubParagraphNoStar
  }
  \newcommand{\xxxSubParagraphStar}[1]{\oldsubparagraph*{#1}\mbox{}}
  \newcommand{\xxxSubParagraphNoStar}[1]{\oldsubparagraph{#1}\mbox{}}
\fi
\makeatother

\usepackage{longtable,booktabs,array}
\usepackage{calc} 
\usepackage{etoolbox}
\makeatletter
\patchcmd\longtable{\par}{\if@noskipsec\mbox{}\fi\par}{}{}
\makeatother
\IfFileExists{footnotehyper.sty}{\usepackage{footnotehyper}}{\usepackage{footnote}}
\makesavenoteenv{longtable}
\usepackage{graphicx}
\makeatletter
\def\maxwidth{\ifdim\Gin@nat@width>\linewidth\linewidth\else\Gin@nat@width\fi}
\def\maxheight{\ifdim\Gin@nat@height>\textheight\textheight\else\Gin@nat@height\fi}
\makeatother
\setkeys{Gin}{width=\maxwidth,height=\maxheight,keepaspectratio}
\makeatletter
\def\fps@figure{htbp}
\makeatother

\makeatletter
\@ifpackageloaded{caption}{}{\usepackage{caption}}
\AtBeginDocument{%
\ifdefined\contentsname
  \renewcommand*\contentsname{Table of contents}
\else
  \newcommand\contentsname{Table of contents}
\fi
\ifdefined\listfigurename
  \renewcommand*\listfigurename{List of Figures}
\else
  \newcommand\listfigurename{List of Figures}
\fi
\ifdefined\listtablename
  \renewcommand*\listtablename{List of Tables}
\else
  \newcommand\listtablename{List of Tables}
\fi
\ifdefined\figurename
  \renewcommand*\figurename{Figure}
\else
  \newcommand\figurename{Figure}
\fi
\ifdefined\tablename
  \renewcommand*\tablename{Table}
\else
  \newcommand\tablename{Table}
\fi
}
\@ifpackageloaded{float}{}{\usepackage{float}}
\floatstyle{ruled}
\@ifundefined{c@chapter}{\newfloat{codelisting}{h}{lop}}{\newfloat{codelisting}{h}{lop}[chapter]}
\floatname{codelisting}{Listing}

\makeatother
\makeatletter
\@ifpackageloaded{caption}{}{\usepackage{caption}}
\@ifpackageloaded{subcaption}{}{\usepackage{subcaption}}
\makeatother

\ifLuaTeX
  \usepackage{selnolig}  
\fi
\usepackage[]{natbib}
\usepackage{bookmark}

\IfFileExists{xurl.sty}{\usepackage{xurl}}{} 
\hypersetup{
  pdftitle={Title},
  pdfauthor={Author 1; Author 2},
  pdfkeywords={3 to 6 keywords, that do not appear in the title},
  colorlinks=true,
  linkcolor={blue},
  filecolor={Maroon},
  citecolor={Blue},
  urlcolor={Blue},
  pdfcreator={LaTeX via pandoc}}

\usepackage{enumitem}
\usepackage{bbm}
\usepackage{algorithm}
\usepackage{algorithmic}

\renewcommand{\baselinestretch}{1.412}

\newcommand{\wh}{\widehat}

\newcommand{\T}{\!\mbox{\scriptsize T}}

\newcommand{\bX}{\mbox{\bf X}}

\newcommand{\bSigma}{\mbox{\boldmath{$\Sigma$}}}

\newcommand{\bx}{\mbox{\bf x}}

\newcommand{\bI}{\mbox{\bf I}}

\newcommand{\RR}{\mathbb{R}}
\newcommand{\cP}{\mathcal{P}}
\newcommand{\cC}{\mathcal{C}}
\newcommand{\cI}{\mathcal{I}}
\newcommand{\eE}{\mathbb{E}}
\newcommand{\II}{\mathbbm{1}}

\newtheorem{theorem}{Theorem}[section]

\newtheorem{lemma}[theorem]{Lemma}

\newtheorem{condition}{Condition}

\newcommand{\anon}{1}

\begin{document}

\def\spacingset#1{\renewcommand{\baselinestretch}%
{#1}\small\normalsize} \spacingset{1}


\if1\anon
{
  \title{\bf Model-free and Distributionally Robust Feature Screening with False Discovery Control for High-Dimensional Heterogeneous Data}
  \author{Cong Cheng
    \\
    Department of Statistics, University of Georgia\\
    and \\
   Runze Li \\
   Department of Statistics, Pennsylvania State University \\
   and \\
    Yuan Ke \\
    Department of Statistics, University of Georgia}
  \maketitle
} \fi

\if0\anon
{
  \bigskip
  \bigskip
  \bigskip
  \begin{center}
    {\LARGE\bf Title}
\end{center}
  \medskip
} \fi

\bigskip
\begin{abstract}
In this paper, we propose a model-free feature screening framework tailored for high-dimensional and heterogeneous datasets, based on a novel distributionally robust dependence measure termed Copula Divergence. The proposed screening method, named CD-Screen, addresses critical limitations of existing feature screening methods, such as restrictive modeling assumptions and sensitivity to heterogeneous feature distributions. CD-Screen ranks features according to their Copula Divergence without relying on a specific regression model or distributional assumptions. Additionally, we introduce CD-FDR, a data-driven procedure to control false discoveries, ensuring accurate and efficient feature selection. Theoretical analyses establish the sure screening and rank consistency properties of CD-Screen, along with asymptotic control of the false discovery rate by CD-FDR. Extensive simulation studies demonstrate the superior performance of our methods compared to traditional screening approaches across diverse scenarios. Furthermore, a real data analysis of the relationship between stock returns and inflation in the United States {illustrates the practical use of our method and provides descriptive evidence on} sector-specific responses to economic changes. 
\end{abstract}

\noindent%
{\it Keywords:} copula divergence, dependence measure, optimal transport, probit transformation
\vfill

\newpage
\spacingset{1.8} 

\section{Introduction}
\label{sec:introduction}

High-dimensional heterogeneous data are omnipresent in diverse modern research fields, including business, economics, engineering, finance, and social science. In many modern applications, it is common to collect datasets comprising a vast number of features whose distributions differ considerably, posing unique challenges for statistical learning \citep{ammar2006analysis}. The ultrahigh dimensionality complicates accurate statistical analyses and effective computational strategies, as most collected features are irrelevant or redundant \citep{wang2017heterogeneous}. Additionally, heterogeneity among features exacerbates the difficulty of correctly specifying regression models due to the risk of mis-specification, potentially resulting in biased estimation, diminished predictive accuracy, and inflated computational cost. Therefore, removing irrelevant features before formal modeling is crucial to mitigate these issues and facilitate reliable downstream analyses.

Feature screening methods have emerged as powerful tools to overcome the challenges posed by high dimensionality. The seminal Sure Independence Screening (SIS) method  \citep{fan2008sure} marked a significant advancement in this area by selecting a subset of relevant features based on marginal Pearson correlations. Recently, several model-free feature screening approaches have been developed \citep[see e.g.][]{zhu2011model, xue2017robust, zhou2018model, liu2022model, zhao2022distribution, tong2023model, chen2023note, tian2025feature}. These methods do not rely on pre-specified regression models, thereby alleviating concerns related to model mis-specification. 
Model-free screening techniques have found wide application across diverse areas, including discriminant analysis \citep{cui2015model}, censored data analysis \citep{zhou2017model}, survival data analysis \citep{lin2018model}.

Despite their growing popularity, existing model-free screening approaches exhibit critical limitations.
First, although model-free methods circumvent issues related to model mis-specification, they often rely on restrictive assumptions, such as identically distributed features or stringent moment conditions \citep{zhu2011model}. These assumptions are typically violated in high-dimensional heterogeneous settings, leading to reduced reliability and compromised performance. 
Second, most screening methods require the selection of a threshold parameter to distinguish active features from inactive ones \citep{guo2023threshold}. In model-based settings, such thresholds can be informed by goodness-of-fit criteria. However, in a model-free context, goodness-of-fit is not well-defined, making threshold selection more challenging \citep{liu2022model}. In practice, conservative thresholds may be chosen to ensure the inclusion of all active features. While this approach increases the chance of capturing relevant features, it also admits many inactive features, thereby inflating the false discovery rate (FDR). 
Balancing the sure screening property with effective FDR control remains a fundamental challenge in model-free feature screening for high-dimensional heterogeneous data. To date, this issue has not been completely addressed in the literature.

\color{black}

Beyond marginal screening, there is also a growing literature on conditional, forward, and joint screening methods. Conditional screening methods \citep[e.g.][]{wu2015conditional, wen2018sure, zhang2018variable, tong2023model} assess feature relevance after accounting for prespecified covariates or selected predictors, while forward and joint screening methods \citep[e.g.][]{zhong2016regularized, zhou2020model, xia2021copula, jiang2024screen} aim to capture variables whose effects may be weak marginally but stronger jointly or conditionally. These approaches can be advantageous in the presence of interaction effects, strong feature correlations, or important conditioning variables. Nevertheless, model-free marginal screening remains an important tool in ultrahigh-dimensional heterogeneous problems, due to its scalability, interpretability, and minimal structural assumptions. The goal of this paper is to strengthen this marginal screening paradigm by developing a dependence measure that is robust to heterogeneous marginal distributions and capable of capturing general marginal dependence beyond linear association.

\color{black}

In this paper, we introduce a novel model-free and distributionally robust feature screening framework for high-dimensional heterogeneous data. We define {Copula Divergence} to measure dependence between two continuous random variables by quantifying the divergence between their copula function and the independence copula. By transforming variables to standard Gaussian, Copula Divergence is robust to marginal distributions, making it well-suited for heterogeneous settings. Based on this measure, we propose CD-Screen, a screening method that ranks features without requiring explicit distributional assumptions or model specification. To address threshold selection and control false discoveries, we further develop CD-FDR, an adaptive procedure that effectively controls the FDR at a pre-specified level. Theoretically, we analyze the statistical properties of Copula Divergence and establish that CD-Screen satisfies sure screening and rank consistency under mild conditions. We also provide asymptotic guarantees showing that CD-FDR controls the FDR at a user-specified level. Extensive numerical examples demonstrate that CD-Screen outperforms traditional methods across various scenarios, and that CD-FDR offers strong finite-sample FDR control.

The rest of the paper is organized as follows.
In Section \ref{sec:CD}, we discuss the limitations of Wasserstein dependence and introduce Copula Divergence as a novel dependence measure. We also present its estimation procedure and theoretical properties. 
Section~\ref{sec:CD_Screen} proposes {CD-Screen}, a model-free feature screening method designed for high-dimensional heterogeneous data, and establishes its sure screening and rank consistency properties under mild conditions. 
In Section~\ref{sec:CD_FDR}, we develop a data-driven threshold selection approach to control FDR for CD-Screen.
Section~\ref{sec:simulations} demonstrates the finite-sample performance of our methods through numerical studies.  Section \ref{sec:conclusion} is the conclusion of the paper. 
Proofs of the main theorems, additional technical details, and supplementary numerical results are provided in the supplementary material.

\section{Copula Divergence }\label{sec:CD}

\subsection{Wasserstein Dependence and Its Limitations}
\label{sec:WD_measure}

Let $X$ and $Y$ be two continuous random variables with a joint cumulative distribution function (CDF) $F_{X,Y}(x,y)$, and marginal CDFs $F_X(x)$ and $F_Y(y)$. $X$ and $Y$ are independent if and only if 
\begin{align*}
    F_{X,Y}(x,y) = F_X(x) F_Y(y), \quad \text{for all possible values } x, y.
\end{align*}
This classical result in probability theory motivates us to measure the dependence between $X$ and $Y$ by computing the optimal transport cost between the probability measure associated with $F_{X,Y}(x,y)$ and the product probability measure associated with $F_X(x) F_Y(y)$.

Let $\mathcal{P}_r\left(\mathbb{R}^d\right)$ be the set of  all Borel probability measures on $\mathbb{R}^d$  with finite moments of order $r \in [1, \infty)$. Let $\Gamma(\mu, \rho)$ be the set of probability measures $\gamma$ on $\mathbb{R}^d \times \mathbb{R}^d$ with marginal measures $\mu, \nu \in \mathcal{P}_r\left(\mathbb{R}^d\right)$, that is
     \begin{equation*}
        \gamma\left(B \times \mathbb{R}^d\right) = \mu(B) \quad \text{and} \quad \gamma\left(\mathbb{R}^d \times B\right) = \nu(B), 
        \quad
        \text{for all Borel sets } B \subseteq \mathbb{R}^d.
    \end{equation*}
The $r$-Wasserstein distance \citep{kantorovich1960mathematical, vaserstein1969markov} between measures $\mu$ and $\nu$ in $\mathcal{P}_r\left(\mathbb{R}^d\right)$ can be defined by
    \begin{equation*}
        \mathcal{W}_r(\mu, \nu) \doteq \left( \inf_{\gamma \in \Gamma(\mu, \nu)} \int_{\mathbb{R}^d \times \mathbb{R}^d} \|u - v\|^r \, \mathrm{d}\gamma(u, v) \right)^{1/r},
    \end{equation*}
where $\|\cdot\|$ is a norm in $\mathbb{R}^d$. In this paper, we mainly consider $r=1,2$ and choose $\|\cdot\|$ as the corresponding $\ell_r$ norm. There is extensive literature on computational methods for the  $r$-Wasserstein distance \citep{santambrogio2015optimal,peyre2019computational}. We detail the specific computation approach adopted in this study in Appendix C in the supplementary material.

With a slight abuse of notation, we denote the probability measures corresponding to $F_{X,Y}(x,y)$ and $F_X(x) F_Y(y)$ by $\gamma_{X,Y}$ and $\mu_{X}\otimes \nu_{Y}$, respectively. It is therefore natural to quantify the dependence of  $(X,Y)$ by the $r$-Wasserstein distance between these two measures. Following the discussions in \cite{mordant2022measuring}, the $r$-Wasserstein Dependence between $X$ and $Y$ can be defined as
\begin{equation}
    WD(X,Y; r) \doteq \mathcal{W}_r\left(\gamma_{X,Y},\,\mu_{X}\otimes \nu_{Y}\right).
    \label{eq:W-dependence}
\end{equation}

\begin{figure}[htbp]
    \centering
    \begin{minipage}{0.32\textwidth} 
        \centering\includegraphics[width=\textwidth]{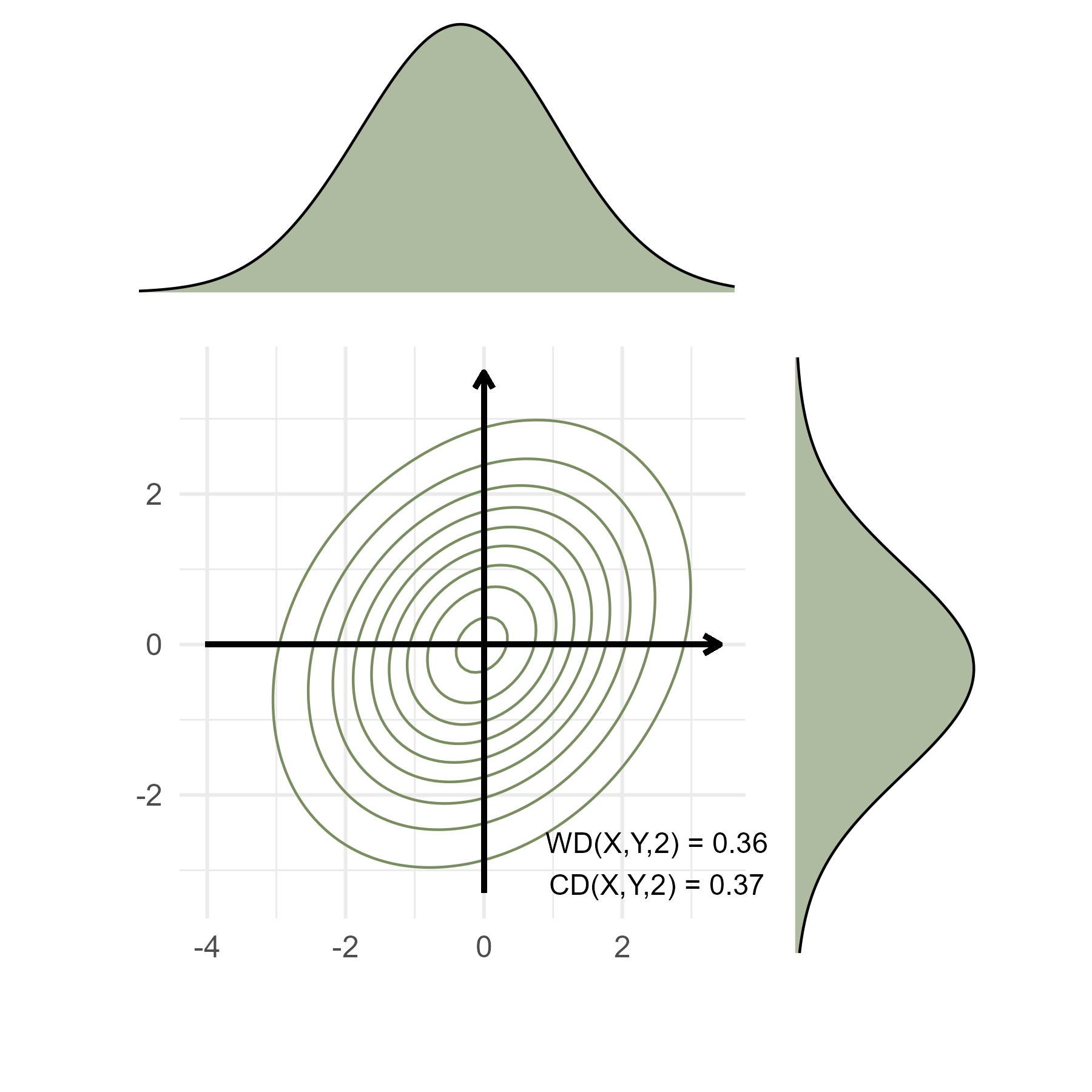}
    \end{minipage}
    \begin{minipage}{0.32\textwidth} 
        \centering
        \includegraphics[width=\textwidth]{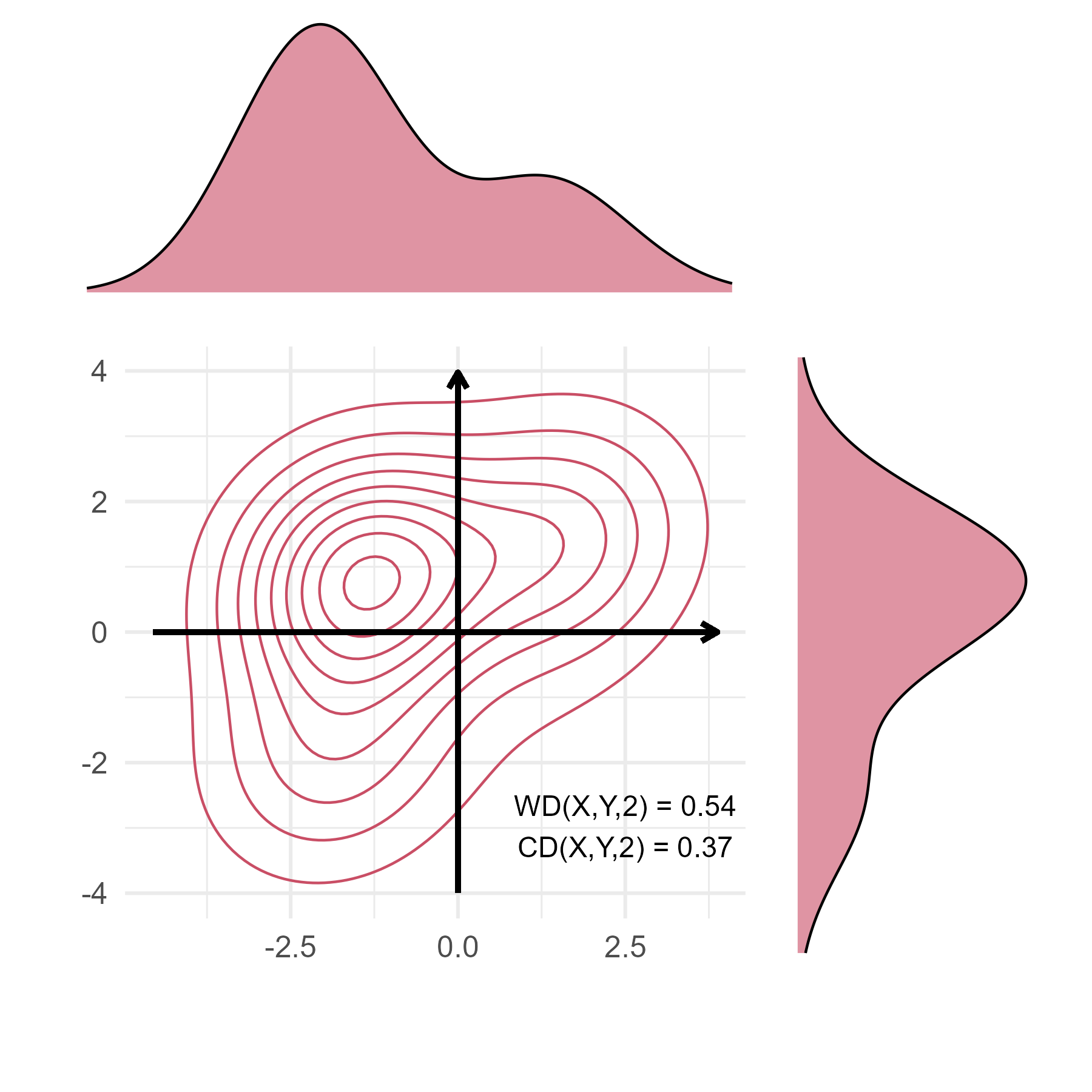}
    \end{minipage}
    \begin{minipage}{0.32\textwidth} 
        \centering
        \includegraphics[width=\textwidth]{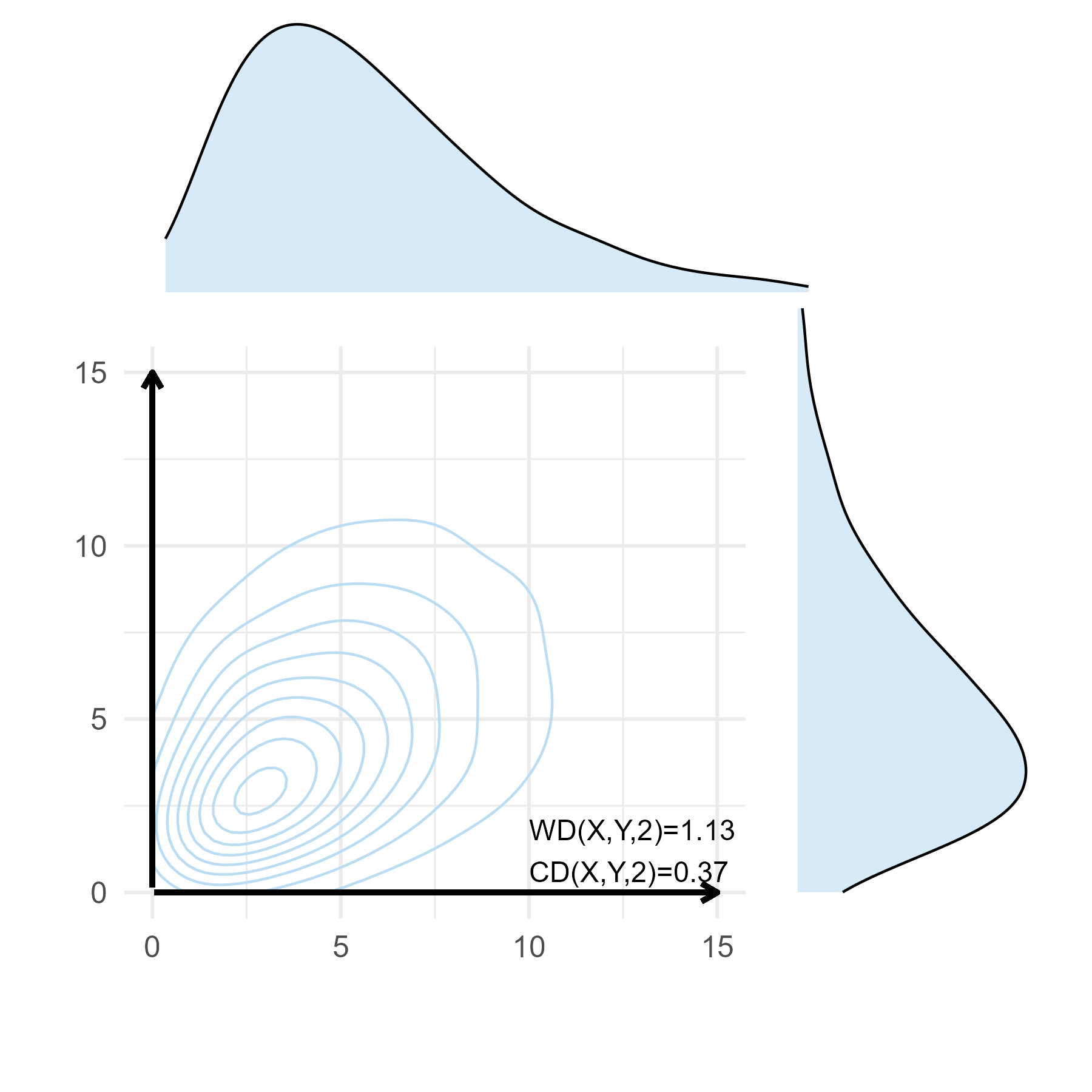}
    \end{minipage}
    \caption{
    Contour plots and marginal densities for three bivariate distributions sharing the same Gaussian copula, with a correlation matrix having unit diagonal entries and off-diagonal entries equal to 0.5. The marginals are standard Gaussian in the left panel; Gaussian mixtures in the middle panel, with $X \sim 0.7N(-1.5,0.6)+0.3N(1.5,0.6)$ and $Y \sim 0.2N(-2,0.5)+0.8N(1,0.7)$; and chi-square with 5 degrees of freedom in the right panel. The reported $WD(X,Y;2)$ and $CD(X,Y;2)$ are Wasserstein Dependence and Copula Divergence with $r=2$, averaged over 20 replications with 5000 samples each.
    }\label{fig:contours}
\end{figure}

The $r$-Wasserstein Dependence and its variants have been studied for various learning tasks \citep[see e.g.][]{ozair2019wasserstein,  catalano2024wasserstein, de2025high}. However, the measure defined in \eqref{eq:W-dependence} has several limitations. First, the scaling of $WD(X,Y; r)$ depends heavily on the marginal distributions of $X$ and $Y$, complicating the comparison of dependence among heterogeneous features and the response variable. In Figure \ref{fig:contours}, we use a toy example to illustrate this limitation of the Wasserstein Dependence.  We simulate three bivariate distributions that share the same copula function but differ in their marginal distributions. Although the underlying dependence structure is identical across all three cases, the computed values of the $2$-Wasserstein Dependence vary substantially. This example highlights the sensitivity of Wasserstein Dependence to marginal distributions, making it unsuitable for comparing dependence strength across heterogeneous features.

Second, computing $WD(X,Y; r)$ requires estimating the joint and marginal distributions from the same sample, causing the `double dipping' issue \citep{kriegeskorte2009circular} and compromising the independence of estimates. Permutation and sample splitting methods \citep{nies2022d} can address this issue but introduce extra computational complexity and reduce effective sample sizes, especially problematic for high-dimensional small-sample scenarios. Third, theoretical properties of empirical $r$-Wasserstein distances typically rely on strong assumptions about marginal distributions. For example, \cite{rippl2016limit} derived limiting distributions under Gaussian assumptions, and several studies \citep{fournier2015rate, del2019central, weed2019sharp, lei2020convergence, del2024central} have examined asymptotic and non-asymptotic properties under restrictive conditions, limiting their applicability to high-dimensional heterogeneous data.


\subsection{Copula Divergence with Probit Transformation}
\label{sec:copula_divergence}

To address the aforementioned limitations, we introduce a novel statistical measure termed Copula Divergence. To begin with, we reformulate the problem of dependence measurement through the lens of divergence between the copula function and the independence copula.

According to Sklar's theorem \citep{sklar1959fonctions}, we can represent the joint CDF $F_{X,Y}(x,y)$ through the copula function of marginal CDFs:
\begin{align*}
    F_{X,Y}(x,y)  =\cP(X\leqslant x, Y\leqslant y) =\cP\left(F_X(X)\leqslant F_X(x), F_Y(Y)\leqslant F_Y(y) \right) 
   \doteq \cC\left(F_X(x), F_Y(y)\right),
\end{align*}
where $\cC(F_X(x), F_Y(y))$ is the copula function of $(X, Y)$. In addition, we denote the independence copula function by $\cI (F_X(x), F_Y(y)) \doteq F_{X}(x)F_{Y}(y)$.  Both copula functions can be viewed as bivariate CDFs whose marginal distributions are uniform on the interval $[0,1]$. Therefore, the dependence between $X$ and $Y$  can be characterized by the divergence between the copula function $\cC(\cdot, \cdot)$ and the independence copula $\cI (\cdot, \cdot)$, which is invariant to the marginal distributions of $X$ and $Y$.

Let $\Phi(\cdot)$ and $\phi(\cdot)$ denote the CDF and density of the standard Gaussian distribution. We apply the Probit transformation to $F_{X}(X)$ and $F_{Y}(Y)$ by defining
\begin{align*}
    S = \Phi^{-1}\left(F_{X}(X)\right) \quad \text{and} \quad T = \Phi^{-1}\left(F_{Y}(Y)\right).
\end{align*}
Because copulas (with continuous marginals) are invariant under monotonic nondecreasing transformations on marginals, the joint CDF of $(S,T)$ can still be represented by the copula function $\cC(\cdot, \cdot)$ as
\begin{align*}
    F_{S,T} (s, t) = \cP(S\leqslant s, T\leqslant t)= \cC(\Phi(s), \Phi(t)), \quad \forall (s,t) \in \RR^2.
\end{align*}

The discussions above motivate us to propose a dependence measure that is robust to the marginal distributions. We define the { Copula Divergence} between $X$ and $Y$ by
\begin{equation}
    CD\left(X,Y;r\right)\doteq \mathcal{W}_r\left(\gamma_{S,T},\,\phi\otimes\phi\right),\label{eq:population_DR-WD}
\end{equation}
where $\gamma_{S,T}$ is the probability measure associated with  $(S,T)$, and $\phi\otimes \phi$ is the joint probability measure of two independent standard Gaussian random variables. The next lemma shows that Copula Divergence is non-negative and invariant under monotonic transformations.

\begin{lemma}\label{lemma:CD}
Let $(X,Y)$ be a pair of random variables with continuous marginals. Let $g_1(\cdot)$ and $g_2(\cdot)$ be two monotonically increasing and invertible functions. Then, we have
\begin{enumerate}[label=(\alph*)]
    \item $CD\left(X,Y;r\right)\geqslant 0$ for any $r\in[1,\infty)$.
    \item $CD\left(X,Y;r\right)=CD\left(g_1(X),g_2(Y);r\right)$ for any $r\in[1,\infty)$.
\end{enumerate}
\end{lemma}

Copula Divergence transforms the marginal distributions of $X$ and $Y$ to the standard Gaussian distribution, thereby eliminating sensitivity to the original marginal distributions. 
This transformation ensures that dependence is assessed solely based on the dependence structure, making it well-suited for high-dimensional feature screening where feature distributions may vary widely. Revisiting the toy example in Figure~\ref{fig:contours}, we observe that the Copula Divergence values are identical across all three scenarios, correctly capturing the dependence between the two random variables regardless of their marginal distributions.
{Furthermore, after the copula transformation, the target marginal distributions are standardized and fixed. Although empirical marginal CDFs are used to construct the rank-based Probit-transformed sample in Section 2.3, the subsequent optimal transport comparison is made against fixed standard Gaussian marginals. This construction therefore mitigates the double-use-of-data concern for both estimation and inference.} This fact improves sample efficiency and computational speed, as it avoids the additional complexity introduced by the sample-splitting technique.


\subsection{Estimation of Copula Divergence }
\label{sec:CD_estimation}

Let $\{(x_{i},y_i)\}_{i=1}^n$ be an i.i.d. random sample observed from $(X,Y)$. The empirical marginal CDFs of $X$ and $Y$ can be calculated by
\begin{align*}
    \wh{F}_{X}(x)=\frac{1}{n} \sum_{i=1}^n \II_{\left\{x_{i} \leqslant x\right\}}
    \quad \text{and} \quad 
    \wh{F}_{Y}(y)=\frac{1}{n} \sum_{i=1}^n \II_{\left\{y_i \leqslant y\right\}},
\end{align*}
where $\II(\cdot)$ is an indicator function. 

The Probit transformed sample of $\{(x_{i},y_i)\}_{i=1}^n$ is defined as
\begin{equation}
    (\wh{s}_{i},\wh{t}_{i})\doteq\left(\Phi^{-1}\left(\frac{n}{n+1}\wh{F}_{X}(x_{i})\right),\Phi^{-1}\left(\frac{n}{n+1}\wh{F}_{Y}(y_i)\right)\right),\label{eq:gaussian_sample}
\end{equation}
which is the empirical counterpart of a Gaussianized random sample
\begin{equation}
    (s_{i},t_{i})\doteq\left(\Phi^{-1}\left(F_{X}(x_{i})\right),\Phi^{-1}\left(F_{Y}(y_i)\right)\right),
    \quad \text{for } i =1, \ \ldots, \ n.
    \label{eq:genuine_gaussian_sample}
\end{equation}

Then, we propose to estimate ${CD}\left(X,Y;r\right)$ by
\begin{equation}
    \wh{CD}\left(X,Y;r\right)\doteq \mathcal{W}_r\left(\wh{\gamma}_{{S},{T}},\,\phi\otimes\phi\right),
    \quad \text{for } r =1, 2,
    \label{eq:sample_DR-WD}
\end{equation}
where $\wh{\gamma}_{{S},{T}}$ represents the probability measure of the Probit random sample \eqref{eq:gaussian_sample}.

In the following two theorems, we introduce the theoretical properties of the Copula Divergence estimator proposed in \eqref{eq:sample_DR-WD} with $r=1$ and $2$, respectively.

\begin{theorem}[Estimation of Copula Divergence with $r=1$]\label{thm:gaussianized_wd_1}
Let $M_1$, $C_1$, $n_1$, and $\varepsilon_1$ be some positive constants. When $n\geqslant n_1$, the following result holds $\forall\varepsilon\in \left(M_1 n^{-1/2}\log n, \ \varepsilon_1 \right)$:
        \begin{equation}
            \cP\left(\left|\wh{CD}\left(X,Y;1\right)-CD\left(X,Y;1\right)\right|\geqslant \varepsilon\right)\leqslant C_1 \exp \left(-\frac{C_1 n \varepsilon^2}{\log^2 n}\right).\nonumber
        \end{equation}
        More specifically, let $K>0$, and $\kappa\in[0,\frac{1}{2})$ be two constants. When $n\geqslant n_1$, we have
        \begin{equation}
            \cP\left(\left|\wh{CD}\left(X,Y;1\right)-CD\left(X,Y;1\right)\right|\geqslant Kn^{-\kappa}\right)\leqslant C_1 \exp{\left(-\frac{C_1  K^2 n^{1-2\kappa}}{\log^2{n}}\right)}.\nonumber
        \end{equation}
\end{theorem}

\begin{theorem}[Estimation of Copula Divergence with $r=2$]\label{thm:gaussianized_wd_2}
Let $M_2$, $C_2$, $n_2$, $\varepsilon_2$, and $\beta\in(0,1)$ be some positive constants.
When $n\geqslant n_2$, the following result holds 
$\forall\varepsilon\in(M_2 n^{-1/4}(\log n )^{3/4}, \ \varepsilon_2)$:
        \begin{equation}
            \cP\left(\left|\wh{CD}\left(X,Y;2\right)-CD\left(X,Y;2\right)\right|\geqslant \varepsilon\right)\leqslant C_2 \left\{ \exp \left(-\frac{C_2 n \varepsilon^4}{\log^2 n}\right)+ \exp \left(-C_2\left(n\varepsilon^2\right)^{\beta}\right) \right\}.\nonumber
        \end{equation}
More specifically, let $K>0$, and $\kappa\in[0,\frac{1}{4})$ be two constants. When $n\geqslant n_2$, the following result holds
        \begin{align}
            & \cP\left(\left|\wh{CD}\left(X,Y;2\right)-CD\left(X,Y;2\right)\right|\geqslant Kn^{-\kappa}\right) \nonumber \\ 
            \leqslant & 
            C_2 \left\{\exp{\left(-\frac{C_2 K^4 n^{1-4\kappa}}{\log^2{n}}\right)}+\exp \left(-C_2 K^{2\beta} n^{\beta(1-2\kappa)}\right) \right\}.\nonumber
        \end{align}
\end{theorem}

Theorems \ref{thm:gaussianized_wd_1} and \ref{thm:gaussianized_wd_2} establish the non-asymptotic properties of the Copula Divergence estimators for \( r = 1 \) and 2, respectively. The upper bounds for the error probabilities decay exponentially, ensuring the fast convergence of the estimators. Notably, these results are derived without imposing restrictive conditions on the moments or distributions of the random variables, highlighting the robustness of Copula Divergence. This robustness paves the way for its application to feature screening in high-dimensional heterogeneous data.

\color{black}

The proposed method can be implemented with either \(r=1\) or \(r=2\). In practice, \(r=1\) is generally more stable in finite samples, which is consistent with the sharper concentration bound in Theorem~2.2 compared with the corresponding result for \(r=2\) in Theorem~2.3. Therefore, we recommend \(r=1\) as the default choice when the sample size is small or moderate, or when finite-sample stability is the primary concern. The choice \(r=2\), based on a squared-distance geometry, may be more responsive to stronger departures from independence and can be used as a sensitivity analysis, especially when the dependence signal is expected to be strong or nonlinear. Computationally, the two choices are similar in our implementation, since they use the same empirical copula transformation and optimal transport computation with different cost functions.

\color{black}

\section{Feature Screening with Copula Divergence}
\label{sec:CD_Screen}
\vspace{-0.1in}

\subsection{Feature Screening Methodology}
\vspace{-0.1in}

The favorable properties of Copula Divergence encourage us to construct a model-free and distributionally robust feature screening method for high-dimensional heterogeneous data.

Let $Y \in \RR$ be a response variable and $\bX=(X_1, \ldots, X_p)^{\T} \in \RR^p$ be a vector of $p$ features. We allow the features to have heterogeneous marginal distributions.
Suppose there exists a sparse set of active features
\begin{equation*}
    \mathcal{A}\doteq\left\{j\in\{1,\cdots,p\}: F_{Y|\mathbf{X}}(y|\mathbf{X}) \text{ functionally depends on } X_j \right\}.
\end{equation*}
Denote by $\mathcal{S}_{\mathcal{A}}$ the cardinality of the active set $\mathcal{A}$. We further assume $\mathcal{S}_{\mathcal{A}} \ll p$. 

We propose to approximate the active set by retaining the features that have high Copula Divergences with the response variable. Specifically, we approximate $\mathcal{A}$ via
\begin{align*}
     \mathcal{A}^\ast (t) \doteq  
     \{
     j\in\{1,\cdots,p\} : CD\left(X_j,Y;r\right) \geqslant t 
     \},
\end{align*}
where $t>0$ is a screening threshold.

Let $\{(\bx_i,y_i)\}_{i=1}^n$ be an i.i.d. random sample drawn from $(\bX,Y)$, where $\bx_i = (x_{i,1}, \ldots, x_{i,p})^{\T}$. We estimate $CD\left(X_j,Y;r\right)$ by $\wh{CD}\left(X_j,Y;r\right)$ as defined in \eqref{eq:sample_DR-WD}. Then the active set is estimated by
\begin{equation}
    \widehat{\mathcal{A}}(t_{n})=\left\{j\in\{1,\cdots,p\}: \wh{CD}\left(X_j,Y;r\right)\geqslant t_{n}\right\},\label{eq:screening set}
\end{equation}
where $t_{n}$ is a screening threshold. We refer to this screening procedure as \emph{Copula Divergence-based Feature Screening (CD-Screen)}, and summarize it in Algorithm \ref{algo:CD-Screen}.

\begin{algorithm}[htbp]
    \caption{Copula Divergence-based Feature Screening (CD-Screen)}
    \label{algo:CD-Screen}
    \begin{algorithmic}
        \STATE \textbf{Input}: An observed sample $\{(\bx_i,y_i)\}_{i=1}^n$, an order $r \in \{1, 2\}$, and a  threshold $t>0$.
        \STATE \textbf{Run}:
        \STATE (1) \parbox[t]{0.85\linewidth}{Calculate the empirical CDF of $Y$ by $\widehat{F}_{Y}(\cdot)=\frac{1}{n} \sum_{i=1}^n \II_{\left\{y_i \leqslant \cdot\right\}}$.}
        \STATE (2) \parbox[t]{0.85\linewidth}{Calculate transformed sample  $\left\{\widehat{t}_i\right\}_{i=1}^n=\left\{\Phi^{-1}\left(\frac{n}{n+1}\widehat{F}_{Y}(y_i)\right)\right\}_{i=1}^n$.}
        \STATE (3) \parbox[t]{0.85\linewidth}{\textbf{For} $j \in \{1,\cdots,p\}$ \textbf{do}:}
        \STATE \hspace{1.5em}(a) \parbox[t]{0.85\linewidth}{Calculate the empirical CDF of $X_j$ by $\widehat{F}_{X_j}(\cdot)=\frac{1}{n} \sum_{i=1}^n \II_{\left\{x_{i,j} \leqslant \cdot\right\}}$.}
        \STATE \hspace{1.5em}(b) \parbox[t]{0.85\linewidth}{Calculate transformed sample $\left\{\widehat{s}_{i,j}\right\}_{i=1}^n=\left\{\Phi^{-1}\left(\frac{n}{n+1}\widehat{F}_{X_j}(x_{i,j})\right)\right\}_{i=1}^n$.}
        \STATE \hspace{1.5em}(c) \parbox[t]{0.85\linewidth}{Estimate Copula Divergence $\widehat{CD} \left(X_j,Y;r\right)$ by \eqref{eq:sample_DR-WD}.}
        \STATE \hspace{1.5em} \parbox[t]{0.85\linewidth}{\textbf{End For}}
        \STATE (4) \parbox[t]{0.85\linewidth}{Find the selected set $\widehat{\mathcal{A}}(t)$ by \eqref{eq:screening set}.}
        
        \STATE \textbf{Output}: Selected set $\widehat{\mathcal{A}}(t)$.
    \end{algorithmic}
\end{algorithm}

\subsection{Theoretical Properties for CD-Screen}

In this subsection, we establish the sure screening property and a stronger rank consistency property for CD-Screen under some mild conditions. We first introduce these conditions. 

\begin{condition}\label{condition:minimum strength} 
Let $c_{1}>0$, $c_{2}>0$,  $\kappa_{1}\in[0,\frac{1}{2})$, and $\kappa_{2}\in[0,\frac{1}{2r})$ be some constants.
\begin{itemize}
    \item[(a)] $\min\limits_{j\in \mathcal{A}}\left\{\left[CD\left(X_j,Y;r\right)\right]^r\right\}\geqslant c_1n^{-\kappa_1}$. 
    \item[(b)] $\min\limits_{j\in \mathcal{A}}\left\{CD\left(X_j,Y;r\right)\right\}-\max\limits_{j\in \mathcal{A}^c}\left\{CD\left(X_j,Y;r\right)\right\}\geqslant c_{2} n^{-\kappa_{2}}$.
\end{itemize}
\end{condition}


Condition~\ref{condition:minimum strength}(a) ensures that the Copula Divergence of the active features and the response variable remains uniformly bounded below and does not decay to zero too rapidly as the sample size $n$ increases. Condition~\ref{condition:minimum strength}(b) imposes an assumption on the signal strength gap between the active and inactive features. 
Since the Copula Divergence is always non-negative, Condition~\ref{condition:minimum strength}(a) is weaker than Condition~\ref{condition:minimum strength}(b). These conditions can be viewed as minimum signal strength assumptions, facilitating the distinction between active and inactive features. In general, Condition~\ref{condition:minimum strength} is very mild, as it permits the minimum signal strength to approach zero as the sample size increases.

\begin{theorem}[Sure screening property]\label{thm:sure_screen_DR-WD-SIS}
    Suppose Condition \ref{condition:minimum strength}(a) is satisfied. Choose $t_{n}= \{c_{t}n^{-\tau}\}^{1/r}$ for some $c_{t}\in (0,c_1)$ and $\tau\geqslant \kappa_1$.  Let $C_1>0$ and $n_1>0$, and $\beta\in(0,1)$ be three constants.  The following results hold for $n\geq n_1$.
        
\begin{itemize}
    \item[(a)] {\bf When $r\mathbf{=1}$}, we have 
    \begin{equation}
            \cP\left(\mathcal{A}\subseteq\wh{\mathcal{A}}(t_n)\right)\geqslant1-\mathcal{O}\left(\mathcal{S}_{\mathcal{A}}\exp{\left(-\frac{C_1n^{1-2\kappa_1}}{\log^2{n}}\right)}\right).\nonumber
        \end{equation}

    \item[(b)]  {\bf When $r\mathbf{=2}$}, we have 
    \begin{equation}
            \cP\left(\mathcal{A}\subseteq\wh{\mathcal{A}}(t_n)\right)\geqslant1-\mathcal{O}\left(\mathcal{S}_{\mathcal{A}}\left[\exp{\left(-\frac{C_1n^{1-2\kappa_1}}{\log^2{n}}\right)}+\exp \left(-C_1n^{\beta(1-\kappa_1)}\right)\right]\right).\nonumber
    \end{equation}    
\end{itemize}
\end{theorem}

Theorem~\ref{thm:sure_screen_DR-WD-SIS} establishes the sure screening property of the proposed CD-Screen procedure under Condition~\ref{condition:minimum strength}(a). The theorem provides explicit probabilistic bounds for $r=1$ and $r=2$, corresponding to different dependence measures used in the screening statistic. In both cases, the probability of missing any active variable decays exponentially fast with the sample size $n$, up to logarithmic factors. Consequently, by choosing the threshold $t_n$ of an appropriate order, CD-Screen achieves the desired sure screening property, ensuring that no relevant features are excluded in the screening stage.

\begin{theorem}[Rank consistency property]\label{theorem:rankconsistency}
Suppose Condition \ref{condition:minimum strength}(b) is satisfied.  Let $C_2>0$, $n_2>0$, and $\beta\in(0,1)$ be three constants.  The following results hold for $n\geq n_2$.
\begin{itemize}
    \item[(a)] {\bf When $r\mathbf{=1}$}, we have 
        \begin{equation}
            \cP\left(\min\limits_{j\in \mathcal{A}}\left\{\wh{CD}\left(X_j,Y;1\right)\right\}-\max\limits_{j\in \mathcal{A}^c}\left\{\wh{CD}\left(X_j,Y;1\right)\right\} > 0\right)\geqslant1-\mathcal{O}\left(p\exp{\left(-\frac{C_2n^{1-2\kappa_2}}{\log^2{n}}\right)}\right).\nonumber
        \end{equation}
    If $\log p=o\left(\frac{n^{1-2\kappa_2}}{\log^2{n}}\right)$, we also have
        \begin{equation*}
            \liminf _{n \to \infty}\left(\min\limits_{j\in \mathcal{A}}\left\{\wh{CD}\left(X_j,Y;1\right)\right\}-\max\limits_{j\in \mathcal{A}^c}\left\{\wh{CD}\left(X_j,Y;1\right)\right\}\right)>0 \text {, almost surely. }
        \end{equation*}
    
    \item[(b)]  {\bf When $r\mathbf{=2}$}, we have 
        \begin{align}
            &\cP\left(\min\limits_{j\in \mathcal{A}}\left\{\wh{CD}\left(X_j,Y;2\right)\right\}-\max\limits_{j\in \mathcal{A}^c}\left\{\wh{CD}\left(X_j,Y;2\right)\right\} > 0\right)\nonumber\\
            \geqslant&1-\mathcal{O}\left(p\left[\exp{\left(-\frac{C_2n^{1-4\kappa_2}}{\log^2{n}}\right)}+\exp \left(-C_2n^{\beta(1-2\kappa_2)}\right)\right]\right).\nonumber
        \end{align}
        If $\log p=o\left(\frac{n^{1-4\kappa_2}}{\log^2{n}}\right)$ and $\log p=o\left(n^{\beta(1-2\kappa_2)}\right)$, we also have
        \begin{equation*}
            \liminf _{n \to \infty}\left(\min\limits_{j\in \mathcal{A}}\left\{\wh{CD}\left(X_j,Y;2\right)\right\}-\max\limits_{j\in \mathcal{A}^c}\left\{\wh{CD}\left(X_j,Y;2\right)\right\}\right)>0 \text {, almost surely. }
        \end{equation*}  
\end{itemize}
\end{theorem}

Under Condition~\ref{condition:minimum strength}(b), Theorem~\ref{theorem:rankconsistency} establishes the rank consistency property, ensuring that features ranked by their Copula Divergence can effectively distinguish active features from inactive ones. Together, these results demonstrate that ranking features based on Copula Divergence and selecting the top-ranked features provides a reliable and robust feature screening method for high-dimensional heterogeneous data.

\section{False Discovery Control for CD-Screen}\label{sec:CD_FDR}
\vspace{-0.1in}


False discovery control is crucial in feature screening to  avoid the inclusion of too many irrelevant features. Effective false discovery control relies on understanding the distribution of the screening statistic under the null hypothesis, where the feature is independent of the response. This becomes particularly challenging in high-dimensional heterogeneous settings. In model-free feature screening, conventional methods, such as information criteria and cross-validation, are inapplicable due to the lack of a well-defined goodness-of-fit measure. Moreover, Wasserstein distance-based metrics lack analytic limiting distributions without strong distributional assumptions. While the knockoff method \citep[e.g.,][]{barber2015controlling, candes2018panning} offers flexibility, it relies on moment conditions that may not hold in complex data, and constructing knockoff features often requires sample splitting, reducing effective sample size and potentially lowering power.

In this section, we introduce a data-driven FDR control procedure for CD-Screen to address the limitations discussed above. Our approach is computationally efficient, robust to heterogeneous feature distributions, and avoids sample splitting. 

When the $j$-th feature $X_j$ is independent of $Y$, the copula function of $(X_j, Y)$ reduces to the independence copula, and the probability measure of $({S}_j, {T})$, denoted by $\gamma_{S_j,T}$, is $\phi \otimes \phi$. As a result, the limiting null distribution of Copula Divergence is identical across all $p$ features, regardless of their marginal distributions.

Denote by $\boldsymbol{\Pi}$ the set of all permutations of $\{1, \dots, n\}$. We define the set of empirical probability measures for two independent standard Gaussian random variables as follows:
\begin{equation}
    \mathcal{H} \doteq \left\{\wh{\gamma}_{{Z}_1,{Z}_2} : \left\{{z}_{1,i},{z}_{2,i}\right\}_{i=1}^n = \left\{\Phi^{-1}\left(\frac{i}{n+1}\right),\Phi^{-1}\left(\frac{\boldsymbol{\pi}(i)}{n+1}\right)\right\}_{i=1}^n, \ \boldsymbol{\pi} \in \boldsymbol{\Pi} \right\}.
\end{equation}
By randomly drawing $m$ empirical measures $\{\wh{\gamma}_{{Z}_1,{Z}_2}^{(k)}\}_{k=1}^m \subset \mathcal{H}$, we calculate
\begin{equation}
    W^{(k)} \doteq \mathcal{W}_{r}(\wh{\gamma}_{{Z}_1,{Z}_2}^{(k)}, \phi \otimes \phi), \quad \text{for} \quad k = 1, \dots, m.\label{eq:assisting_statistics}
\end{equation}
We then construct the empirical CDF of Copula Divergence under the null hypothesis as
\begin{equation*}
    \wh{F}_{W}(w) \doteq \frac{1}{m}\sum_{k=1}^{m} \II_{\left\{W^{(k)}\leqslant w\right\}}.
\end{equation*}
This empirical CDF serves as a unified estimator for the unknown limiting distribution of Copula Divergence under the null hypothesis for all heterogeneous features in CD-Screen.

To utilize the empirical CDF under the null, we define a statistic for feature $X_j$ as
\begin{align}
    \wh{U}_j \doteq \wh{F}_{W}(\wh{CD}(X_j,Y;r)) - \frac{1}{2}, \quad \text{for } j = 1, \dots, p.\label{eq:statistics_fdr}
\end{align}
For a given screening threshold $t$, the estimated active set is 
\begin{equation}
    \wh{\mathcal{A}}(t) = \left\{j \in \{1, \dots, p\} : \wh{U}_j \geqslant t \right\}. \nonumber
\end{equation}
Accordingly, the false discovery proportion (FDP) of $\wh{\mathcal{A}}(t)$ is defined as
\begin{align}
    \operatorname{FDP}(t) \doteq \frac{\#\{j \in \mathcal{A}^c : \wh{U}_j \geqslant t\}}{\#\{j : \wh{U}_j \geqslant t\}},\label{eq:FDP}
\end{align}
where $\#\{\cdot\}$ denotes the cardinality of a set, and $\mathcal{A}^c$ is the complement of the active set $\mathcal{A}$. Throughout this paper, we adopt the convention that $0/0 = 0$.  Furthermore, the FDR is defined as the expectation of the FDP, i.e. $\operatorname{FDR}(t) \doteq \eE[\operatorname{FDP}(t)]$.

It is straightforward to verify that $\widehat{U}_j$ is approximately symmetric about zero when $j \in \mathcal{A}^c$. Combined with the sparsity assumption $\mathcal{S}_{\mathcal{A}} \ll p$, we approximate the unknown numerator in \eqref{eq:FDP} by
\begin{align*}
    \#\{j \in \mathcal{A}^c : \wh{U}_j \geqslant t\} & \approx
        \#\{j \in \mathcal{A}^c : \wh{U}_j \leqslant -t\} 
        \leqslant
         \#\{j : \wh{U}_j \leqslant -t\}.
\end{align*}
Accordingly, we estimate $\operatorname{FDP}(t)$ as
\begin{equation}
    \widehat{\operatorname{FDP}}(t) = \frac{\#\{j : \wh{U}_j \leqslant -t\}}{\#\{j : \wh{U}_j \geqslant t\}}.\nonumber
    \label{eq:FDP_estimate}
\end{equation}

To control FDR at a pre-specified level $\alpha \in (0,1)$, we select the threshold $T_{\alpha}$ as
\begin{equation}
    T_{\alpha} = \inf \left\{ t > 0 : \frac{1 + \#\{j : \wh{U}_j \leqslant -t\}}{\#\{j : \wh{U}_j \geqslant t\}} \leqslant \alpha \right\},
    \label{eq:threshold}
\end{equation}
where the constant $1$ in the numerator ensures the theoretical property of FDR control, as established in Theorem \ref{thm:FDR_control} below. The estimated active set is then given by
\begin{equation}
    \wh{\mathcal{A}}(T_{\alpha}) = \left\{ j \in \{1, \dots, p\} : \wh{U}_j \geqslant T_{\alpha} \right\}.
    \label{eq:selected set by threshold}
\end{equation}
If no solution exists for \eqref{eq:threshold}, we set $\wh{\mathcal{A}}(T_{\alpha})$ to be an empty set. 
We refer to this FDR control procedure for CD-Screen as CD-FDR and summarize it in Algorithm~\ref{algo:CD-FDR}.
\begin{algorithm}[htbp]
    \caption{FDR control for CD-Screen (CD-FDR)}
    \label{algo:CD-FDR}
    \begin{algorithmic}
        \STATE \textbf{Input}: An observed sample $\{(\bx_i,y_i)\}_{i=1}^n$, an order $r \in \{1, 2\}$, the number of empirical measures $m$, and a desired FDR level $\alpha \in (0,1)$.
        \STATE \textbf{Run}:
        \STATE (1) \parbox[t]{0.85\linewidth}{Generate $m$ empirical samples of size $n$ from all permutations of $\{1,\ldots,n\}$.}
        \STATE (2) \parbox[t]{0.85\linewidth}{Construct $m$ empirical measures $\{\wh{\gamma}_{{Z}_1,{Z}_2}^{(k)}\}_{k=1}^m$.}
        \STATE (3) \parbox[t]{0.85\linewidth}{Calculate statistics $W^{(k)}$ by \eqref{eq:assisting_statistics}.}
        \STATE (4) \parbox[t]{0.85\linewidth}{Calculate empirical CDF of CD under $H_0$ by $\wh{F}_{W}(w) \doteq \frac{1}{m}\sum_{k=1}^{m} \II_{\left\{W^{(k)}\leqslant w\right\}}$.}
        \STATE (5) \parbox[t]{0.85\linewidth}{\textbf{For} $j \in \{1,\cdots,p\}$ \textbf{do}:}
        \STATE \hspace{1.5em} (a) \parbox[t]{0.85\linewidth}{Calculate $\widehat{CD} \left(X_j,Y;r\right)$ by \eqref{eq:sample_DR-WD}.}
        \STATE \hspace{1.5em} (b) \parbox[t]{0.85\linewidth}{Calculate statistics $\wh{U}_j$ by \eqref{eq:statistics_fdr}.}
        \STATE \hspace{1.5em} \parbox[t]{0.85\linewidth}{\textbf{End For}}
        \STATE (6) \parbox[t]{0.85\linewidth}{Sort $\left|\wh{U}_j\right|$'s from the smallest to the largest as $t_1,\cdots,t_p$. Define $t_0=0$
        and $t_{p+1}=\infty$.}\vspace{1.5em}
        \STATE (7) \parbox[t]{0.85\linewidth}{Set $k=0$.}
        \STATE (8) \parbox[t]{0.85\linewidth}{\textbf{While} $\frac{1+\#\{j: \wh{U}_j\leqslant -t_k\}}{\#\{j:\wh{U}_j\geqslant t_k\}}>\alpha$ \textbf{and} $k\leqslant p$ \textbf{do}:}
        \STATE \hspace{1.5em} \parbox[t]{0.85\linewidth}{$k=k+1$.}
        \STATE (9) \parbox[t]{0.85\linewidth}{Let $T_\alpha = t_k$ and find the selected set $\widehat{\mathcal{A}}(T_{\alpha})$ by \eqref{eq:selected set by threshold}.}

        \STATE \textbf{Output}: Threshold $T_\alpha$, and the selected set $\widehat{\mathcal{A}}(T_\alpha)$.
    \end{algorithmic}
\end{algorithm}

Next, we introduce a condition and present a theorem demonstrating that the CD-FDR procedure asymptotically controls FDR at the pre-specified level $\alpha \in (0,1)$.

\begin{condition}\label{asmpt:weak dependence}
    Let $p_{0} = \#\{\mathcal{A}^c\}$ be the cardinality of the inactive set, and define $B_j \doteq \II_{\left\{\wh{U}_j < 0\right\}}$ and $S^c \doteq \sum_{j \in \mathcal{A}^c} B_j$. We assume that as $n \to \infty$, $p_{0} \to \infty$ and $\operatorname{Var}[S^c] = o(p_{0}^{2})$.
\end{condition}
Condition~\ref{asmpt:weak dependence} imposes a mild assumption on the dependence structure among inactive features. It holds in most practical scenarios, except in rare cases where the Copula Divergence of inactive features exhibits constant pairwise correlations or forms a finite number of groups with homogeneous within-cluster correlations. This condition is commonly adopted in the literature; see \cite{dai2023false} and references therein.

\begin{theorem}[Asymptotic FDR control]\label{thm:FDR_control} Suppose Conditions \ref{condition:minimum strength} and \ref{asmpt:weak dependence} hold. Let $m\to\infty$ and $m=o(n!)$, as $n\to\infty$. Then, the active set $\wh{\mathcal{A}}(T_{\alpha})$ selected by CD-FDR satisfies
\begin{align*}
    \limsup_{n\rightarrow\infty}\operatorname{FDR}\left(T_{\alpha} \right) \leqslant \alpha, \quad \forall \alpha\in(0,1). 
\end{align*}
\end{theorem}

Theorem~\ref{thm:FDR_control} suggests that \(m\) should not be chosen too large. In addition to increased computational cost, an excessively large \(m\) may exhaust the set of permutations and lead to near-identical empirical measures, thereby compromising the effectiveness
 of  \(\widehat{U}_j\).

\section{Numerical Studies}\label{sec:simulations}
\vspace{-0.1in}

\subsection{Feature Screening Performance}\label{subsec:screen_perform}
\vspace{-0.1in}

In this subsection, we evaluate the empirical performance of the proposed CD-Screen method through several simulated high-dimensional sparse regression models with heterogeneous features. We denote CD-Screen with $r=1$ and $2$ by CD-1 and CD-2, respectively. We use Wasserstein Dependence-based feature screening with $r=1$ and $2$ as two baseline methods, and refer to them as WD-1 and WD-2, respectively. 

Additionally, we compare the performance of CD-Screen with several well-received feature screening methods, including Sure Independence Screening (SIS; \citealp{fan2008sure}), Distance Correlation-based screening (DC-SIS; \citealp{li2012feature}), bias-corrected Distance Correlation-based screening (bcDC-SIS; \citealp{szekely2014partial}), and the Martingale Difference Correlation-based screening (MDC-SIS; \citealp{shao2014martingale}). 


We consider the following three regression models. For each model, we set $n=200$, $p \in \{1000, 2000\}$, and simulate 200 replications.

\begin{description}

    \item[Model 1]\label{exp:screening.1} \textbf{(Linear model): } $Y=5X_1+3X_{12}+4X_{26}+6X_{39}+\sqrt{20}\varepsilon$, where $\varepsilon \overset{\text{i.i.d.}}{\sim} N(0,1)$. This model has $4$ active features.

    \item[Model 2]\label{exp:screening.2} \textbf{(Additive nonlinear model): } $Y=5X_1+2\sin\left(\frac{\pi}{2}X_{2}\right)+2\left|X_{3}\right|+2\exp(5X_{4})+\varepsilon$, where $\varepsilon \overset{\text{i.i.d.}}{\sim} N(0,1)$. This model has $4$ active features.

    \item[Model 3] \label{exp:screening.3}\textbf{(Non-additive nonlinear model): } $Y=4X_1+3\log\left(\left|\frac{X_{2}}{1-X_1}\right|\right)\sin\left(2\pi\left|X_{3}\right|\right)+4\left|X_3\right|\exp\left(5X_{4}+5X_{5}\right)+ \varepsilon$, where $\varepsilon \overset{\text{i.i.d.}}{\sim} N(0,1)$. This model has $5$ active variables.
    
\end{description}

Furthermore, we generate the heterogeneous feature vector  $\mathbf{X}=\left(X_1, \cdots, X_{p}\right)^{\T}$ as follows. We define $F_0$ as the standard Gaussian distribution, $F_1$ as the Student's $t$ distribution with degrees of freedom 5, $F_2(|x|)\sim\frac{1}{2}\operatorname{Pareto} (k=3, \alpha=1)$ is a symmetric Pareto distribution with shape parameter 3 and scale parameter 1, and $F_3(|x|)\sim\frac{1}{2}\operatorname{Weibull(k=1.5,\lambda=1)}$ is a symmetric Weibull distribution with shape parameter 1.5 and scale parameter 1.  We set $\mathbf{X}^*=\left(X^*_1, \cdots, X^*_{p}\right)^{\T}$ with $X^*_i\sim F_{i\mod 4}$ for $i=1,\ \cdots, \ p$. We then generate $\mathbf{X}=\bSigma^{\frac{1}{2}}\mathbf{X}^*$, where $\bSigma=\left(\sigma_{i j}\right)_{p \times p}$ with $\sigma_{i j}=0.5^{|i-j|}$. We also consider generating $\mathbf{X}\sim MVN(\mathbf{0}_{p}, \bSigma)$ to simulate a homogeneous feature vector. The corresponding simulation results are presented in Appendix A.1 in the supplementary material.

In each replication, we rank the features in descending order according to the screening criteria of the aforementioned eight screening approaches. Then, we record the minimum model size that includes all active features specified in the model. The screening performance is measured by quantiles (at levels $5\%$, $25\%$, $50\%$, $75\%$, $95\%$) of the minimum model size over 200 replications.

\color{black}

The simulation results are reported in Tables~\ref{tab:sim_model_1}--\ref{tab:sim_model_3}. In the sparse linear setting of Model~1, most methods perform well, except those based on Wasserstein Dependence. In the heterogeneous nonlinear setting of Model~2, CD-Screen is the only method that consistently recovers all active features while maintaining a moderate model size. In the more challenging non-additive nonlinear setting of Model~3, competing methods either miss active features or fail to reduce dimensionality effectively, whereas CD-Screen remains robust. The poor performance of Wasserstein Dependence-based methods across all models further illustrates their limitations, as discussed in Section~\ref{sec:WD_measure}. Overall, these results demonstrate the advantages of CD-Screen for model-free screening in high-dimensional heterogeneous data.

The results in Tables~1--3 also reveal an important trade-off of the proposed copula-based screening method. In Model~1, which is a sparse linear model, SIS slightly outperforms CD-Screen. This is expected because SIS is tailored to linear marginal signals and directly uses raw-scale Pearson correlations. By contrast, CD-Screen transforms each variable through its empirical marginal distribution and therefore removes marginal-scale information; this robustness may lead to some efficiency loss when the true relationship is well approximated by a simple linear model. In Models~2 and 3, however, the active variables enter the response through nonlinear and non-additive structures, for which marginal linear correlations can be weak or misleading. Copula Divergence measures distributional dependence after marginal standardization and can therefore capture a broader class of nonlinear relationships. This explains the stronger performance of CD-Screen in the nonlinear settings.

To further assess robustness beyond the main simulation settings, we also conduct an additional experiment with non-additive noise and irregular covariate marginals, including multimodal, bounded, and rounded distributions. The results, reported in Appendix~A.1, show that CD-Screen continues to recover the active features with very small model sizes, further supporting its robustness in heterogeneous high-dimensional settings.

\color{black}

\begin{table}[htbp]
\centering
\setlength{\tabcolsep}{3pt} 
\caption{{\bf Simulation results for Model 1}: Quantiles of minimum model size that include all 4 active features over 200 replications. }\label{tab:sim_model_1}
\small
\begin{tabular}{lccccc ccccc}
\toprule
 & \multicolumn{5}{c}{$p=1000$} & \multicolumn{5}{c}{$p=2000$}\\
 \cmidrule(lr){2-6} \cmidrule(lr){7-11}
 Quantile & 5\% & 25\%& 50\%& 75\%  & 95\%   & 5\% & 25\%& 50\%& 75\%  & 95\% \\
\midrule
SIS      & 4.00 &  6.00 &  8.00 & 15.00 & 48.60
         & 4.00  &  6.00 &  8.00 &   17.00 & 98.20\\
DC-SIS   & 4.00 &  6.00 &  9.00 & 16.00 & 108.15
         & 4.00  &  5.00 &   9.00 & 22.00 & 162.80\\
bcDC-SIS & 4.00  & 6.00 &  9.00 & 16.00 & 118.20
         & 4.00 &   5.00 &   9.00 &  23.00 & 178.85\\
MDC-SIS  & 4.00 &  6.00 &  8.00&  16.25 & 93.20
         & 4.00   & 6.00  &  9.00 & 20.00 & 128.10\\
WD-1     & 370.70 & 559.75 & 706.00 & 868.75 & 957.55
         & 859.10 & 1183.00 & 1441.00 & 1688.75 & 1922.15\\
WD-2     & 403.95 & 557.25 & 693.5 & 848.00 & 957.10
         & 902.90 & 1145.00 & 1427.50 & 1680.25 & 1929.50\\
CD-1     & 4.00  & 6.00 & 10.00 & 26.00 & 241.05
         & 4.00 &  6.00 &  10.00 &  43.25 & 372.15\\
CD-2     & 4.00 & 6.75 &  11.00 &  38.00 & 294.30
         & 4.00  &  6.00 &  12.50 & 65.50 & 469.35\\
\bottomrule
\end{tabular}
\end{table}

\begin{table}[htbp]
\centering
\setlength{\tabcolsep}{3pt} 
\caption{{\bf Simulation results for Model 2}: Quantiles of minimum model size that include all 4 active features over 200 replications. }\label{tab:sim_model_2}
\small
\begin{tabular}{lccccc ccccc}
\toprule
 & \multicolumn{5}{c}{$p=1000$} & \multicolumn{5}{c}{$p=2000$}\\
 \cmidrule(lr){2-6} \cmidrule(lr){7-11}
 Quantile & 5\% & 25\%& 50\%& 75\%  & 95\%   & 5\% & 25\%& 50\%& 75\%  & 95\% \\
\midrule
SIS      & 58.70 & 390.50 & 663.50 & 835.75 & 970.00
         & 139.60 & 654.50  & 1299.00 & 1703.25 & 1962.10\\
DC-SIS   & 55.60 &254.50 & 487.00 & 709.50 & 895.50
         & 109.55 & 503.75 & 1054.00 & 1499.75 & 1857.70\\
bcDC-SIS & 25.90 & 165.75 & 414.50 & 674.50 & 957.10
         & 101.55 & 442.50 & 993.00 & 1439.50 & 1837.50\\
MDC-SIS  & 38.80 & 310.50 & 614.00 & 787.00 & 992.50
         & 108.00 & 728.50 & 1196.50 & 1636.00 & 1953.10\\
WD-1     & 285.95 & 356.75 & 443.50 & 549.25 & 743.25
         & 573.35 & 720.75 & 868.00 & 1129.00 & 1481.10\\
WD-2     & 452.50 & 669.00 & 851.00 & 932.00 & 989.15
         & 886.85 & 1358.50 & 1633.50 & 1808.25 & 1973.15\\
CD-1     & 4.00 & 4.00 & 4.00 & 5.00 & 5.00
         & 4.00 & 4.00 & 4.00 & 5.00 & 5.00\\
CD-2     & 4.00 & 4.00 & 4.00 & 4.00 & 5.00
         & 4.00 & 4.00 & 4.00 & 5.00 & 5.00\\
\bottomrule
\end{tabular}
\end{table}

\begin{table}[htbp]
\centering
\setlength{\tabcolsep}{3pt} 
\caption{{\bf Simulation results for Model 3}: Quantiles of minimum model size that include all 5 active features over 200 replications. }\label{tab:sim_model_3}
\small
\begin{tabular}{lccccc ccccc}
\toprule
 & \multicolumn{5}{c}{$p=1000$} & \multicolumn{5}{c}{$p=2000$}\\
 \cmidrule(lr){2-6} \cmidrule(lr){7-11}
 Quantile & 5\% & 25\%& 50\%& 75\%  & 95\%   & 5\% & 25\%& 50\%& 75\%  & 95\% \\
\midrule
SIS      & 101.75 & 448.00 & 721.50 & 852.50 & 978.15
         & 359.90 &1098.00 & 1470.50 & 1753.50 & 1947.05\\
DC-SIS   & 95.10 & 432.25 & 696.50 & 840.75 & 961.05
         & 288.55 & 1027.00 & 1463.00 & 1685.50 &1957.10\\
bcDC-SIS & 129.80 & 432.25 & 696.00 & 840.75 & 961.05
         & 359.25 & 1004.00 & 1335.50 & 1695.75 & 1946.05\\
MDC-SIS  & 106.85 & 550.50& 726.50 & 883.50 & 982.05
         & 400.25 & 1081.25 & 1434.00 & 1755.25 & 1972.00\\
WD-1     & 545.65 & 775.75 & 877.50 & 943.75 & 982.05
         & 1060.20 & 1518.75 & 1759.00 & 1907.50 & 1986.10\\
WD-2     & 549.50 & 751.75 & 876.50 & 947.25 & 986.05
         & 1129.85 & 1562.50 & 1766.00 & 1893.75 & 1979.05\\
CD-1     & 5.00 & 5.00 & 6.00 & 6.00 & 7.00
         & 5.00 & 5.00 & 6.00 & 6.00 & 7.00\\
CD-2     & 5.00 & 5.00 & 6.00 & 6.00 & 7.00
         & 5.00 & 5.00 & 6.00 & 6.00 & 6.05\\
\bottomrule
\end{tabular}
\end{table}

\subsection{FDR Control Performance}\label{subsec:fdr_perform}
\vspace{-0.1in}

In this subsection, we evaluate the finite-sample performance of the proposed CD-FDR method, as outlined in Algorithm \ref{algo:CD-FDR}. We set $m=3000$ and aim to control FDR at pre-specified levels $\alpha \in \{0.15, 0.20, 0.25, 0.30\}$. For each simulated model, we consider sample size $n=400$ and dimensionality $p \in \{1000, 2000\}$, conducting $100$ replications.

We design a challenging non-additive and nonlinear high-dimensional sparse regression model as Model 4 below. The heterogeneous feature vector $\mathbf{X}$ is generated following the procedure described in Section \ref{subsec:screen_perform}, with the exception that we set $\bSigma = \bI_p$. Additional simulation results for a linear model are provided in Appendix A.2 in the supplementary material.

\begin{description}

    \item[Model 4] \label{exp:FDR.3}\textbf{(Non-additive nonlinear model): } $Y=6X_1+10\exp(0.5X_2)+7X_3^2+7\log\left(0.5\left|\frac{X_{4}}{1-X_4}\right|\right)+6X_5+10\exp(0.5X_6)+7X_7^2+7\log\left(0.5\left|\frac{X_{8}}{1-X_8}\right|\right)+\varepsilon$, where $\varepsilon \overset{\text{i.i.d.}}{\sim} N(0,1)$. This model has $8$ active features.
    
\end{description}

For each replication, we implement the proposed CD-FDR procedure to screen active features while controlling the FDR at the designated levels. For comparison, we consider two competing methods. The first method, denoted by CD-BH, employs the same Copula Divergence statistic but replaces the FDR control procedure in Section~\ref{sec:CD_FDR} with the classical Benjamini--Hochberg method \citep{benjamini1995controlling}. This provides a baseline for assessing the advantages of our proposed FDR control strategy. The second method, denoted by PC-FDR, is the model-free feature screening approach with data-driven FDR control proposed by \cite{liu2022model}, representing a state-of-the-art model-free feature screening method with FDR control.

We compute the false discovery proportion (FDP) for each replication as the ratio of false discoveries to total discoveries. The FDR is then estimated by averaging the FDPs over the $100$ replications, and the estimate is denoted by $\widehat{\mathrm{FDR}}$. The screening performance is further evaluated by the median selected model size ($|\widehat{\mathcal{S}}|$) over 100 replications, and by the proportion of replications in which each active feature is selected ($\mathcal{P}_i$, $i=1,\dots,8$).

The simulation results, summarized in Tables~\ref{tab:FDR_results_1000_5} and \ref{tab:FDR_results_2000_5}, demonstrate that the proposed CD-FDR method effectively controls the FDR at the pre-specified levels across  all scenarios. Moreover, CD-FDR consistently identifies active features in the majority of replications, yielding high selection accuracy and strong empirical screening performance under high-dimensional and heterogeneous conditions.

In contrast, both CD-BH and PC-FDR fail to control the FDR below the designated levels when $\alpha$ is small and/or when the dimensionality $p$ is large. This underperformance is expected, as neither competing method is specifically designed for heterogeneous high-dimensional data. When heterogeneity is present, the null $p$-values are no longer uniformly distributed, violating the assumptions required for valid BH control. For PC-FDR, although knockoff features are introduced to aid FDR control, the construction of valid knockoffs relies on a second-order approximation that presumes homogeneous covariance structures. When this assumption is violated, the resulting knockoff statistics no longer provide reliable symmetry properties, leading to inflated FDR in heterogeneous settings.

\begin{table}[htbp]
\centering
\caption{{\bf Simulation results for Model 4 when $p=1000$.}}
\label{tab:FDR_results_1000_5}
\footnotesize
\begin{tabular}{c c c c c c c c c c c}
\toprule
$\alpha$ & $|\widehat S|$ & $P_1$ & $P_2$ & $P_3$ & $P_4$ & $P_5$ & $P_6$ & $P_7$ & $P_8$ & $\widehat{\mathrm{FDR}}$ \\
\midrule

\multicolumn{11}{c}{\textbf{CD-FDR with $r=1$}} \\
0.15 & 9   & 0.79 & 0.81 & 0.76 & 0.80 & 0.80 & 0.81 & 0.76 & 0.80 & 0.16 \\
0.20 & 9.5 & 0.85 & 0.88 & 0.79 & 0.85 & 0.88 & 0.88 & 0.79 & 0.87 & 0.20 \\
0.25 & 10  & 0.93 & 0.95 & 0.90 & 0.92 & 0.93 & 0.95 & 0.91 & 0.94 & 0.25 \\
0.30 & 11  & 0.94 & 0.96 & 0.92 & 0.95 & 0.96 & 0.96 & 0.91 & 0.95 & 0.30 \\

\midrule
\multicolumn{11}{c}{\textbf{CD-FDR with $r=2$}} \\
0.15 & 8   & 0.71 & 0.71 & 0.67 & 0.61 & 0.64 & 0.71 & 0.61 & 0.67 & 0.13 \\
0.20 & 8.5 & 0.77 & 0.77 & 0.71 & 0.67 & 0.69 & 0.77 & 0.67 & 0.73 & 0.18 \\
0.25 & 10  & 0.92 & 0.92 & 0.84 & 0.78 & 0.84 & 0.92 & 0.87 & 0.88 & 0.26 \\
0.30 & 11  & 0.95 & 0.95 & 0.90 & 0.89 & 0.89 & 0.95 & 0.88 & 0.90 & 0.29 \\

\midrule
\multicolumn{11}{c}{\textbf{CD-BH with $r=1$}} \\
0.15 & 9  & 0.97 & 1.00 & 0.90 & 0.97 & 0.99 & 1.00 & 0.92 & 0.99 & 0.16 \\
0.20 & 10 & 0.97 & 1.00 & 0.96 & 0.97 & 0.99 & 1.00 & 0.93 & 0.99 & 0.22 \\
0.25 & 10 & 0.99 & 1.00 & 0.96 & 0.99 & 0.99 & 1.00 & 0.94 & 0.99 & 0.25 \\
0.30 & 11 & 0.99 & 1.00 & 0.99 & 0.99 & 0.99 & 1.00 & 0.96 & 0.99 & 0.29 \\

\midrule
\multicolumn{11}{c}{\textbf{CD-BH with $r=2$}} \\
0.15 & 9   & 0.99 & 1.00 & 0.89 & 0.91 & 0.89 & 0.99 & 1.00 & 0.93 & 0.17 \\
0.20 & 9   & 0.99 & 1.00 & 0.92 & 0.91 & 0.91 & 0.99 & 1.00 & 0.93 & 0.21 \\
0.25 & 10  & 0.99 & 1.00 & 0.93 & 0.92 & 0.93 & 0.99 & 1.00 & 0.95 & 0.25 \\
0.30 & 11  & 1.00 & 1.00 & 0.96 & 0.93 & 0.95 & 1.00 & 1.00 & 0.96 & 0.31 \\

\midrule
\multicolumn{11}{c}{\textbf{PC-FDR}} \\
0.15 & 9  & 0.88 & 0.88 & 0.82 & 0.87 & 0.88 & 0.88 & 0.84 & 0.88 & 0.21 \\
0.20 & 10 & 0.89 & 0.91 & 0.84 & 0.89 & 0.90 & 0.91 & 0.86 & 0.90 & 0.23 \\
0.25 & 10 & 0.89 & 0.92 & 0.84 & 0.89 & 0.91 & 0.92 & 0.87 & 0.91 & 0.24 \\
0.30 & 10 & 0.89 & 0.92 & 0.84 & 0.89 & 0.91 & 0.92 & 0.87 & 0.87 & 0.25 \\

\bottomrule
\end{tabular}
\end{table}

\begin{table}[htbp]
\centering
\caption{{\bf Simulation results for Model 4 when $p=2000$.}}
\label{tab:FDR_results_2000_5}
\footnotesize
\begin{tabular}{c c c c c c c c c c c}
\toprule
$\alpha$ & $|\widehat S|$ & $P_1$ & $P_2$ & $P_3$ & $P_4$ & $P_5$ & $P_6$ & $P_7$ & $P_8$ & $\widehat{\mathrm{FDR}}$ \\
\midrule

\multicolumn{11}{c}{\textbf{CD-FDR with $r=1$}} \\
0.15 &  9  & 0.69 & 0.70 & 0.65 & 0.68 & 0.68 & 0.70 & 0.65 & 0.69 & 0.16 \\
0.20 & 10  & 0.72 & 0.74 & 0.71 & 0.70 & 0.73 & 0.74 & 0.71 & 0.74 & 0.20 \\
0.25 & 10  & 0.84 & 0.85 & 0.82 & 0.84 & 0.84 & 0.85 & 0.81 & 0.82 & 0.28 \\
0.30 & 11  & 0.85 & 0.90 & 0.86 & 0.89 & 0.90 & 0.90 & 0.85 & 0.89 & 0.30 \\

\midrule
\multicolumn{11}{c}{\textbf{CD-FDR with $r=2$}} \\
0.15 & 7.5 & 0.57 & 0.58 & 0.53 & 0.50 & 0.58 & 0.58 & 0.53 & 0.49 & 0.13 \\
0.20 &  8  & 0.64 & 0.65 & 0.60 & 0.55 & 0.65 & 0.65 & 0.60 & 0.57 & 0.16 \\
0.25 &  9  & 0.83 & 0.84 & 0.78 & 0.71 & 0.84 & 0.84 & 0.78 & 0.77 & 0.23 \\
0.30 & 10  & 0.85 & 0.86 & 0.81 & 0.75 & 0.85 & 0.86 & 0.78 & 0.78 & 0.26 \\

\midrule
\multicolumn{11}{c}{\textbf{CD-BH with $r=1$}} \\
0.15 & 10 & 0.98 & 1.00 & 0.93 & 0.97 & 0.95 & 1.00 & 0.91 & 0.98 & 0.20 \\
0.20 & 10 & 0.98 & 1.00 & 0.94 & 0.97 & 0.99 & 1.00 & 0.91 & 0.99 & 0.25 \\
0.25 & 11 & 0.99 & 1.00 & 0.96 & 0.98 & 0.98 & 1.00 & 0.95 & 0.98 & 0.30 \\
0.30 & 11 & 0.97 & 1.00 & 0.97 & 0.96 & 0.98 & 1.00 & 0.95 & 0.98 & 0.30 \\

\midrule
\multicolumn{11}{c}{\textbf{CD-BH with $r=2$}} \\
0.15 &  9   & 0.99 & 1.00 & 0.93 & 0.82 & 0.99 & 1.00 & 0.89 & 0.85 & 0.17 \\
0.20 &  9   & 0.97 & 1.00 & 0.91 & 0.86 & 1.00 & 1.00 & 0.90 & 0.86 & 0.20 \\
0.25 & 10   & 1.00 & 1.00 & 0.95 & 0.87 & 1.00 & 1.00 & 0.91 & 0.91 & 0.25 \\
0.30 & 10.5 & 1.00 & 1.00 & 0.95 & 0.87 & 1.00 & 1.00 & 0.93 & 0.89 & 0.30 \\

\midrule
\multicolumn{11}{c}{\textbf{PC-FDR}} \\
0.15 & 10.5 & 0.93 & 0.94 & 0.88 & 0.94 & 0.92 & 0.94 & 0.88 & 0.94 & 0.29 \\
0.20 & 12   & 0.94 & 0.95 & 0.89 & 0.95 & 0.93 & 0.95 & 0.89 & 0.95 & 0.31 \\
0.25 & 12   & 0.94 & 0.95 & 0.89 & 0.95 & 0.94 & 0.96 & 0.89 & 0.95 & 0.31 \\
0.30 & 12   & 0.94 & 0.95 & 0.89 & 0.95 & 0.96 & 0.96 & 0.89 & 0.95 & 0.31 \\

\bottomrule
\end{tabular}
\end{table}

\subsection{Real Data Analysis}
\label{sec:appl}
\vspace{-0.1in}

This study aims to identify a sparse subset of stocks from the S\&P 500 index whose monthly returns exhibit {notable marginal associations} with the U.S. inflation rate from January 2020 to December 2023. This period was marked by significant inflation volatility due to the COVID-19 pandemic and subsequent recovery policies \citep{jaltuszyk2022inflation, ball2022understanding}. Understanding inflation dynamics during this time is vital, given its broad impact on economic stability, monetary policy, consumer behavior, and market confidence. Stock returns reflect investor expectations and market sentiment, making them sensitive indicators of macroeconomic changes, including inflation \citep{thorbecke2020impact}. Financial and consumer sectors, in particular, often respond sharply to shifts in inflation expectations. However, stock returns are inherently heterogeneous due to firm-specific and industry-level differences, posing challenges for traditional statistical analyses that assume specific regression models and homogeneous features. Such models risk bias and poor inference under mis-specification. A model-free feature screening approach can address these issues, and enable the reliable identification of influential stocks without restrictive assumptions.

In this study, the response variable is the monthly inflation rate in the United States from January 2020 to December 2023 \footnote{The inflation rate data is publicly available at Federal Reserve Economic Data: \url{https://fred.stlouisfed.org}. Inflation is computed from the monthly CPI data (FRED series ID: CPIAUCNS) using month-to-month percentage changes. }.  The explanatory variables consist of monthly stock returns for 489 stocks that remained continuously listed in the S\&P 500 index throughout this period \footnote{The S\&P 500 data is publicly available at \url{https://finance.yahoo.com/quote/\%5EGSPC/history}. }. To facilitate the interpretation of results, these stocks are categorized into 11 sectors, as detailed in Appendix B in the supplementary material.

\begin{figure}[htbp]
    \centering
    \begin{minipage}{0.425\textwidth} 
        \centering\includegraphics[width=\textwidth]{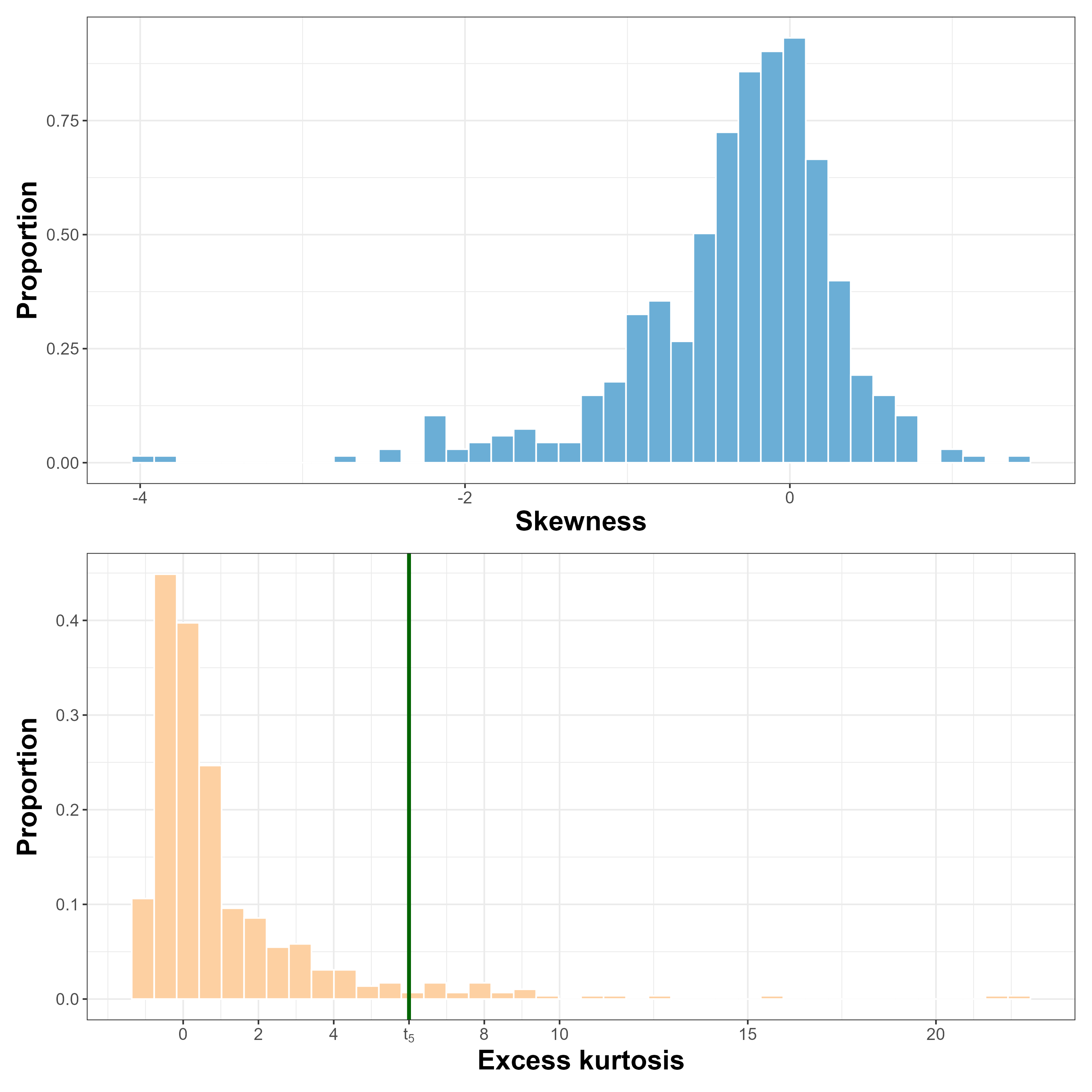}
    \end{minipage}
    \begin{minipage}{0.525\textwidth} 
        \centering
        \includegraphics[width=\textwidth]{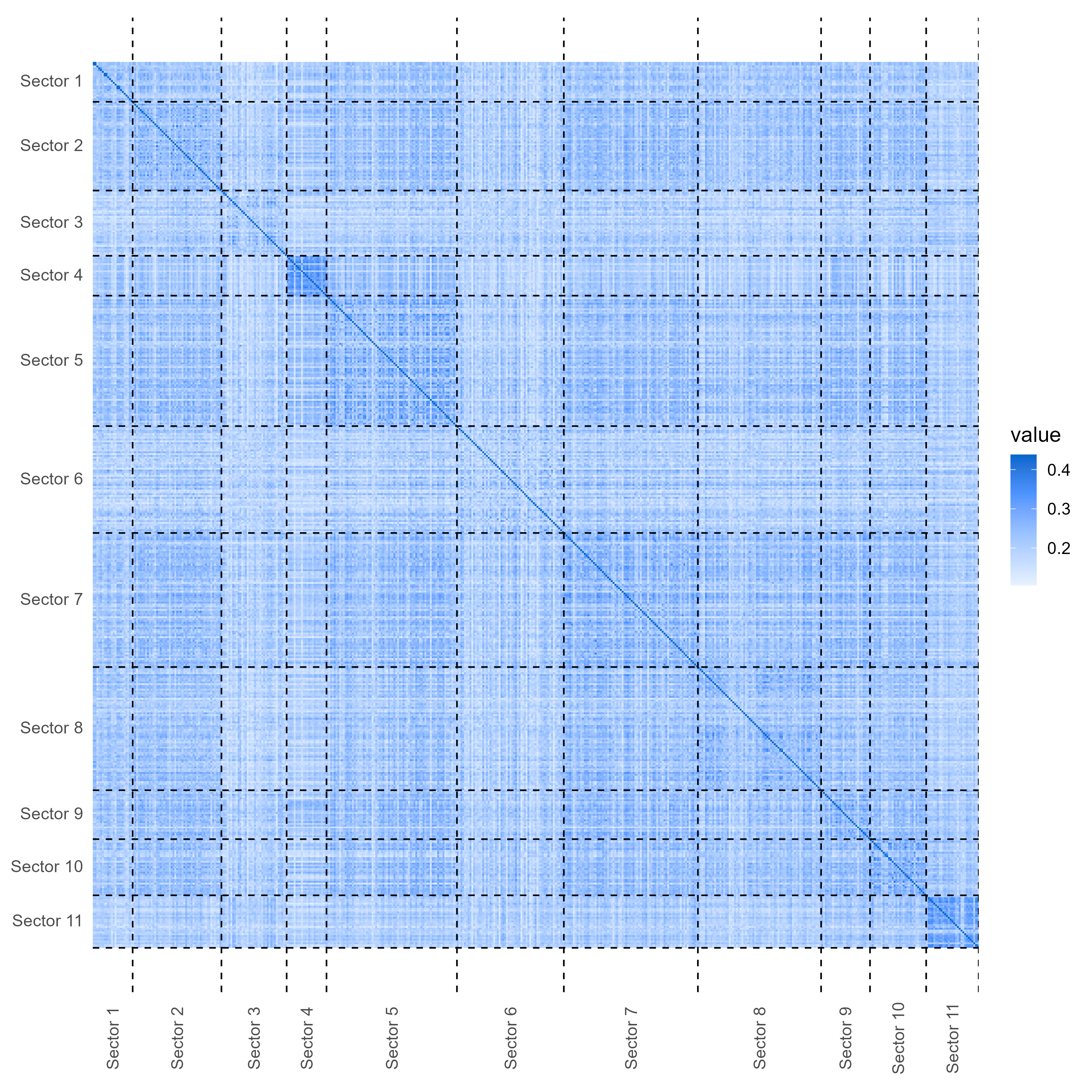}
    \end{minipage}
    
    \caption{The left panel presents the histograms of skewness and excess kurtosis for the monthly log returns of 489 stocks, which highlights the heterogeneity in the data. The right panel visualizes the pairwise dependence among features (sorted by sectors).}\label{fig:hist_cor}
\end{figure}

We examine the distributional properties of the stock returns by computing sample skewness and excess kurtosis for each feature. The histograms in Figure~\ref{fig:hist_cor} illustrate these characteristics, indicating that most stock returns exhibit notable skewness and positive excess kurtosis. This finding highlights the significant heterogeneity in the data, with many features deviating substantially from a Gaussian distribution. Additionally, we assess dependencies among features by calculating pairwise Copula Divergence with $r = 2$. {Although $r=1$ is recommended as a stable default for small or moderate samples, we use $r=2$ in this illustrative application because the goal is to explore potentially stronger and nonlinear marginal dependence patterns in a highly heterogeneous financial dataset.} The resulting dependency structure, visualized in the right panel of Figure~\ref{fig:hist_cor}, reveals strong interdependencies among stock returns, particularly within sectors. In summary, the dataset includes a sample size of $n = 48$ months and $p = 489$ features. The features are characterized by considerable heterogeneity and strong dependence among features.

We apply the CD-FDR procedure with $r = 2$ and $m = 360$ to identify stocks whose marginal dependence measures are strongly associated with the inflation rate, targeting an FDR level of $\alpha = 0.20$. To account for sampling variability in estimating the empirical null distribution of the Copula Divergence, we repeat the procedure 200 times. The average number of selected stocks across these repetitions is approximately $6.125$, which motivates us to select the top six stocks ranked by their Copula Divergence with inflation. Additionally, we report the top ten stocks within each sector in Appendix~B of the supplementary material.

\begin{figure}
    \centering
    \includegraphics[width=\linewidth]{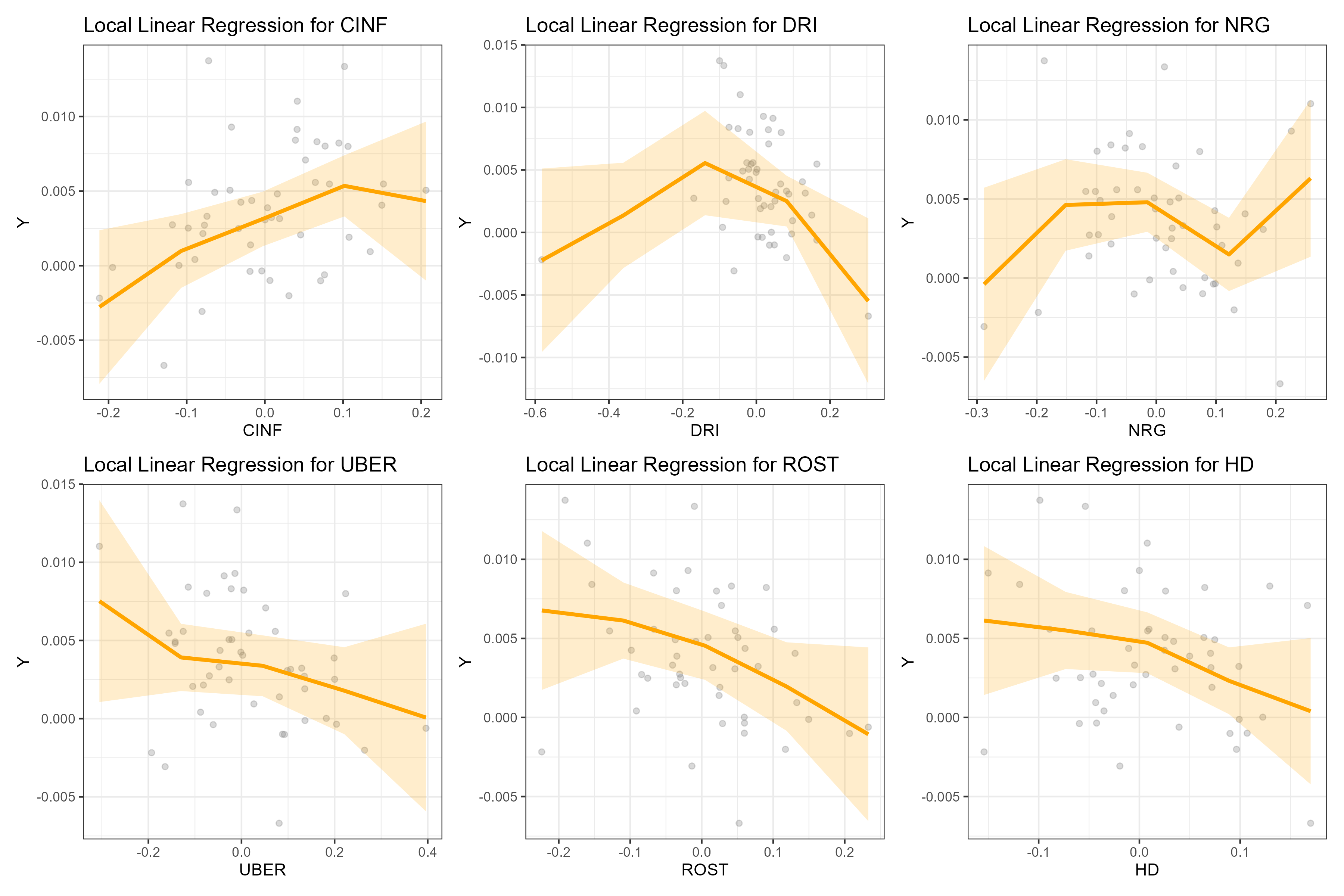}
    \caption{Fitted local linear regressions between monthly inflation rate and monthly log return of 6 selected stocks. Each subplot represents a different stock. The orange lines depict the fitted local linear regression, and the shaded areas represent the 95\% confidence intervals.}
    \label{fig:inf_llr}
\end{figure}

To further examine the marginal regression relationship of the selected features, we fit univariate local linear regressions \citep{fan1992variable, fan1993local} of the response on each selected stock return. The bandwidth is chosen by 5-fold cross-validation. The resulting fitted curves and confidence intervals are presented in Figure~\ref{fig:inf_llr}, providing additional insight into the nature of these marginal associations.

{We emphasize that this real-data analysis is intended primarily as an illustration of the proposed screening procedure in a high-dimensional heterogeneous setting. Because the sample covers a relatively short time period and the analysis is descriptive, the selected stocks should be interpreted as variables exhibiting strong marginal associations with inflation during this period, rather than as providing definitive empirical conclusions or causal explanations. The economic discussion below is therefore meant to provide context for the selected variables, not to establish formal economic relationships.}

The six selected stocks are Cincinnati Financial (CINF), Darden Restaurants (DRI), NRG Energy (NRG), Uber Technologies (UBER), Ross Stores (ROST), and Home Depot (HD). They all {exhibit notable marginal associations with U.S. inflation}, each influenced by distinct economic factors. 

Darden Restaurants (DRI), Ross Stores (ROST), and Home Depot (HD), all among the top Consumer Discretionary stocks, show strong associations with inflation, reflecting the sector’s exposure to fluctuations in household consumption, cost pass-through dynamics, and broader macroeconomic conditions affecting retail and services \citep{dekimpe2023retailing}. Our method consistently highlights Consumer Discretionary companies, with several of the top-ranked stocks belonging to this sector. These include Aptiv (APTV), which reflects demand in housing renovation and automotive supply chains; and travel-related firms such as Royal Caribbean (RCL) and Expedia Group (EXPE), whose performances are tightly linked to post-pandemic recovery and inflation-induced shifts in household budgets. Homebuilders like NVR (NVR) and Lennar (LEN) similarly appear at the top, consistent with the sector’s responsiveness to interest rate cycles and broader macroeconomic conditions.

Cincinnati Financial (CINF), from the Financials sector, shows a positive correlation with inflation due to rising interest rates and increased demand for life insurance during the pandemic \citep{li2024impact}. In contrast, Uber Technologies (UBER), classified under Industrials but reliant on discretionary spending, exhibits negative relationships with inflation. Higher inflation reduces consumers' disposable income and confidence, dampening demand for non-essential goods and services. These effects have become more pronounced in the post-COVID period.

{Thus, the selected stocks provide descriptive evidence of heterogeneous sector-level associations with inflation during the study period. These results illustrate how the proposed model-free screening and FDR control procedure can identify interpretable marginal relationships in high-dimensional, heterogeneous datasets.}

\color{black}
To further evaluate the practical impact of the proposed screening procedure, we conduct an additional downstream prediction comparison using the real-data example. The results, reported in Appendix~A.3, show that applying CD-FDR before fitting a random forest model leads to lower test prediction error than using all features or features selected by SIS, while retaining a smaller and more interpretable set of predictors.

\color{black}

\color{black}

\section{Conclusion and Discussion}
\label{sec:conclusion}

In this paper, we introduce a model-free feature screening framework for high-dimensional heterogeneous data. The proposed framework addresses limitations of existing screening methods, including restrictive modeling assumptions, sensitivity to marginal distributions, and the double-use-of-data concern for both estimation and inference. We propose CD-Screen, which ranks features using Copula Divergence, and CD-FDR, an adaptive procedure for false discovery rate control. We establish key statistical properties of Copula Divergence and show that CD-Screen satisfies sure screening and rank consistency, while CD-FDR asymptotically controls the FDR at a desired level. Simulations demonstrate strong finite-sample performance under various settings. An economic case study on stock returns and U.S. inflation from 2020 to 2023 further illustrates the practical utility of the proposed methods.

The robustness of CD-Screen comes from its copula-based construction, which removes marginal-scale information and focuses on dependence structure. As a result, when covariates are homogeneous, light-tailed, and well-behaved, and when the relationship is close to linear, specialized methods such as SIS may be more statistically efficient. Rank-based copula estimation may also have non-negligible finite-sample variability with small samples, ties, rounded measurements, and tail-concentrated dependence. Thus, CD-Screen and CD-FDR should be viewed as robust model-free tools for heterogeneous high-dimensional data, rather than uniformly superior replacements for methods tailored to simpler homogeneous models.

\color{black}

\noindent{\textbf{Acknowledgments}} \\
The authors express their gratitude to the editor, associate editor, and two anonymous reviewers for their constructive comments.

\noindent{\textbf{Disclosure Statement}} \\
The authors report that there are no competing interests to declare.

\noindent{\textbf{Funding}} \\
Yuan Ke's research was supported by the U.S. National Science Foundation through grants 2210468, 2243044, 2324389, and 2514399, as well as the U.S. National Institutes of Health through grant R01 HL172291. Runze Li's research was supported by an U.S. National Science Foundation grant DMS 2514400 and two U.S. National Institutes of Health grants R01 GM163244 and R01 AI192205. The content is solely the responsibility of the authors and does not necessarily represent the official views of the NSF or the NIH.

\setstretch{1.4}
\bibliography{ref.bib}

@inproceedings{sklar1959fonctions,
  title={Fonctions de r{\'e}partition {\`a} n dimensions et leurs marges},
  author={Sklar, M},
  booktitle={Annales de l'ISUP},
  volume={8},
  number={3},
  pages={229--231},
  year={1959}
}

@article{fan2008sure,
author = {Fan, Jianqing and Lv, Jinchi},
title = {Sure independence screening for ultrahigh dimensional feature space},
journal = {Journal of the Royal Statistical Society: Series B (Statistical Methodology)},
volume = {70},
number = {5},
pages = {849-911},
doi = {https://doi.org/10.1111/j.1467-9868.2008.00674.x},
year = {2008}
}

@article{li2012feature,
  title={Feature screening via distance correlation learning},
  author={Li, Runze and Zhong, Wei and Zhu, Liping},
  journal={Journal of the American Statistical Association},
  volume={107},
  number={499},
  pages={1129--1139},
  year={2012},
  publisher={Taylor \& Francis}
}

@article{liu2022model,
  title={Model-free feature screening and FDR control with knockoff features},
  author={Liu, Wanjun and Ke, Yuan and Liu, Jingyuan and Li, Runze},
  journal={Journal of the American Statistical Association},
  volume={117},
  number={537},
  pages={428--443},
  year={2022},
  publisher={Taylor \& Francis}
}

@article{mai2023coordinatewise,
author = {Qing Mai and Di He and Hui Zou},
title = {Coordinatewise Gaussianization: Theories and Applications},
journal = {Journal of the American Statistical Association},
volume = {118},
number = {544},
pages = {2329--2343},
year = {2023},
publisher = {ASA Website},
doi = {10.1080/01621459.2022.2044825},
URL = {
        https://doi.org/10.1080/01621459.2022.2044825
},
eprint = { 
        https://doi.org/10.1080/01621459.2022.2044825
}
}

@inbook{McDiarmid_1989, place={Cambridge}, series={London Mathematical Society Lecture Note Series}, title={On the method of bounded differences}, booktitle={Surveys in Combinatorics, 1989: Invited Papers at the Twelfth British Combinatorial Conference}, publisher={Cambridge University Press}, author={McDiarmid, Colin}, editor={Siemons, J.Editor}, year={1989}, pages={148–188}, collection={London Mathematical Society Lecture Note Series}}

@article{abramovich2006adapting,
author = {Felix Abramovich and Yoav Benjamini and David L. Donoho and Iain M. Johnstone},
title = {{Adapting to unknown sparsity by controlling the false discovery rate}},
volume = {34},
journal = {The Annals of Statistics},
number = {2},
publisher = {Institute of Mathematical Statistics},
pages = {584 -- 653},
year = {2006},
doi = {10.1214/009053606000000074},
URL = {https://doi.org/10.1214/009053606000000074}
}

@article{shao2014martingale,
author = {Xiaofeng Shao and Jingsi Zhang},
title = {Martingale Difference Correlation and Its Use in High-Dimensional Variable Screening},
journal = {Journal of the American Statistical Association},
volume = {109},
number = {507},
pages = {1302--1318},
year = {2014},
publisher = {ASA Website},
doi = {10.1080/01621459.2014.887012}
}

@article{fournier2015rate,
  title={On the rate of convergence in Wasserstein distance of the empirical measure},
  author={Fournier, Nicolas and Guillin, Arnaud},
  journal={Probability Theory and Related Fields},
  volume={162},
  number={3},
  pages={707--738},
  year={2015},
  publisher={Springer}
}

@article{weed2019sharp,
  title={Sharp asymptotic and finite-sample rates of convergence of empirical measures in Wasserstein distance},
  author={Weed, Jonathan and Bach, Francis},
  journal={Bernoulli},
  volume={25},
  number={4A},
  pages={2620--2648},
  year={2019},
  publisher={JSTOR}
}

@article{del2019central,
author = {Eustasio del Barrio and Jean-Michel Loubes},
title = {{Central limit theorems for empirical transportation cost in general dimension}},
volume = {47},
journal = {The Annals of Probability},
number = {2},
publisher = {Institute of Mathematical Statistics},
pages = {926 -- 951},
year = {2019},
doi = {10.1214/18-AOP1275}
}

@article{barber2015controlling,
author = {Rina Foygel Barber and Emmanuel J. Cand{\`e}s},
title = {{Controlling the false discovery rate via knockoffs}},
volume = {43},
journal = {The Annals of Statistics},
number = {5},
publisher = {Institute of Mathematical Statistics},
pages = {2055 -- 2085},
year = {2015},
doi = {10.1214/15-AOS1337}
}

@article{szekely2014partial,
author = {G{\'a}bor J. Sz{\'e}kely and Maria L. Rizzo},
title = {{Partial distance correlation with methods for dissimilarities}},
volume = {42},
journal = {The Annals of Statistics},
number = {6},
publisher = {Institute of Mathematical Statistics},
pages = {2382 -- 2412},
year = {2014},
doi = {10.1214/14-AOS1255}
}

@article{rippl2016limit,
  title={Limit laws of the empirical Wasserstein distance: Gaussian distributions},
  author={Rippl, Thomas and Munk, Axel and Sturm, Anja},
  journal={Journal of Multivariate Analysis},
  volume={151},
  pages={90--109},
  year={2016},
  publisher={Elsevier}
}

@inproceedings{del2024central,
  title={Central limit theorems for general transportation costs},
  author={del Barrio, Eustasio and Gonz{\'a}lez-Sanz, Alberto and Loubes, Jean-Michel},
  booktitle={Annales de l'Institut Henri Poincare (B) Probabilites et statistiques},
  volume={60},
  number={2},
  pages={847--873},
  year={2024},
  organization={Institut Henri Poincar{\'e}}
}

@article{lei2020convergence,
  title={Convergence and concentration of empirical measures under Wasserstein distance in unbounded functional spaces},
  author={Lei, Jing},
  journal={Bernoulli},
  volume={26},
  number={1},
  pages={767--798},
  year={2020}
}

@article{ozair2019wasserstein,
  title={Wasserstein dependency measure for representation learning},
  author={Ozair, Sherjil and Lynch, Corey and Bengio, Yoshua and Van den Oord, Aaron and Levine, Sergey and Sermanet, Pierre},
  journal={Advances in Neural Information Processing Systems},
  volume={32},
  year={2019}
}

@article{de2025high,
  title={High-dimensional copula-based Wasserstein dependence},
  author={De Keyser, Steven and Gijbels, Ir{\`e}ne},
  journal={Computational Statistics \& Data Analysis},
  volume={204},
  pages={108096},
  year={2025},
  publisher={Elsevier}
}

@article{mordant2022measuring,
  title={Measuring dependence between random vectors via optimal transport},
  author={Mordant, Gilles and Segers, Johan},
  journal={Journal of Multivariate Analysis},
  volume={189},
  pages={104912},
  year={2022},
  publisher={Elsevier}
}

@article{nies2022d,
  title={Transport Dependency: Optimal Transport Based Dependency Measures},
  author={Nies, Thomas Giacomo and Staudt, Thomas and Munk, Axel},
  journal={Contributions to the Theory of Statistical Optimal Transport},
  pages={137},
  year={2022},
  publisher={University of G{\"o}ttingen}
}

@article{hartmann2020semi,
  title={Semi-discrete optimal transport: a solution procedure for the unsquared Euclidean distance case},
  author={Hartmann, Valentin and Schuhmacher, Dominic},
  journal={Mathematical Methods of Operations Research},
  volume={92},
  number={1},
  pages={133--163},
  year={2020},
  publisher={Springer}
}

@article{dieci2024solving,
  title={Solving semi-discrete optimal transport problems: star shapedeness and Newton’s method},
  author={Dieci, Luca and Omarov, Daniyar},
  journal={Numerical Algorithms},
  pages={1--56},
  year={2024},
  publisher={Springer}
}

@article{sobol1967distribution,
title = {On the distribution of points in a cube and the approximate evaluation of integrals},
journal = {USSR Computational Mathematics and Mathematical Physics},
volume = {7},
number = {4},
pages = {86-112},
year = {1967},
issn = {0041-5553},
doi = {https://doi.org/10.1016/0041-5553(67)90144-9},
url = {https://www.sciencedirect.com/science/article/pii/0041555367901449},
author = {I.M Sobol'}
}

@article{fan1993local,
 ISSN = {00905364, 21688966},
  author = {Jianqing Fan},
 journal = {The Annals of Statistics},
 number = {1},
 pages = {196--216},
 publisher = {Institute of Mathematical Statistics},
 title = {Local Linear Regression Smoothers and Their Minimax Efficiencies},
 urldate = {2025-12-05},
 volume = {21},
 year = {1993}
}

@article{ball2022understanding,
  title={Understanding US inflation during the COVID-19 era},
  author={Ball, Laurence and Leigh, Daniel and Mishra, Prachi},
  journal={Brookings Papers on Economic Activity},
  volume={2022},
  number={2},
  pages={1--80},
  year={2022},
  publisher={Johns Hopkins University Press}
}

@article{catalano2024wasserstein,
  title={A Wasserstein index of dependence for random measures},
  author={Catalano, Marta and Lavenant, Hugo and Lijoi, Antonio and Pr{\"u}nster, Igor},
  journal={Journal of the American Statistical Association},
  volume={119},
  number={547},
  pages={2396--2406},
  year={2024},
  publisher={Taylor \& Francis}
}

@article{cui2015model,
  title={Model-free feature screening for ultrahigh dimensional discriminant analysis},
  author={Cui, Hengjian and Li, Runze and Zhong, Wei},
  journal={Journal of the American Statistical Association},
  volume={110},
  number={510},
  pages={630--641},
  year={2015},
  publisher={Taylor \& Francis}
}

@article{zhou2017model,
  title={Model-free feature screening for ultrahigh dimensional censored regression},
  author={Zhou, Tingyou and Zhu, Liping},
  journal={Statistics and Computing},
  volume={27},
  pages={947--961},
  year={2017},
  publisher={Springer}
}

@article{lin2018model,
  title={Model-free feature screening for high-dimensional survival data},
  author={Lin, Yuanyuan and Liu, Xianhui and Hao, Meiling},
  journal={Science China Mathematics},
  volume={61},
  pages={1617--1636},
  year={2018},
  publisher={Springer}
}

@inproceedings{wang2017heterogeneous,
  title={Heterogeneous data and big data analytics},
  author={Wang, Lidong},
  booktitle={Automatic Control and Information Sciences},
  volume={3},
  number={1},
  pages={8--15},
  year={2017}
}

@article{ammar2006analysis,
  title={Analysis of heterogeneous data in ultrahigh dimensions},
  author={Ammar, Reda A and Demurjian Sr, SA and Greenshields, Ian R and Pattipati, K and Rajasekaran, Sanguthevar},
  journal={Emergent Information Technologies and Enabling Policies for Counter-Terrorism},
  pages={105--124},
  year={2006},
  publisher={Wiley Online Library}
}

@article{zhu2011model,
  title={Model-free feature screening for ultrahigh-dimensional data},
  author={Zhu, Li-Ping and Li, Lexin and Li, Runze and Zhu, Li-Xing},
  journal={Journal of the American Statistical Association},
  volume={106},
  number={496},
  pages={1464--1475},
  year={2011},
  publisher={Taylor \& Francis}
}

@article{guo2023threshold,
  title={Threshold selection in feature screening for error rate control},
  author={Guo, Xu and Ren, Haojie and Zou, Changliang and Li, Runze},
  journal={Journal of the American Statistical Association},
  volume={118},
  number={543},
  pages={1773--1785},
  year={2023},
  publisher={Taylor \& Francis}
}

@article{xue2017robust,
  title={A robust model-free feature screening method for ultrahigh-dimensional data},
  author={Xue, Jingnan and Liang, Faming},
  journal={Journal of Computational and Graphical Statistics},
  volume={26},
  number={4},
  pages={803--813},
  year={2017},
  publisher={Taylor \& Francis}
}

@article{tong2023model,
  title={Model-free conditional feature screening with FDR control},
  author={Tong, Zhaoxue and Cai, Zhanrui and Yang, Songshan and Li, Runze},
  journal={Journal of the American Statistical Association},
  volume={118},
  number={544},
  pages={2575--2587},
  year={2023},
  publisher={Taylor \& Francis}
}

@article{kantorovich1960mathematical,
  title={Mathematical methods of organizing and planning production},
  author={Kantorovich, Leonid V},
  journal={Management Science},
  volume={6},
  number={4},
  pages={366--422},
  year={1960},
  publisher={INFORMS}
}

@article{vaserstein1969markov,
  title={Markov processes over denumerable products of spaces, describing large systems of automata},
  author={Vaserstein, Leonid Nisonovich},
  journal={Problemy Peredachi Informatsii},
  volume={5},
  number={3},
  pages={64--72},
  year={1969},
  publisher={Russian Academy of Sciences, Branch of Informatics, Computer Equipment and~…}
}

@article{candes2018panning,
  title={Panning for gold:‘model-X’knockoffs for high dimensional controlled variable selection},
  author={Cand\`es, Emmanuel and Fan, Yingying and Janson, Lucas and Lv, Jinchi},
  journal={Journal of the Royal Statistical Society Series B: Statistical Methodology},
  volume={80},
  number={3},
  pages={551--577},
  year={2018},
  publisher={Oxford University Press}
}

@book{santambrogio2015optimal,
  title={Optimal transport for applied mathematicians},
  author={Santambrogio, Filippo},
  volume={87},
  year={2015},
  publisher={Springer}
}

@article{peyre2019computational,
  title={Computational optimal transport: With applications to data science},
  author={Peyr{\'e}, Gabriel and Cuturi, Marco and others},
  journal={Foundations and Trends{\textregistered} in Machine Learning},
  volume={11},
  number={5-6},
  pages={355--607},
  year={2019},
  publisher={Now Publishers, Inc.}
}

@article{kriegeskorte2009circular,
  title={Circular analysis in systems neuroscience: the dangers of double dipping},
  author={Kriegeskorte, Nikolaus and Simmons, W Kyle and Bellgowan, Patrick SF and Baker, Chris I},
  journal={Nature Neuroscience},
  volume={12},
  number={5},
  pages={535--540},
  year={2009},
  publisher={Nature Publishing Group}
}

@article{dai2023false,
  title={False discovery rate control via data splitting},
  author={Dai, Chenguang and Lin, Buyu and Xing, Xin and Liu, Jun S},
  journal={Journal of the American Statistical Association},
  volume={118},
  number={544},
  pages={2503--2520},
  year={2023},
  publisher={Taylor \& Francis}
}

@article{nguyen2023quasi,
  title={Quasi-monte carlo for 3d sliced wasserstein},
  author={Nguyen, Khai and Bariletto, Nicola and Ho, Nhat},
  journal={arXiv preprint arXiv:2309.11713},
  year={2023}
}

@Manual{transport, 
    title = {{transport}: Computation of Optimal Transport Plans and
      Wasserstein Distances},
    author = {Dominic Schuhmacher and Björn Bähre and Nicolas Bonneel
      and Carsten Gottschlich and Valentin Hartmann and Florian
      Heinemann and Bernhard Schmitzer and Jörn Schrieber},
    year = {2024}
  }

@article{schuhmacher2020computation,
  title={Computation of Optimal Transport Plans and Wasserstein Distances},
  author={Schuhmacher, D and B{\"a}hre, B and Gottschlich, C and Hartmann, V and Heinemann, F and Schmitzer, B},
  journal={R package},
  year={2020}
}

@article{benjamini1995controlling,
  title={Controlling the false discovery rate: a practical and powerful approach to multiple testing},
  author={Benjamini, Yoav and Hochberg, Yosef},
  journal={Journal of the Royal statistical society: series B (Methodological)},
  volume={57},
  number={1},
  pages={289--300},
  year={1995},
  publisher={Wiley Online Library}
}

@article{dekimpe2023retailing,
  title={Retailing in times of soaring inflation: What we know, what we don't know, and a research agenda},
  author={Dekimpe, Marnik G and van Heerde, Harald J},
  journal={Journal of Retailing},
  volume={99},
  number={3},
  pages={322--336},
  year={2023},
  publisher={Elsevier}
}

@article{jaltuszyk2022inflation,
  title={Inflation, the global financial crisis, and COVID-19 pandemic},
  author={Ja{\l}tuszyk, Grzegorz},
  journal={Journal of Management and Financial Sciences},
  number={46},
  pages={9--19},
  year={2022},
  publisher={Szko{\l}a G{\l}{\'o}wna Handlowa w Warszawie}
}

@article{thorbecke2020impact,
  title={The impact of the COVID-19 pandemic on the US economy: Evidence from the stock market},
  author={Thorbecke, Willem},
  journal={Journal of Risk and Financial Management},
  volume={13},
  number={10},
  pages={233},
  year={2020},
  publisher={MDPI}
}

@article{li2024impact,
  title={The Impact of Interest Rates on Consumer Spending Behavior},
  author={Li, Bohan},
  journal={Academic Journal of Business \& Management},
  volume={6},
  number={12},
  pages={77--83},
  year={2024},
  publisher={Francis Academic Press}
}

@article{fan1992variable,
  title={Variable bandwidth and local linear regression smoothers},
  author={Fan, Jianqing and Gijbels, Irene},
  journal={The Annals of Statistics},
  pages={2008--2036},
  year={1992},
  publisher={JSTOR}
}

@article{bonneel2015sliced,
  title={Sliced and radon wasserstein barycenters of measures},
  author={Bonneel, Nicolas and Rabin, Julien and Peyr{\'e}, Gabriel and Pfister, Hanspeter},
  journal={Journal of Mathematical Imaging and Vision},
  volume={51},
  number={1},
  pages={22--45},
  year={2015},
  publisher={Springer}
}

@article{chen2023note,
  title={A note of feature screening via a rank-based coefficient of correlation},
  author={Chen, Li-Pang},
  journal={Biometrical Journal},
  volume={65},
  number={6},
  pages={2100373},
  year={2023},
  publisher={Wiley Online Library}
}

@article{tian2025feature,
  title={Feature Screening for Ultrahigh Dimensional Mixed Data via Wasserstein Distance},
  author={Tian, Bing and Wang, Hong},
  journal={International Statistical Review},
  volume={93},
  number={2},
  pages={267--287},
  year={2025},
  publisher={Wiley Online Library}
}

@article{zhou2018model,
  title={Model-free feature screening for ultrahigh dimensional data through a modified blum-kiefer-rosenblatt correlation},
  author={Zhou, Yeqing and Zhu, Liping},
  journal={Statistica Sinica},
  volume={28},
  number={3},
  pages={1351--1370},
  year={2018},
  publisher={JSTOR}
}

@article{zhao2022distribution,
  title={Distribution-free and model-free multivariate feature screening via multivariate rank distance correlation},
  author={Zhao, Shaofei and Fu, Guifang},
  journal={Journal of Multivariate Analysis},
  volume={192},
  pages={105081},
  year={2022},
  publisher={Elsevier}
}

@article{wu2015conditional,
  title={Conditional quantile screening in ultrahigh-dimensional heterogeneous data},
  author={Wu, Yuanshan and Yin, Guosheng},
  journal={Biometrika},
  volume={102},
  number={1},
  pages={65--76},
  year={2015},
  publisher={Oxford University Press}
}

@article{zhang2018variable,
  title={Variable screening for ultrahigh dimensional heterogeneous data via conditional quantile correlations},
  author={Zhang, Shucong and Zhou, Yong},
  journal={Journal of Multivariate Analysis},
  volume={165},
  pages={1--13},
  year={2018},
  publisher={Elsevier}
}

@article{jiang2024screen,
  title={Screen then select: a strategy for correlated predictors in high-dimensional quantile regression},
  author={Jiang, Xuejun and Liang, Yakun and Wang, Haofeng},
  journal={Statistics and Computing},
  volume={34},
  number={3},
  pages={112},
  year={2024},
  publisher={Springer}
}

@article{zhong2016regularized,
  title={Regularized quantile regression and robust feature screening for single index models},
  author={Zhong, Wei and Zhu, Liping and Li, Runze and Cui, Hengjian},
  journal={Statistica Sinica},
  volume={26},
  number={1},
  pages={69},
  year={2016}
}

@article{wen2018sure,
  title={Sure independence screening adjusted for confounding covariates with ultrahigh dimensional data},
  author={Wen, Canhong and Pan, Wenliang and Huang, Mian and Wang, Xueqin},
  journal={Statistica Sinica},
  volume={28},
  number={1},
  pages={293--317},
  year={2018},
  publisher={JSTOR}
}

@article{zhou2020model,
  title={Model-free forward screening via cumulative divergence},
  author={Zhou, Tingyou and Zhu, Liping and Xu, Chen and Li, Runze},
  journal={Journal of the American Statistical Association},
  volume={115},
  number={531},
  pages={1393--1405},
  year={2020},
  publisher={Taylor \& Francis}
}

@article{xia2021copula,
  title={Copula-based partial correlation screening},
  author={Xia, Xiaochao and Li, Jialiang},
  journal={Statistica Sinica},
  volume={31},
  number={1},
  pages={421--447},
  year={2021},
  publisher={JSTOR}
}

\end{document}


\def\spacingset#1{\renewcommand{\baselinestretch}%
{#1}\small\normalsize} \spacingset{1}

\title{\bf
Supplementary Material for ``Model-free and Distributionally Robust Feature Screening with False Discovery Control for High-Dimensional Heterogeneous Data''
}
\author{}
\date{}
\maketitle

This document serves as the Supplementary Material for the paper 
\textit{``Model-free and Distributionally Robust Feature Screening with False Discovery Control for High-Dimensional Heterogeneous Data.''}
It is organized into four appendices. Appendix~A reports additional numerical results that complement the numerical studies presented in Section~5 of the main document. Appendix~B outlines computational and implementation details for the $r$-Wasserstein Distance, Wasserstein Dependence, and Copula Divergence. Appendix~C collects several auxiliary lemmas and theorems that are used to establish the main theoretical results. Appendix~D provides the complete proofs of all theoretical results stated in the main manuscript.

\appendix


\section{Additional Numerical Results}

\setcounter{table}{0}
\renewcommand{\thetable}{A.\arabic{table}}
\setcounter{figure}{0}
\renewcommand{\thefigure}{A.\arabic{figure}}

\subsection{Additional Simulation Results for Section 5.1}

In this appendix, we present additional simulation results for Models 1--3 introduced in Section~5.1. To simulate a homogeneous feature vector, we generate $\mathbf{X} \sim \text{MVN}(\mathbf{0}_p, \bSigma)$. For each model, we set $n=200$ and $p \in \{1000, 2000\}$, conducting 200 replications. All other simulation settings follow those described in Section~5.1. The supplement simulation results for Models  1--3 with homogeneous features are presented in Table \ref{tab:sup_model_1} to Table \ref{tab:sup_model_3}, respectively.

\begin{table}[H]
\centering
\setlength{\tabcolsep}{3pt} 
\caption{Supplement simulation results for Model 1 with homogeneous features.}\label{tab:sup_model_1}
\small
\begin{tabular}{lccccc ccccc}
\toprule
 & \multicolumn{5}{c}{$p=1000$} & \multicolumn{5}{c}{$p=2000$}\\
 \cmidrule(lr){2-6} \cmidrule(lr){7-11}
 Quantile & 5\% & 25\%& 50\%& 75\%  & 95\%   & 5\% & 25\%& 50\%& 75\%  & 95\% \\
\midrule
SIS      & 4.00 & 4.00 & 6.00 & 8.00 & 17.25
         & 4.00 & 5.00 & 6.00 & 8.00 & 21.15\\
DC-SIS   & 4.00 & 4.00 & 6.00 & 8.00 & 36.25
         & 4.00 & 5.00 & 6.00 & 10.00 & 53.10\\
bcDC-SIS & 4.00 & 4.00 & 6.00 & 8.00 & 37.15
         & 4.00 & 5.00 & 6.00 & 10.00 & 50.30\\
MDC-SIS  & 4.00 & 4.00 & 6.00 & 7.00 & 25.10
         & 4.00 & 5.00 & 6.00 & 8.00 & 31.00\\
WD-1     & 55.95 & 195.25 & 401.50 & 635.25 & 864.10
         & 98.55 & 345.75 & 794.00 & 1397.75 & 1793.65\\
WD-2     & 32.00 & 142.25 & 302.50 & 556.50 & 855.70
         & 64.75 & 335.75 & 689.50 & 1255.25 & 1776.15\\
CD-1     & 4.00 & 4.00 & 6.00 & 9.00 & 67.45
         & 4.00 & 5.00 & 6.00 & 14.25 & 159.90\\
CD-2     & 4.00 & 4.00 & 6.00 & 11.00 & 90.25
         & 4.00 & 5.00 & 6.00 & 14.00 & 119.35\\
\bottomrule
\end{tabular}
\end{table}

\begin{table}[H]
\centering
\setlength{\tabcolsep}{3pt} 
\caption{Supplement simulation results for Model 2 with homogeneous features.}\label{tab:sup_model_2}
\small
\begin{tabular}{lccccc ccccc}
\toprule
 & \multicolumn{5}{c}{$p=1000$} & \multicolumn{5}{c}{$p=2000$}\\
 \cmidrule(lr){2-6} \cmidrule(lr){7-11}
 Quantile & 5\% & 25\%& 50\%& 75\%  & 95\%   & 5\% & 25\%& 50\%& 75\%  & 95\% \\
\midrule
SIS      & 119.95 & 406.50 & 622.00 & 833.50 & 969.00
         & 196.35 & 596.00 & 1192.00 & 1604.50 & 1945.05\\
DC-SIS   & 103.05 & 303.50 & 541.50 & 742.75 & 935.25
         & 137.70 & 500.25 & 936.00 & 1403.50 & 1870.45\\
bcDC-SIS & 44.95 & 217.00 & 463.00 &735.50 & 926.15
         & 74.50 & 409.50 & 974.50 & 1444.50 & 1852.35\\
MDC-SIS  & 76.95 & 344.75 & 587.50 & 806.50 & 974.45
         & 113.70 & 576.75 & 1179.00 & 1658.25 & 1933.25\\
WD-1     & 20.80 & 105.75 & 225.00 & 382.50 & 784.65
         & 28.95 & 173.75 & 457.50 & 900.75 & 1661.60\\
WD-2     & 399.95 & 701.25 & 845.50 & 933.25 & 984.10
         & 915.50 & 1372.25 & 1685.50 & 1864.75 & 1973.10\\
CD-1     & 4.00 & 4.00 & 4.00 & 5.00 & 5.00
         & 4.00 & 4.00 & 4.00 & 5.00 & 5.00\\
CD-2     & 4.00 & 4.00 & 4.00 & 5.00 & 5.00
         & 4.00 & 4.00 & 4.00 & 5.00 & 5.00\\
\bottomrule
\end{tabular}
\end{table}

\begin{table}[H]
\centering
\setlength{\tabcolsep}{3pt} 
\caption{Supplement simulation results for Model 3 with homogeneous features.}\label{tab:sup_model_3}
\small
\begin{tabular}{lccccc ccccc}
\toprule
 & \multicolumn{5}{c}{$p=1000$} & \multicolumn{5}{c}{$p=2000$}\\
 \cmidrule(lr){2-6} \cmidrule(lr){7-11}
 Quantile & 5\% & 25\%& 50\%& 75\%  & 95\%   & 5\% & 25\%& 50\%& 75\%  & 95\% \\
\midrule
SIS      & 109.00 & 390.50 & 671.00 & 847.75 & 965.05 
         & 189.60 & 835.50 & 1372.50 & 1683.00 & 1944.05\\
DC-SIS   & 100.70 & 390.50 & 631.00 & 829.50 & 965.10
         & 184.20 & 830.25 & 1274.50 & 1671.25 & 1939.45\\
bcDC-SIS & 101.90 & 375.25 & 596.00 & 830.00 & 948.25
         & 157.35 & 704.75 & 1241.00 & 1661.00 & 1885.15\\
MDC-SIS  & 99.45 & 377.00 & 634.00 & 843.25 & 960.05
         & 120.35 & 798.25 & 1291.50 & 1650.50 & 1895.45\\
WD-1     & 311.15 & 672.00 & 838.00 & 926.50 & 984.05
         & 721.30 & 1381.75 & 1635.00 & 1826.00 & 1960.20\\
WD-2     & 572.30 & 787.00 & 863.00 & 948.00 & 992.05
         & 1145.30 & 1517.75 & 1782.50 & 1875.00 & 1978.05 \\
CD-1     & 5.00 &  5.00 &  6.00 &  7.00 & 23.30
          & 5.00 & 5.00 & 6.00 & 7.00 & 13.05\\
CD-2     & 5.00 &  5.00  & 6.00 &  6.00 & 13.15 
         & 5.00 & 5.00 & 6.00 & 6.00 & 20.05\\
\bottomrule
\end{tabular}
\end{table}

To further evaluate the robustness of the proposed method, we consider an additional simulation setting, denoted as Model~\(2'\). This model extends Model~2 by incorporating both a non-additive noise structure and more irregular marginal distributions for the covariates.

The response is generated from
\[
Y
=
5X_1
+
2\sin\left(\frac{\pi}{2}X_2\right)
+
2|X_3|
+
2\exp(0.5X_4)
+
(1+|X_1|)\varepsilon,
\]
where \(\varepsilon\sim N(0,1)\). The multiplicative factor \((1+|X_1|)\) makes the noise level depend on the covariate \(X_1\), thereby introducing heteroscedasticity and a non-additive error structure. The active set is \(\mathcal{S}=\{1,2,3,4\}\).

To assess robustness to irregular marginal distributions, the covariates are independently generated from
\[
X_j \sim F_{j \bmod 3}, \qquad j=1,\ldots,p,
\]
where
\[
F_0:\quad X \sim 0.5N(-2,0.5^2)+0.5N(2,0.5^2),
\]
is a two-component Gaussian mixture distribution,
\[
F_1:\quad X \sim \mathrm{Unif}(-2,2),
\]
is a bounded uniform distribution, and
\[
F_2:\quad X=\mathrm{round}(Z,1)+10^{-6}E,
\qquad Z\sim N(0,1),\quad E\sim N(0,1),
\]
is a rounded Gaussian distribution with a small continuous jitter to avoid exact ties. Thus, this setting includes multimodal, bounded, and rounded covariate marginals. We consider \(p=1000\) and \(p=2000\), and repeat each experiment 200 times.

Table~\ref{tab:sim_model_2prime} reports the quantiles of the minimum model size required to include all four active features. The proposed CD-Screen methods, CD-1 and CD-2, continue to perform strongly under this more challenging setting. For both \(p=1000\) and \(p=2000\), the median minimum model size is 4, which is exactly the true number of active features, and the 95\% quantile is only 5. This indicates that CD-Screen almost always ranks all active features at the very top.

In contrast, the competing methods require much larger model sizes to recover all active features. When \(p=1000\), the median minimum model sizes for SIS, DC-SIS, bcDC-SIS, and MDC-SIS range from 266.50 to 389.00. Their performance further deteriorates when \(p=2000\), with median minimum model sizes ranging from 546.50 to 781.00. The Wasserstein Dependence-based methods perform even worse, requiring median model sizes of 548.00 and 878.50 for \(p=1000\), and 1093.50 and 1757.00 for \(p=2000\), respectively. These results show that the competing methods either fail to rank the active features sufficiently high or include a large number of irrelevant variables before recovering all active features.

Overall, this additional simulation confirms that the proposed CD-Screen procedure remains robust in the presence of non-additive noise and irregular covariate marginals. The results further support the use of the proposed copula-based screening framework for high-dimensional heterogeneous data.

\begin{table}[htbp]
\centering
\setlength{\tabcolsep}{3pt}
\caption{\textbf{Simulation results for Model 2$^\prime$}: Quantiles of the minimum model size that includes all four active features over 200 replications.}
\label{tab:sim_model_2prime}
\small
\resizebox{\linewidth}{!}{%
\begin{tabular}{lccccc ccccc}
\toprule
 & \multicolumn{5}{c}{$p=1000$} & \multicolumn{5}{c}{$p=2000$}\\
\cmidrule(lr){2-6} \cmidrule(lr){7-11}
Quantile & 5\% & 25\% & 50\% & 75\% & 95\%
         & 5\% & 25\% & 50\% & 75\% & 95\% \\
\midrule
SIS      & 9.90 & 128.00 & 389.00 & 735.00 & 949.95
         & 13.85 & 264.25 & 781.00 & 1461.00 & 1897.00\\
DC-SIS   & 9.00 & 102.75 & 266.50 & 566.50 & 895.05
         & 12.85 & 200.50 & 546.50 & 1137.00 & 1777.80\\
bcDC-SIS & 6.95 & 98.75 & 334.50 & 542.25 & 851.90
         & 13.80 & 198.75 & 672.50 & 1105.50 & 1698.65\\
MDC-SIS  & 12.75 & 162.75 & 379.00 & 663.00 & 894.10
         & 20.40 & 310.00 & 743.00 & 1295.75 & 1798.00\\
WD-1     & 351.95 & 424.00 & 548.00 & 751.25 & 926.40
         & 704.45 & 839.75 & 1093.50 & 1510.25 & 1852.75\\
WD-2     & 513.35 & 771.25 & 878.50 & 942.00 & 986.35
         & 1052.10 & 1535.75 & 1757.00 & 1888.00 & 1973.50\\
CD-1     & 4.00 & 4.00 & 4.00 & 5.00 & 5.00
         & 4.00 & 4.00 & 4.00 & 5.00 & 5.00\\
CD-2     & 4.00 & 4.00 & 4.00 & 5.00 & 5.00
         & 4.00 & 4.00 & 4.00 & 5.00 & 5.00\\
\bottomrule
\end{tabular}%
}
\end{table}

\color{black}

\subsection{Additional Simulation Results for for Section 5.2}\label{app:simulation_screen}

In this appendix, we present additional simulation results Section~5.2 with a linear regression model. We set $n=400$ and $p \in \{1000, 2000\}$, conducting 100 replications. All other simulation settings follow those described in Section~5.2. The supplement simulation results are presented in Table \ref{tab:FDR_results_1000_4} and \ref{tab:FDR_results_2000_4}. 

\begin{description}
    \item[Model 5]\label{exp:FDR.2} \textbf{(Linear model): } $Y=8X_1+9X_2+9X_3+9X_{4}+8X_5+8X_6+9X_7+9X_8+9X_{9}+8X_{10}+\varepsilon$, where $\varepsilon \overset{\text{i.i.d.}}{\sim} N(0,1)$. This model has $10$ active features. 
\end{description}

\begin{table}[H]
\centering
\caption{FDR control performances for Model 5 when $p=1000$.}\label{tab:FDR_results_1000_4}
\footnotesize 
\begin{tabular}{c c c c c c c c c c c c c}
\toprule
$\alpha$ & $|\widehat S|$ & $P_1$ & $P_2$ & $P_3$ & $P_4$ & $P_5$ & $P_6$ & $P_7$ & $P_8$ & $P_9$ & $P_{10}$ & $\widehat{\mathrm{FDR}}$ \\
\midrule

\multicolumn{13}{c}{\textbf{CD-FDR with r=1}} \\
0.15 & 10 & 0.79 & 0.81 & 0.78 & 0.77 & 0.80 & 0.81 & 0.80 & 0.76 & 0.81 & 0.81 & 0.13 \\
0.20 & 11 & 0.86 & 0.88 & 0.87 & 0.85 & 0.88 & 0.88 & 0.86 & 0.84 & 0.88 & 0.88 & 0.16 \\
0.25 & 12 & 0.92 & 0.93 & 0.91 & 0.90 & 0.92 & 0.93 & 0.92 & 0.89 & 0.93 & 0.93 & 0.25 \\
0.30 & 14 & 0.97 & 0.98 & 0.97 & 0.94 & 0.96 & 0.98 & 0.97 & 0.94 & 0.98 & 0.98 & 0.29 \\

\midrule
\multicolumn{13}{c}{\textbf{CD-FDR with r=2}} \\
0.15 & 11 & 0.82 & 0.82 & 0.80 & 0.80 & 0.82 & 0.82 & 0.80 & 0.78 & 0.82 & 0.82 & 0.15 \\
0.20 & 12 & 0.91 & 0.91 & 0.90 & 0.89 & 0.91 & 0.91 & 0.90 & 0.90 & 0.91 & 0.91 & 0.20 \\
0.25 & 13 & 0.98 & 0.98 & 0.96 & 0.96 & 0.98 & 0.98 & 0.96 & 0.95 & 0.98 & 0.98 & 0.25 \\
0.30 & 14 & 0.99 & 0.99 & 0.97 & 0.96 & 0.98 & 0.99 & 0.98 & 0.96 & 0.99 & 0.99 & 0.28 \\

\midrule
\multicolumn{13}{c}{\textbf{BH-CD-FDR with r=1}} \\
0.15 & 12 & 0.98 & 1.00 & 0.99 & 0.97 & 0.99 & 1.00 & 0.98 & 0.95 & 1.00 & 1.00 & 0.16 \\
0.20 & 12.5 & 0.98 & 1.00 & 0.99 & 0.96 & 0.99 & 1.00 & 0.99 & 0.95 & 1.00 & 1.00 & 0.20 \\
0.25 & 13 & 0.98 & 1.00 & 0.99 & 0.97 & 0.99 & 1.00 & 0.99 & 0.96 & 1.00 & 1.00 & 0.25 \\
0.30 & 14 & 0.99 & 1.00 & 0.99 & 0.97 & 1.00 & 1.00 & 0.99 & 0.97 & 1.00 & 1.00 & 0.29 \\

\midrule
\multicolumn{13}{c}{\textbf{BH-CD-FDR with r=2}} \\
0.15 & 12 & 1.00 & 1.00 & 0.98 & 0.98 & 0.99 & 1.00 & 0.98 & 0.97 & 1.00 & 1.00 & 0.18 \\
0.20 & 12 & 1.00 & 1.00 & 0.98 & 0.98 & 0.99 & 1.00 & 0.98 & 0.98 & 1.00 & 1.00 & 0.21 \\
0.25 & 13.5 & 1.00 & 1.00 & 0.99 & 0.98 & 1.00 & 1.00 & 0.99 & 0.98 & 1.00 & 1.00 & 0.27 \\
0.30 & 14 & 1.00 & 1.00 & 0.98 & 0.98 & 0.99 & 1.00 & 0.99 & 0.98 & 1.00 & 1.00 & 0.31 \\

\midrule
\multicolumn{13}{c}{\textbf{PCSIS}} \\
0.15 & 10 & 1.00 & 1.00 & 1.00 & 0.98 & 1.00 & 1.00 & 0.98 & 0.98 & 1.00 & 1.00 & 0.06 \\
0.20 & 10 & 1.00 & 1.00 & 1.00 & 0.99 & 1.00 & 1.00 & 0.99 & 0.99 & 1.00 & 1.00 & 0.06 \\
0.25 & 10 & 1.00 & 1.00 & 1.00 & 0.99 & 1.00 & 1.00 & 0.99 & 0.99 & 1.00 & 1.00 & 0.06 \\
0.30 & 10 & 1.00 & 1.00 & 1.00 & 0.99 & 1.00 & 1.00 & 0.99 & 0.99 & 1.00 & 1.00 & 0.07 \\

\bottomrule
\end{tabular}
\end{table}

\begin{table}[H]
\centering
\caption{FDR control performances for Model 5 when $p=2000$.}
\label{tab:FDR_results_2000_4}
\footnotesize
\begin{tabular}{c c c c c c c c c c c c c}
\toprule
$\alpha$ & $|\widehat S|$ & $P_1$ & $P_2$ & $P_3$ & $P_4$ & $P_5$ & $P_6$ & $P_7$ & $P_8$ & $P_9$ & $P_{10}$ & $\widehat{\mathrm{FDR}}$ \\
\midrule

\multicolumn{13}{c}{\textbf{CD-FDR with r=1}} \\
0.15 & 10 & 0.65 & 0.71 & 0.69 & 0.67 & 0.70 & 0.71 & 0.70 & 0.68 & 0.70 & 0.71 & 0.17 \\
0.20 & 11 & 0.83 & 0.85 & 0.84 & 0.80 & 0.85 & 0.85 & 0.85 & 0.80 & 0.84 & 0.85 & 0.20 \\
0.25 & 13 & 0.88 & 0.91 & 0.89 & 0.86 & 0.90 & 0.91 & 0.91 & 0.89 & 0.91 & 0.91 & 0.26 \\
0.30 & 13 & 0.88 & 0.92 & 0.91 & 0.87 & 0.91 & 0.92 & 0.92 & 0.90 & 0.92 & 0.92 & 0.27 \\

\midrule
\multicolumn{13}{c}{\textbf{CD-FDR with r=2}} \\
0.15 & 11 & 0.69 & 0.69 & 0.69 & 0.62 & 0.68 & 0.69 & 0.69 & 0.67 & 0.69 & 0.69 & 0.16 \\
0.20 & 12 & 0.83 & 0.84 & 0.82 & 0.78 & 0.83 & 0.84 & 0.84 & 0.80 & 0.84 & 0.84 & 0.21 \\
0.25 & 13 & 0.93 & 0.94 & 0.94 & 0.86 & 0.93 & 0.94 & 0.94 & 0.93 & 0.94 & 0.94 & 0.26 \\
0.30 & 14 & 0.94 & 0.96 & 0.96 & 0.87 & 0.95 & 0.96 & 0.96 & 0.90 & 0.96 & 0.96 & 0.30 \\

\midrule
\multicolumn{13}{c}{\textbf{BH-CD-FDR with r=1}} \\
0.15 & 12 & 0.96 & 1.00 & 0.97 & 0.92 & 0.99 & 1.00 & 1.00 & 0.97 & 1.00 & 1.00 & 0.18 \\
0.20 & 12 & 0.97 & 1.00 & 0.98 & 0.92 & 0.99 & 1.00 & 1.00 & 0.98 & 0.99 & 0.99 & 0.23 \\
0.25 & 14 & 0.98 & 1.00 & 0.98 & 0.93 & 0.98 & 1.00 & 1.00 & 0.98 & 0.99 & 0.99 & 0.27 \\
0.30 & 14 & 0.98 & 1.00 & 0.98 & 0.94 & 0.98 & 1.00 & 1.00 & 0.98 & 0.99 & 0.99 & 0.30 \\

\midrule
\multicolumn{13}{c}{\textbf{BH-CD-FDR with r=2}} \\
0.15 & 11.5 & 0.97 & 1.00 & 0.98 & 0.91 & 0.99 & 1.00 & 1.00 & 0.99 & 1.00 & 1.00 & 0.18 \\
0.20 & 13 & 0.98 & 1.00 & 0.99 & 0.92 & 0.99 & 1.00 & 1.00 & 0.99 & 1.00 & 1.00 & 0.24 \\
0.25 & 13.5 & 0.98 & 1.00 & 0.99 & 0.92 & 0.99 & 1.00 & 1.00 & 0.99 & 1.00 & 1.00 & 0.27 \\
0.30 & 14 & 0.98 & 1.00 & 0.98 & 0.93 & 0.98 & 1.00 & 1.00 & 0.98 & 0.99 & 1.00 & 0.32 \\

\midrule
\multicolumn{13}{c}{\textbf{PCSIS}} \\
0.15 & 10 & 0.98 & 1.00 & 0.99 & 0.94 & 1.00 & 1.00 & 1.00 & 0.99 & 1.00 & 1.00 & 0.06 \\
0.20 & 10 & 0.98 & 1.00 & 1.00 & 0.96 & 1.00 & 1.00 & 1.00 & 1.00 & 1.00 & 1.00 & 0.06 \\
0.25 & 10 & 0.98 & 1.00 & 1.00 & 0.97 & 1.00 & 1.00 & 1.00 & 1.00 & 1.00 & 1.00 & 0.07 \\
0.30 & 10 & 0.99 & 1.00 & 1.00 & 0.97 & 1.00 & 1.00 & 1.00 & 1.00 & 1.00 & 1.00 & 0.07 \\

\bottomrule
\end{tabular}
\end{table}


\subsection{Additional Real Data Analysis Results for Section 5.3}

In this appendix, we present additional results for Section 5.3 in the main document. Table \ref{tab:sectors} contains the 11 sectors of S\&P 500 stocks that we used in our analysis. Tables \ref{top10_CD} -- \ref{top10_industrials} list the top 10 stocks in each sector that have the highest Copula Divergence values.

\begin{table}[htbp]
\centering
\caption{The 11 Sectors of S\&P 500 stocks}
\label{tab:sectors}
\begin{tabular}{lc|lc}  
\toprule  
Index & Sector name & Index & Sector name\
\\ \midrule  
Sector 1 & Communication Services 
& Sector 7 & Industrials
\\
Sector 2 & Consumer Discretionary
& Sector 8 & Information Technology
\\
Sector 3 & Consumer Staples
& Sector 9 & Materials
\\
Sector 4 & Energy
& Sector 10 & Real Estate
\\
Sector 5 & Financials
& Sector 11 & Utilities
\\
Sector 6 & Health Care & & \\
\bottomrule  
\end{tabular}
\end{table}

\begin{table}[htbp]
\centering
\caption{Top 10 stocks in Consumer Discretionary sector.}
\begin{tabular}{c|cc}
\hline
Rank & Code & Name \\ \hline
1  & DRI  & Darden Restaurants \\
2  & ROST & Ross Stores \\
3  & HD   & Home Depot (The) \\
4  & NVR  & NVR, Inc. \\
5  & RCL  & Royal Caribbean Group \\
6  & EXPE & Expedia Group \\
7  & DHI  & D. R. Horton \\
8  & GM   & General Motors \\
9  & LEN  & Lennar \\
10 & APTV & Aptiv \\
\hline
\end{tabular}
\label{top10_CD}
\end{table}

\begin{table}[htbp]
\centering
\caption{Top 10 stocks in Financial sector.}
\begin{tabular}{c|cc}
\hline
Rank & Code & Name \\ \hline
1  & CINF & Cincinnati Financial \\
2  & SYF  & Synchrony Financial \\
3  & MS   & Morgan Stanley \\
4  & ERIE & Erie Indemnity \\
5  & COF  & Capital One \\
6  & GS   & Goldman Sachs \\
7  & ACGL & Arch Capital Group \\
8  & RF   & Regions Financial Corporation \\
9  & EG   & Everest Group \\
10 & MKTX & MarketAxess \\
\hline
\end{tabular}
\label{top10_financial}
\end{table}

\begin{table}[htbp]
\centering
\caption{Top 10 stocks in Utilities sector.}
\begin{tabular}{c|cc}
\hline
Rank & Code & Name \\ \hline
1  & NRG & NRG Energy \\
2  & PCG & PG\&E Corporation \\
3  & SRE & Sempra \\
4  & VST & Vistra Corp. \\
5  & AES & AES Corporation \\
6  & PNW & Pinnacle West \\
7  & DUK & Duke Energy \\
8  & CNP & CenterPoint Energy \\
9  & EIX & Edison International \\
10 & PPL & PPL Corporation \\
\hline
\end{tabular}
\label{top10_utilities}
\end{table}

\begin{table}[htbp]
\centering
\caption{Top 10 stocks in Industrials sector.}
\begin{tabular}{c|cc}
\hline
Rank & Code & Name \\ \hline
1  & UBER & Uber \\
2  & BLDR & Builders FirstSource \\
3  & PAYC & Paycom \\
4  & LUV  & Southwest Airlines \\
5  & DAY  & Dayforce \\
6  & FAST & Fastenal \\
7  & NSC  & Norfolk Southern Railway \\
8  & IR   & Ingersoll Rand \\
9  & SNA  & Snap-on \\
10 & PNR  & Pentair \\
\hline
\end{tabular}
\label{top10_industrials}
\end{table}


To further assess the practical impact of the proposed screening procedure, we conducted an additional prediction comparison based on the real-data example in Section~5.3. Since the data consist of monthly time series observations, we used a forward-looking train-test split to preserve the temporal ordering. Specifically, the first 38 observations were used as the training set, and the remaining 10 observations were used as the test set.

We considered random forest as the downstream prediction model, motivated by the nonlinear marginal relationships observed in Figure~3. Three approaches were compared: (i) random forest using all features without screening, denoted by RF; (ii) random forest after screening by SIS, denoted by RF + SIS; and (iii) random forest after screening by the proposed CD-FDR method, denoted by RF + CD-FDR. For SIS, following the common recommendation in the sure screening literature, we retained the top $\lceil n_{\rm train}/\log(n_{\rm train}) \rceil$ variables ranked by absolute marginal correlation. Since $n_{\rm train}=38$, this gives $d_{\rm SIS}=10$ selected variables. The proposed CD-FDR method selected 6 variables, which were then used for the subsequent prediction analysis.

\begin{table}[htbp]
\centering
\caption{Prediction performance comparison for the real-data example.}
\label{tab:prediction-performance}
\begin{tabular}{lcc}
\toprule
Method & RMSE & MAE \\
\midrule
RF & 0.00299 & 0.00240 \\
RF + CD-FDR & 0.00285 & 0.00220 \\
RF + SIS & 0.00297 & 0.00242 \\
\bottomrule
\end{tabular}
\end{table}

Table~\ref{tab:prediction-performance} reports the test prediction errors. The random forest model combined with the proposed CD-FDR screening method achieves the smallest RMSE and MAE among the three competing approaches. Compared with fitting random forest using all features, RF + CD-FDR reduces the RMSE from 0.00299 to 0.00285 and the MAE from 0.00240 to 0.00220. It also outperforms RF + SIS, which yields an RMSE of 0.00297 and an MAE of 0.00242. These results suggest that CD-FDR can remove irrelevant or noisy features while retaining predictors that are useful for downstream prediction. Therefore, in this real-data example, the proposed screening procedure not only yields a smaller and more interpretable set of selected variables, but also improves predictive performance.

\color{black}


\section{Computation of $r$-Wasserstein Distance, Wasserstein Dependence, and Copula Divergence}\label{app:compute}

In this paper, we adopt the R package \texttt{transport} \citep{schuhmacher2020computation} to compute the $r$-Wasserstein distance between two empirical samples. In Sections~5.2 and~5.3, we further employ the sliced Wasserstein distance \citep{bonneel2015sliced} option  to accelerate computation without compromising numerical stability.

The computation of Wasserstein Dependence requires estimating the joint and marginal distributions using the same dataset. Following \cite{nies2022d}, two primary estimation strategies have been proposed. The first approach relies on sample splitting: given a sample of size $2n$, $\{(x_i, y_i)\}_{i=1}^{2n}$, the first half is used to estimate the joint distribution and the second half to estimate the marginal distributions. Although conceptually straightforward, this method suffers from reduced sample efficiency. An alternative approach is the permutation-based estimator, in which the permuted sample $\{(x_i, y_{\sigma(i)})\}$ serves as a surrogate for the independent distribution, where $\sigma$ denotes a permutation over $\{1,\dots,2n\}$. However, this estimator is known to perform poorly in practice due to the variability introduced by random permutations.

The estimation of Copula Divergence is based on the semi-discrete Wasserstein distance. Several algorithms have been proposed for approximating this distance \citep[see, e.g.,][]{hartmann2020semi,dieci2024solving}. In this work, we construct a two-dimensional low-discrepancy sequence \citep{sobol1967distribution} of size $n$ on $[0,1]^2$, followed by Probit transformations to approximate the target measure $\phi\otimes\phi$. This approach induces a Quasi-Monte Carlo approximation to the Wasserstein functional, as suggested in \cite{nguyen2023quasi}. Consequently, computing the divergence reduces to evaluating the discrete Wasserstein distance between the Gaussianized sample $\{(\widehat{s}_{i}, \widehat{t}_{i})\}_{i=1}^n$ and the low-discrepancy reference points $\{(s^*_{i}, t^*_{i})\}_{i=1}^n$, given by
\begin{align*}
    \mathcal{W}_{r}(\widehat{f}_{\widehat{S},\widehat{T}}, [\phi\otimes\phi]_n)
    =\left\{\inf_{\boldsymbol{\pi}\in\boldsymbol{\Pi}}
    \left[\frac{1}{n}\sum_{i=1}^{n}
    \left\|\begin{pmatrix}
        s^*_i\\[2pt]
        t^*_i
    \end{pmatrix}
    -
    \begin{pmatrix}
        \widehat{s}_{\boldsymbol{\pi}(i)}\\[2pt]
        \widehat{t}_{\boldsymbol{\pi}(i)}
    \end{pmatrix}\right\|^{r}
    \right]\right\}^{1/r},
\end{align*}
where $[\phi\otimes\phi]_n$ denotes the empirical measure associated with $\{(s^*_{i},t^*_{i})\}_{i=1}^n$, $\widehat{f}_{\widehat{S},\widehat{T}}$ represents the empirical distribution of the Gaussianized sample, and $\boldsymbol{\Pi}$ denotes the set of all permutations of $\{1,\dots,n\}$.


\section{Auxiliary Lemmas and Theorems}

In this appendix, we provide some auxiliary lemmas and theorems to pave the way for the proof of the major theoretical results in the main document.

\setcounter{equation}{0}
\renewcommand{\theequation}{D.\arabic{equation}}

\begin{lemma}\label{prop:McDiarmid}
\citep[cf. Lemma 1.2 in][]{McDiarmid_1989}. 
Let $(z_1,\dots,z_n)$ be independent random variables, and let $g$ be a measurable function such that
\[
    \lvert g(\mathbf{x}) - g(\tilde{\mathbf{x}}) \rvert \le c_i,
\]
whenever $\mathbf{x},\tilde{\mathbf{x}}\in\mathbb{R}^n$ differ only in their $i$-th coordinate.  
Then for any $\varepsilon>0$,
\[
    \cP\!\left(\lvert g(z_1,\dots,z_n) - \eE g(z_1,\dots,z_n) \rvert \ge \varepsilon\right)
    \le 2\exp\!\left(-2\frac{\varepsilon^2}{\sum_{i=1}^n c_i^2}\right).
\]
\end{lemma}

\begin{lemma}\label{prop:Abramovich}
\citep[cf. Lemmas 12.1 and 12.3 in][]{abramovich2006adapting}.  
For $z \ge 1$ and $0<\eta\le 0.01$, we have
\[
    \frac{\phi(z)}{2z} \le 1-\Phi(z) \le \frac{\phi(z)}{z}
    \quad \text{and} \quad
    \Phi^{-1}(1-\eta) \le \sqrt{2\log(\eta^{-1})}.
\]
\end{lemma}

\begin{lemma}\label{prop:gaussianized_bound} 
\citep[cf. Proposition 3 in][]{mai2023coordinatewise}.  
For any sample $\{\xi_i\}_{i=1}^n$ with empirical CDF $\widehat{F}_\xi(\cdot)$, there exist positive constants $C$ and $n_0$, independent of $\xi$ and $n$, such that for all $n\ge n_0$ and all $t\in\mathbb{R}$,
\[
    \Phi^{-1}\!\left(\frac{n}{n+1}\widehat{F}_\xi(t)\right) \le C\sqrt{\log n}.
\]
\end{lemma}

The proofs of Lemmas~\ref{prop:McDiarmid}–\ref{prop:gaussianized_bound} are available in their original references and are therefore omitted here.


\begin{lemma}\label{lemma:abs_gaussianized_diff} 
\citep[Modified from Lemma 5 in][]{mai2023coordinatewise}. 
Let $\{z_i\}_{i=1}^n\overset{\text{i.i.d.}}{\sim} N(0,1)$ and define
\[
z_i^*\doteq z_i\II_{\{\lvert z_i\rvert\le\sqrt{2\log n}\}}+\sign(z_i)\sqrt{2\log n}\II_{\{\lvert z_i\rvert>\sqrt{2\log n}\}}. 
\]
Then there exist positive constants $M$, $C$, $n_0$, and $\varepsilon_0$, independent of $n$, such that for all $n\ge n_0$ and $\varepsilon\in(Mn^{-1},\varepsilon_0)$,
\begin{equation}
    \cP\Bigg(\frac{1}{n}\sum_{i=1}^n\lvert z_i^*-z_i\rvert\ge\varepsilon\Bigg)\le C\exp\!\left(-\frac{Cn\varepsilon^2}{\log n}\right). \nonumber
\end{equation}

\begin{proof}
Define 
\[
g(\mathbf{x})\doteq\frac{1}{n}\sum_{i=1}^n \lvert x_i\rvert \II_{\{\lvert x_i\rvert\le\sqrt{2\log n}\}},\qquad \mathbf{x}=(x_1,\dots,x_n)\in\mathbb{R}^n.
\]
Consider $\mathbf{x},\tilde{\mathbf{x}}\in\mathbb{R}^n$ differing only in the $i$-th coordinate. Without loss of generality, let 
$\mathbf{x}=(x_1,\dots,x_i, \ldots, x_n)$ and $\tilde{\mathbf{x}}=(x_1, \ldots, \tilde{x}_i,\dots,x_n)$. Then
\[
    \lvert g(\mathbf{x})-g(\tilde{\mathbf{x}})\rvert
    =\Bigg\lvert \frac{1}{n}\lvert x_i\rvert\II_{\{\lvert x_i\rvert\le\sqrt{2\log n}\}}
     -\frac{1}{n}\lvert \tilde{x}_i\rvert\II_{\{\lvert\tilde{x}_i\rvert\le\sqrt{2\log n}\}}\Bigg\rvert
     \le C\frac{\sqrt{\log n}}{n},
\]
for some positive constant $C$ independent of $n$. Since $z_i\overset{\text{i.i.d.}}{\sim}N(0,1)$, applying Lemma~\ref{prop:McDiarmid} yields
\begin{equation}
    \cP\!\left(\Bigg\lvert\frac{1}{n}\!\sum_{i=1}^n\!\lvert z_i\rvert\II_{\{\lvert z_i\rvert\le\sqrt{2\log n}\}}
    -\eE[\lvert z_i\rvert\II_{\{\lvert z_i\rvert\le\sqrt{2\log n}\}}]\Bigg\rvert\!
    \ge\varepsilon\right)
    \le C\exp\!\left(-\frac{Cn\varepsilon^2}{\log n}\right),
    \label{eq:trucated_gaussian_abs_central}
\end{equation}
for any $\varepsilon>0$ and constant $C$ independent of $n$.

Next, observe that
\begin{equation}
    \eE[\lvert z_i\rvert\II_{\{\lvert z_i\rvert\le\sqrt{2\log n}\}}]
    =\int_{-\sqrt{2\log n}}^{\sqrt{2\log n}}\!\lvert x\rvert\phi(x)\,dx
    =\sqrt{\frac{2}{\pi}}-\sqrt{\frac{2}{\pi}}\frac{1}{n},
    \label{eq:trucated_gaussian_abs_exp}
\end{equation}
whenever $n\ge3$. Combining \eqref{eq:trucated_gaussian_abs_central} and \eqref{eq:trucated_gaussian_abs_exp}, and assuming 
$\varepsilon>\frac{2}{n}\sqrt{\frac{2}{\pi}}$, we obtain
\begin{align}
    & \cP\!\left(\Bigg\lvert\frac{1}{n}\!\sum_{i=1}^n\!\lvert z_i\rvert\II_{\{\lvert z_i\rvert\le\sqrt{2\log n}\}}
    -\sqrt{\frac{2}{\pi}}\Bigg\rvert\ge\varepsilon\right)
    \nonumber\\
    &=\cP\!\left(\Bigg\lvert\frac{1}{n}\!\sum_{i=1}^n\!\lvert z_i\rvert\II_{\{\lvert z_i\rvert\le\sqrt{2\log n}\}}
    -\eE[\lvert z_i\rvert\II_{\{\lvert z_i\rvert\le \sqrt{2\log n}\}}]
    -\sqrt{\frac{2}{\pi}}\frac{1}{n}\Bigg\rvert\ge\varepsilon\right)
    \nonumber\\
    &\le\cP\!\left(\Bigg\lvert\frac{1}{n}\!\sum_{i=1}^n\!\lvert z_i\rvert\II_{\{\lvert z_i\rvert\le\sqrt{2\log n}\}}
    -\eE[\lvert z_i\rvert\II_{\{\lvert z_i\rvert\le\sqrt{2\log n}\}}]\Bigg\rvert\ge\frac{\varepsilon}{2}\right)
    \nonumber\\
    &\le C\exp\!\left(-\frac{Cn\varepsilon^2}{\log n}\right).
    \label{eq:abs_gaussianized_diff_2_res}
\end{align}

Finally,
\begin{align}
    \cP\!\left(\frac{1}{n}\sum_{i=1}^n\lvert z_i^*-z_i\rvert\ge\varepsilon\right)
    &=\cP\!\left(\frac{1}{n}\sum_{i=1}^n(\lvert z_i\rvert-\sqrt{2\log n})
      \II_{\{\lvert z_i\rvert>\sqrt{2\log n}\}}\ge\varepsilon\right)\nonumber\\
    &\le\cP\!\left(\frac{1}{n}\sum_{i=1}^n\lvert z_i\rvert\II_{\{\lvert z_i\rvert>\sqrt{2\log n}\}}\ge\varepsilon\right)\nonumber\\
    &=\cP\!\left(\Bigg\lvert\frac{1}{n}\sum_{i=1}^n\lvert z_i\rvert
    -\frac{1}{n}\sum_{i=1}^n\lvert z_i\rvert\II_{\{\lvert z_i\rvert\le\sqrt{2\log n}\}}\Bigg\rvert\ge\varepsilon\right)\nonumber\\
    &\le \cP\!\left(\Bigg\lvert\frac{1}{n}\sum_{i=1}^n\lvert z_i\rvert-\sqrt{\frac{2}{\pi}}\Bigg\rvert\ge\frac{\varepsilon}{2}\right)
    +\cP\!\left(\Bigg\lvert\frac{1}{n}\sum_{i=1}^n\lvert z_i\rvert\II_{\{\lvert z_i\rvert\le\sqrt{2\log n}\}}
      -\sqrt{\frac{2}{\pi}}\Bigg\rvert\ge\frac{\varepsilon}{2}\right)\label{eq:abs_gaussianized_diff_2}\\
    &\le C\exp\!\left(-\frac{Cn\varepsilon^2}{\log n}\right),\nonumber
\end{align}
where the first term of \eqref{eq:abs_gaussianized_diff_2} follows from Lemma~2 in \cite{mai2023coordinatewise} with $\varepsilon\in(0,\varepsilon_0)$ for some constant $\varepsilon_0$ independent of $n$, and the second term of \eqref{eq:abs_gaussianized_diff_2} follows from \eqref{eq:abs_gaussianized_diff_2_res}.
\end{proof}
\end{lemma}

\begin{lemma}\label{lemma:square_gaussianized_diff} 
\citep[cf. Lemma 12 in][]{mai2023coordinatewise}.  
Under the same conditions as Lemma~\ref{lemma:abs_gaussianized_diff}, there exist positive constants $M$, $C$, $n_0$, and $\varepsilon_0$, all independent of $n$, such that for every $n\ge n_0$ and $\varepsilon\in(M\frac{\sqrt{\log n}}{n},\varepsilon_0)$,
\[
    \cP\!\left(\frac{1}{n}\sum_{i=1}^n (z_i^*-z_i)^2\ge\varepsilon\right)
    \le C\exp\!\left(-\frac{Cn\varepsilon^2}{\log^2 n}\right).
\]

\begin{proof}
The argument parallels Lemma~\ref{lemma:abs_gaussianized_diff}.  
Define  
\[
g(\mathbf{x})\doteq \frac{1}{n}\sum_{i=1}^n x_i^2\II_{\{\lvert x_i\rvert\le\sqrt{2\log n}\}},\qquad \mathbf{x}=(x_1,\ldots,x_n)\in\mathbb{R}^n.
\]
Let $\mathbf{x},\tilde{\mathbf{x}}\in\mathbb{R}^n$ differ only in the $i$-th coordinate; without loss of generality,  
$\mathbf{x}=(x_1,\ldots,x_i, \ldots, x_n)$ and $\tilde{\mathbf{x}}=(x_1, \ldots, \tilde{x}_i,\ldots,x_n)$. Then
\[
    \big\lvert g(\mathbf{x})-g(\tilde{\mathbf{x}})\big\rvert
    =\Bigg\lvert \frac{1}{n}x_i^2\II_{\{\lvert x_i\rvert\le\sqrt{2\log n}\}}
          -\frac{1}{n}\tilde{x}_i^2\II_{\{\lvert\tilde{x}_i\rvert\le\sqrt{2\log n}\}}\Bigg\rvert
    \le C\frac{\log n}{n},
\]
for some constant $C$ independent of $n$.  
Since $z_i\overset{\text{i.i.d.}}{\sim}N(0,1)$, Lemma~\ref{prop:McDiarmid} gives
\begin{equation}
\label{eq:trucated_gaussian_square_central}
    \cP\!\left(\Bigg\lvert\frac{1}{n}\sum_{i=1}^n z_i^2\II_{\{\lvert z_i\rvert\le\sqrt{2\log n}\}}
    -\eE[z_i^2\II_{\{\lvert z_i\rvert\le\sqrt{2\log n}\}}]\Bigg\rvert\ge\varepsilon\right)
    \le C\exp\!\left(-\frac{Cn\varepsilon^2}{\log^2 n}\right).
\end{equation}

Moreover,
\begin{equation}
\label{eq:trucated_gaussian_square_exp}
    \eE[z_i^2\II_{\{\lvert z_i\rvert\le\sqrt{2\log n}\}}]
    =\int_{-\sqrt{2\log n}}^{\sqrt{2\log n}} x^2\phi(x)\,dx
    =1-\frac{2}{\sqrt{\pi}}\frac{\sqrt{\log n}}{n}-2\Phi(-\sqrt{2\log n}).
\end{equation}
For $n\ge100$, applying Lemma~\ref{prop:Abramovich} yields
\[
\Phi(\sqrt{2\log n})\ge1-\frac{1}{n},\qquad
\Phi(-\sqrt{2\log n})\le\frac{1}{n},\qquad
1\ge\eE[z_i^2\II_{\{\lvert z_i\rvert\le\sqrt{2\log n}\}}]
   \ge 1-\Big(\frac{2}{\sqrt{\pi}}+2\Big)\frac{\sqrt{\log n}}{n}.
\]
Thus, if $M\ge 2\big(\frac{2}{\sqrt{\pi}}+2\big)$ and $\varepsilon>M\frac{\sqrt{\log n}}{n}$, then by \eqref{eq:trucated_gaussian_square_central}--\eqref{eq:trucated_gaussian_square_exp},
\begin{align}
\label{eq:square_gaussianized_diff_2_res}
    \cP\!\left(\Bigg\lvert\frac{1}{n}\sum_{i=1}^n z_i^2\II_{\{\lvert z_i\rvert\le\sqrt{2\log n}\}}-1\Bigg\rvert\ge\varepsilon\right)
    &\le
    \cP\!\left(\Bigg\lvert\frac{1}{n}\sum_{i=1}^n z_i^2\II_{\{\lvert z_i\rvert\le\sqrt{2\log n}\}}
    -\eE[z_i^2\II_{\{\lvert z_i\rvert\le\sqrt{2\log n}\}}]\Bigg\rvert\ge\frac{\varepsilon}{2}\right)\nonumber\\
    &\le C\exp\!\left(-\frac{Cn\varepsilon^2}{\log^2 n}\right).
\end{align}

Finally,
\begin{align}
    \cP\!\left(\frac{1}{n}\sum_{i=1}^n(z_i^*-z_i)^2\ge\varepsilon\right)
    &=\cP\!\left(\frac{1}{n}\sum_{i=1}^n\big(\lvert z_i\rvert-\sqrt{2\log n}\big)^2\II_{\{\lvert z_i\rvert>\sqrt{2\log n}\}}\ge\varepsilon\right)\nonumber\\
    &\le\cP\!\left(\frac{1}{n}\sum_{i=1}^n z_i^2\II_{\{\lvert z_i\rvert>\sqrt{2\log n}\}}\ge\varepsilon\right)\nonumber\\
    &=\cP\!\left(\Bigg\lvert\frac{1}{n}\sum_{i=1}^n z_i^2
    -\frac{1}{n}\sum_{i=1}^n z_i^2\II_{\{\lvert z_i\rvert\le\sqrt{2\log n}\}}\Bigg\rvert\ge\varepsilon\right)\nonumber\\
    &\le\cP\!\left(\Bigg\lvert\frac{1}{n}\sum_{i=1}^n z_i^2 -1\Bigg\rvert\ge\frac{\varepsilon}{2}\right) +\cP\!\left(\Bigg\lvert\frac{1}{n}\sum_{i=1}^n z_i^2\II_{\{\lvert z_i\rvert\le\sqrt{2\log n}\}}-1\Bigg\rvert\ge\frac{\varepsilon}{2}\right)\label{eq:square_gaussianized_diff_2}\\
    &\le C\exp\!\left(-\frac{Cn\varepsilon^2}{\log^2 n}\right),\nonumber
\end{align}
where the first term of \eqref{eq:square_gaussianized_diff_2} follows from Lemma~2 in \cite{mai2023coordinatewise} for $\varepsilon\in(0,\varepsilon_0)$ with $\varepsilon_0$ independent of $n$, and the second term of \eqref{eq:square_gaussianized_diff_2} follows from \eqref{eq:square_gaussianized_diff_2_res}.
\end{proof}
\end{lemma}

\begin{theorem}\label{thm:gaussianized_transform_diff_abs} 
\citep[cf. Theorem 1 in][]{mai2023coordinatewise}.  
Let $\xi_i\overset{\text{i.i.d.}}{\sim}F_\xi(\cdot)$, $i=1,\dots,n$, be a continuous random sample with empirical CDF $\widehat{F}_\xi(\cdot)$. Define  
\[
z_i\doteq\Phi^{-1}\!\left(F_\xi(\xi_i)\right),\qquad 
\widehat{z}_i\doteq\Phi^{-1}\!\left(\frac{n}{n+1}\widehat{F}_\xi(\xi_i)\right),\qquad i=1,\dots,n.
\]
Then there exist positive constants $M$, $C$, $n_0$, and $\varepsilon_0$, independent of both $\xi$ and $n$, such that for all $n\ge n_0$ and $\varepsilon\in(M\frac{\log n}{\sqrt{n}},\varepsilon_0)$,
\[
    \cP\!\left(\frac{1}{n}\sum_{i=1}^n\lvert\widehat{z}_i-z_i\rvert\ge\varepsilon\right)
    \le C\exp\!\left(-\frac{Cn\varepsilon^2}{\log n}\right).
\]

\begin{proof}
The result follows directly from Lemma~8 in \cite{mai2023coordinatewise}, together with Lemma~\ref{lemma:abs_gaussianized_diff}, using the fact that 
\( z_i \overset{\text{i.i.d.}}{\sim}N(0,1) \), \( i=1,\dots,n \).
\end{proof}
\end{theorem}

\begin{theorem}\label{thm:gaussianized_transform_diff_square} 
\citep[cf. Lemma 14 in][]{mai2023coordinatewise}.  
Under the same conditions as Theorem~\ref{thm:gaussianized_transform_diff_abs}, there exist positive constants $M$, $C$, $n_0$, and $\varepsilon_0$, independent of both $\xi$ and $n$, such that for all $n\ge n_0$ and $\varepsilon\in(M\frac{(\log n)^{3/2}}{\sqrt{n}},\varepsilon_0)$,
\[
    \cP\!\left(\frac{1}{n}\sum_{i=1}^n(\widehat{z}_i-z_i)^2\ge\varepsilon\right)
    \le C\exp\!\left(-\frac{Cn\varepsilon^2}{\log^2 n}\right).
\]

\begin{proof}
As observed, $z_i\overset{\text{i.i.d.}}{\sim}N(0,1)$.  
Using the same definition of $z_i^*$ as in Lemma~\ref{lemma:abs_gaussianized_diff}, for $n\ge n_0$,
\begin{align}
    \frac{1}{n}\sum_{i=1}^n(\widehat{z}_i-z_i)^2
    &\le \frac{1}{n}\sum_{i=1}^n\lvert\widehat{z}_i-z_i\rvert\Big(\lvert\widehat{z}_i-z_i^*\rvert+\lvert z_i^*-z_i\rvert\Big)\nonumber\\
    &\le C\frac{\sqrt{\log n}}{n}\sum_{i=1}^n\lvert\widehat{z}_i-z_i\rvert
       +\frac{1}{n}\sum_{i=1}^n\lvert\widehat{z}_i-z_i\rvert\,\lvert z_i^*-z_i\rvert
       \label{eq:gaussianized_transform_diff_square_prop3}\\
    &\le C\frac{\sqrt{\log n}}{n}\sum_{i=1}^n\lvert\widehat{z}_i-z_i\rvert
       +\frac{1}{2n}\sum_{i=1}^n(\widehat{z}_i-z_i)^2
       +\frac{1}{2n}\sum_{i=1}^n(z_i^*-z_i)^2,\nonumber
\end{align}
where \eqref{eq:gaussianized_transform_diff_square_prop3} follows from Lemma~\ref{prop:gaussianized_bound}.  
Rearranging yields
\[
    \frac{1}{2n}\sum_{i=1}^n(\widehat{z}_i-z_i)^2
    \le C\frac{\sqrt{\log n}}{n}\sum_{i=1}^n\lvert\widehat{z}_i-z_i\rvert
       +\frac{1}{n}\sum_{i=1}^n(z_i^*-z_i)^2,
\]
and the desired conclusion follows by applying Lemma~\ref{lemma:square_gaussianized_diff} and Theorem~\ref{thm:gaussianized_transform_diff_abs}.
\end{proof}
\end{theorem}

\begin{lemma}\label{lemma:B_expectation}
    Under the same setup as in Theorems~2.2 and~2.3, suppose $X\perp Y$.  
    Define $\widehat{U}\doteq \widehat{F}_{W}(\widehat{CD}(X,Y;r))-\tfrac{1}{2}$ and $B\doteq\II_{\{\widehat{U}<0\}}$, constructed as in Section~4.  
    Let $m\to\infty$ and $m=o(n!)$ as $n\to\infty$. Then
    \[
        \eE[B]\to\tfrac{1}{2}\qquad \text{as }n\to\infty.
    \]
    \begin{proof}
        By definition,
        \[
            \eE[B]=\cP(\widehat{U}<0)
            =\cP\!\left(\frac{\#\{i\in[m]:W_i\le\mathcal{W}_{r}(\widehat{f}_{\widehat{S},\widehat{T}},\phi\otimes\phi)\}}{m}<\tfrac{1}{2}\right).
        \]
        By construction, $\{W_i\}_{i=1}^m$ and $\mathcal{W}_{r}(\widehat{f}_{\widehat{S},\widehat{T}},\phi\otimes \phi)$ are $m+1$ i.i.d.\ random variables uniformly distributed over $n!$ possible values.  
        Define the event 
        \[
            \mathscr{B}=\{\text{no }W_i\text{ equals }\mathcal{W}_{r}(\widehat{f}_{\widehat{S},\widehat{T}},\phi\otimes\phi)\}.
        \]
        Since $\cP(\mathscr{B})=(1-\tfrac{1}{n!})^m\to1$ whenever $m=o(n!)$, we obtain the following results using $m\to\infty$ as $n\to\infty$,
        \begin{align}
            \eE[B]
            &\ge\cP\!\left(\frac{\#\{i\in[m]: W_i<\mathcal{W}_{r}(\widehat{f}_{\widehat{S},\widehat{T}},\phi\otimes\phi)\}}{m}<\tfrac{1}{2}\Bigg|\mathscr{B}\right)\cP(\mathscr{B})\nonumber\\
            &=\cP\!\left(\sum_{i=1}^{m}\II_{\{W_i<\mathcal{W}_{r}(\widehat{f}_{\widehat{S},\widehat{T}},\phi\otimes\phi)\}}<\tfrac{m}{2}\Bigg|\mathscr{B}\right)\cP(\mathscr{B})\nonumber\\
            &=\sum_{i=1}^{\lfloor(m-1)/2\rfloor}\binom{m}{i}\Big(\tfrac{1}{2}\Big)^m\,\cP(\mathscr{B})
             \to \tfrac{1}{2},
             \label{eq:binom_express}
        \end{align}
        where \eqref{eq:binom_express} follows because $\II_{\{W_i<\mathcal{W}_{r}(\widehat{f}_{\widehat{S},\widehat{T}},\phi\otimes\phi)\}} \overset{\text{i.i.d.}}{\sim}\mathrm{Bernoulli}(\tfrac{1}{2})$ conditional on $\mathscr{B}$.

        Conversely,
        \begin{align}
            \eE[B]
            &=\cP\!\left(\tfrac{\#\{i\in[m]:W_i<\mathcal{W}_{r}(\widehat{f}_{\widehat{S},\widehat{T}},\phi\otimes\phi)\}}{m}<\tfrac{1}{2}\Bigg|\mathscr{B}\right)\cP(\mathscr{B})\nonumber\\
            &\quad+\cP\!\left(\tfrac{\#\{i\in[m]:W_i\le\mathcal{W}_{r}(\widehat{f}_{\widehat{S},\widehat{T}},\phi\otimes\phi)\}}{m}<\tfrac{1}{2}\Bigg|\mathscr{B}^c\right)\cP(\mathscr{B}^c)\nonumber\\
            &\le\cP\!\left(\tfrac{\#\{i\in[m]:W_i<\mathcal{W}_{r}(\widehat{f}_{\widehat{S},\widehat{T}},\phi\otimes\phi)\}}{m}<\tfrac{1}{2}\Bigg|\mathscr{B}\right)\cP(\mathscr{B})
               +\cP(\mathscr{B}^c)\to\tfrac{1}{2}.
        \end{align}
        Combining both limits gives $\eE[B]\to\tfrac{1}{2}$ as $n\to\infty$.
    \end{proof}
\end{lemma}

\section{Proofs of Theoretical Results in the Main Document}

\setcounter{equation}{0}
\renewcommand{\theequation}{E.\arabic{equation}}

In this appendix, we provide the proofs of the theoretical results presented in the main document.

\subsection{Proof of Lemma 2.1}
For any $r\in[1,\infty)$, the nonnegativity of $CD(X,Y;r)\ge0$ follows directly from its definition.

To show invariance of $CD(X,Y;r)$ under monotone increasing and invertible marginal transformations, it suffices to verify that the transformed joint distribution $F_{S,T}(s,t)$ remains unchanged.  
Let $X^* \doteq g_1(X)$ and $Y^* \doteq g_2(Y)$, where $g_1(\cdot)$ and $g_2(\cdot)$ are monotone increasing and invertible.  
Define
\[
S^*\doteq \Phi^{-1}(F_{X^*}(X^*)),\qquad 
T^*\doteq \Phi^{-1}(F_{Y^*}(Y^*)).
\]
Since $g_1$ and $g_2$ are monotone,  
\[
F_{X^*}(x^*)=F_X(g_1^{-1}(x^*)),\qquad
F_{Y^*}(y^*)=F_Y(g_2^{-1}(y^*)).
\]
Hence,
\begin{align*}
    F_{S^*,T^*}(s,t)
    &=\cP(S^*\le s,\;T^*\le t)\\
    &=\cP\big(\Phi^{-1}(F_{X^*}(X^*))\le s,\;\Phi^{-1}(F_{Y^*}(Y^*))\le t\big)\\
    &=\cP\big(\Phi^{-1}(F_X(X))\le s,\;\Phi^{-1}(F_Y(Y))\le t\big)\\
    &=\cP(S\le s,\;T\le t)=F_{S,T}(s,t).
\end{align*}
The desired conclusion follows immediately from the definition of $CD(X,Y;r)$.

\qed

\subsection{Proof of Theorems \ref{thm:gaussianized_wd_1} and \ref{thm:gaussianized_wd_2}}

Let $\widehat{f}_{S,T}$ denote the empirical measure of the genuine Gaussianized sample \eqref{eq:genuine_gaussian_sample}.  
For any $r\ge1$, by the triangle inequality and the nonnegativity of the $r$-Wasserstein distance,
\begin{align}
     \big\lvert\widehat{CD}(X,Y;r)-CD(X,Y;r)\big\rvert
     &=\big\lvert \mathcal{W}_r(\widehat{f}_{\widehat{S},\widehat{T}},\phi\otimes\phi)
           -\mathcal{W}_r(f_{S,T},\phi\otimes\phi)\big\rvert \nonumber\\
     &\le \mathcal{W}_r(\widehat{f}_{\widehat{S},\widehat{T}},f_{S,T}) \nonumber\\
     &\le \mathcal{W}_r(\widehat{f}_{\widehat{S},\widehat{T}},\widehat{f}_{S,T})
         +\mathcal{W}_r(\widehat{f}_{S,T},f_{S,T}). \label{eq:wd_tri_split}
\end{align}
We bound the two terms on the right-hand side separately.

\begin{enumerate}
    \item \textbf{Bounding $\mathcal{W}_r(\widehat{f}_{\widehat{S},\widehat{T}},\widehat{f}_{S,T})$.}
    Let $\boldsymbol{\Pi}$ denote the set of all permutations of $\{1,\dots,n\}$. Then,
    \begin{align}
        \mathcal{W}_r^r(\widehat{f}_{\widehat{S},\widehat{T}},\widehat{f}_{S,T})
        &=\inf_{\boldsymbol{\pi}\in\boldsymbol{\Pi}}
          \Bigg\{\frac{1}{n}\sum_{i=1}^n
          \Bigg\|\!
            \begin{pmatrix}s_i\\t_i\end{pmatrix}
            -
            \begin{pmatrix}\widehat{s}_{\boldsymbol{\pi}(i)}\\\widehat{t}_{\boldsymbol{\pi}(i)}\end{pmatrix}
          \!\Bigg\|_q^r\Bigg\}\nonumber\\
        &\le\frac{1}{n}\sum_{i=1}^n
          \Bigg\|
            \begin{pmatrix}s_i\\t_i\end{pmatrix}
            -
            \begin{pmatrix}\widehat{s}_i\\\widehat{t}_i\end{pmatrix}
          \!\Bigg\|_q^r\nonumber\\
        &\le\frac{1}{n}\sum_{i=1}^n(|s_i-\widehat{s}_i|+|t_i-\widehat{t}_i|)^r \label{eq:gaussianized_wd_l_q}\\
        &\le C_r\bigg[\frac{1}{n}\sum_{i=1}^n|s_i-\widehat{s}_i|^r
                    +\frac{1}{n}\sum_{i=1}^n|t_i-\widehat{t}_i|^r\bigg], \label{eq:gaussianized_wd_p_expand}
    \end{align}
    where \eqref{eq:gaussianized_wd_l_q} uses $\|x\|_q\le\|x\|_1$ for $q\ge1$, and \eqref{eq:gaussianized_wd_p_expand} follows from
    $|a+b|^r\le C_r(|a|^r+|b|^r)$ with $C_r=2^{(r-1)_+}$.

    \vspace{0.15cm}
    \noindent\underline{$r=1$ case.}  
    By Theorem~\ref{thm:gaussianized_transform_diff_abs}, there exist constants $M,C,n_0,\varepsilon_0$, independent of the of $(X,Y)$ and $n$, such that for all $n\ge n_0$ and $\varepsilon\in(M\frac{\log n}{\sqrt{n}},\varepsilon_0)$,
    \begin{equation}
        \cP\!\left(\mathcal{W}_{1}(\widehat{f}_{\widehat{S},\widehat{T}},\widehat{f}_{S,T})\ge \varepsilon\right)
        \le C\exp\!\left(-\frac{Cn\varepsilon^2}{\log n}\right). \label{eq:wd_samples_1}
    \end{equation}

    \vspace{0.15cm}
    \noindent\underline{$r=2$ case.}  
    By Theorem~\ref{thm:gaussianized_transform_diff_square}, there exist constants $M,C,n_0,\varepsilon_0>0$, independent of $(X,Y)$ and $n$, such that for all $n\ge n_0$ and $\varepsilon\in(M\frac{(\log n)^{3/4}}{n^{1/4}},\varepsilon_0)$,
    \begin{equation}
        \cP\!\left(\mathcal{W}_{2}(\widehat{f}_{\widehat{S},\widehat{T}},\widehat{f}_{S,T})\ge\varepsilon\right)
        \le C\exp\!\left(-\frac{Cn\varepsilon^4}{\log^2 n}\right). \label{eq:wd_samples_2}
    \end{equation}

    \item \textbf{Bounding $\mathcal{W}_r(\widehat{f}_{S,T},f_{S,T})$.}
    We apply a similar argument as in Theorem~2 of \cite{fournier2015rate}.  
    Since both $S$ and $T$ are standard Gaussian, all required conditions in \cite{fournier2015rate} are satisfied.

    \vspace{0.12cm}
    \noindent\underline{$r=1$ case.}  
    Condition (1) in \cite{fournier2015rate} holds with $\alpha=2$ and any $\gamma<1/8$.  
    Thus, for $n\ge1$ and $\varepsilon\in(0,1)$,
    \begin{equation}
        \cP\!\left(\mathcal{W}_{1}(\widehat{f}_{S,T},f_{S,T})\ge\varepsilon\right)
        \le C\exp\!\left(-\frac{Cn\varepsilon^2}{\log^2(2+1/\varepsilon)}\right). \label{eq:wd_sample_population_1}
    \end{equation}

    \vspace{0.12cm}
    \noindent\underline{$r=2$ case.}  
    Condition (2) holds for all $\alpha\in(0,2)$ and $\gamma>0$.  
    Hence for any $\alpha\in(0,2)$ and $\delta\in(0,\alpha)$, there exists $C>0$ such that for $n\ge1$ and $\varepsilon\in(0,1)$,
    \begin{equation}
        \cP(\mathcal{W}_{2}(\widehat{f}_{S,T},f_{S,T})\ge\varepsilon)
        \le C\exp(-Cn\varepsilon^4)+C\exp\!\left(-C(n\varepsilon^2)^{\frac{\alpha-\delta}{2}}\right). \label{eq:wd_sample_population_2}
    \end{equation}
\end{enumerate}

\noindent\textbf{Conclusion.}  
When $r=1$ and $\varepsilon>M\frac{\log n}{\sqrt{n}}$, one has $\log^2(2+1/\varepsilon)\le C\log^2 n$.  
Combining \eqref{eq:wd_samples_1}, \eqref{eq:wd_sample_population_1}, and \eqref{eq:wd_tri_split},
\[
    \cP\!\left(\big\lvert\widehat{CD}(X,Y;1)-CD(X,Y;1)\big\rvert\ge\varepsilon\right)
    \le C\exp\!\left(-\frac{Cn\varepsilon^2}{\log^2 n}\right).
\]

When $r=2$, combining \eqref{eq:wd_samples_2}, \eqref{eq:wd_sample_population_2}, and \eqref{eq:wd_tri_split}, for any $\beta=\tfrac{\alpha-\delta}{2}\in(0,1)$,
\[
    \cP\!\left(\big\lvert\widehat{CD}(X,Y;2)-CD(X,Y;2)\big\rvert\ge\varepsilon\right)
    \le C\exp\!\left(-\frac{Cn\varepsilon^4}{\log^2 n}\right)
     +C\exp\!\left(-C(n\varepsilon^2)^{\beta}\right).
\]

The second assertion follows immediately from the first.

\qed

\subsection{Proof of Theorem \ref{thm:sure_screen_DR-WD-SIS}}

Under Condition~\ref{condition:minimum strength}(a), we have
\[
    t_n \le c_t^{1/r}n^{-\kappa_1/r} < c_1^{1/r}n^{-\kappa_1/r}
    \le \min_{j\in\mathcal{A}}CD(X_j,Y;r).
\]
Therefore,
\begin{align*}
    \cP(\mathcal{A}\not\subseteq\widehat{\mathcal{A}}(t_n))
    &=\cP\!\left(\bigcup_{j\in\mathcal{A}}\{j\notin\widehat{\mathcal{A}}(t_n)\}\right)\\
    &\le\sum_{j\in\mathcal{A}}\cP\!\left(\widehat{CD}(X_j,Y;r)\le t_n\right)\\
    &\le\sum_{j\in\mathcal{A}}
       \cP\!\Big(\lvert\widehat{CD}(X_j,Y;r)-CD(X_j,Y;r)\rvert
                \ge CD(X_j,Y;r)-t_n\Big)\\
    &\le\sum_{j\in\mathcal{A}}
       \cP\!\Big(\lvert\widehat{CD}(X_j,Y;r)-CD(X_j,Y;r)\rvert
                \ge(c_1^{1/r}-c_t^{1/r})n^{-\kappa_1/r}\Big).
\end{align*}

Choose $n\ge n_1$. We analyze $\cP(\mathcal{A}\subseteq\widehat{\mathcal{A}}(t_n)) = 1-\cP(\mathcal{A}\not\subseteq\widehat{\mathcal{A}}(t_n))$ using Theorems~\ref{thm:gaussianized_wd_1} and~\ref{thm:gaussianized_wd_2}.

\begin{enumerate}
    \item \textbf{$r=1$ case.}
    \[
        \cP(\mathcal{A}\subseteq\widehat{\mathcal{A}}(t_n))
        \ge1 - C_1\mathcal{S}_{\mathcal{A}}
               \exp\!\left(-\frac{C_1n^{1-2\kappa_1}}{\log^2 n}\right).
    \]

    \item \textbf{$r=2$ case.}  
    For any $\beta\in(0,1)$,
    \[
        \cP(\mathcal{A}\subseteq\widehat{\mathcal{A}}(t_n))
        \ge1-\mathcal{S}_{\mathcal{A}}\Big[
            C_1\exp\!\left(-\frac{C_1n^{1-2\kappa_1}}{\log^2 n}\right)
           +C_1\exp\!\big(-C_1n^{\beta(1-\kappa_1)}\big)
        \Big].
    \]
\end{enumerate}

\qed

\subsection{Proof of Theorem \ref{theorem:rankconsistency}}

Under Condition~\ref{condition:minimum strength}(b), we have
\begin{align}
    &\cP\!\left(\min_{j\in\mathcal{A}}\widehat{CD}(X_j,Y;r)
               -\max_{j\in\mathcal{A}^c}\widehat{CD}(X_j,Y;r)\le 0\right)\nonumber\\
    \le&\cP\!\Bigg(\min_{j\in\mathcal{A}}\widehat{CD}(X_j,Y;r)
               -\max_{j\in\mathcal{A}^c}\widehat{CD}(X_j,Y;r)\nonumber\\
    \le&
          \min_{j\in\mathcal{A}}CD(X_j,Y;r)
          -\max_{j\in\mathcal{A}^c}CD(X_j,Y;r)
          -c_2n^{-\kappa_2}\Bigg)\nonumber\\
    =&\cP\!\Big(
         [\min_{j\in\mathcal{A}}CD(X_j,Y;r)-\min_{j\in\mathcal{A}}\widehat{CD}(X_j,Y;r)]\nonumber\\
        &+
         [\max_{j\in\mathcal{A}^c}\widehat{CD}(X_j,Y;r)
         -\max_{j\in\mathcal{A}^c}CD(X_j,Y;r)]
         \ge c_{2}n^{-\kappa_{2}}
        \Big).\nonumber
\end{align}

Let
\[
    j_1\doteq\argmin_{j\in\mathcal{A}}\widehat{CD}(X_j,Y;r),\qquad
    j_2\doteq\argmax_{j\in\mathcal{A}^c}\widehat{CD}(X_j,Y;r).
\]
Then
\[
    CD(X_{j_1},Y;r)\ge\min_{j\in\mathcal{A}}CD(X_j,Y;r),\qquad
    CD(X_{j_2},Y;r)\le\max_{j\in\mathcal{A}^c}CD(X_j,Y;r),
\]
and thus
\begin{align}
    &\cP\!\left(\min_{j\in\mathcal{A}}\widehat{CD}(X_j,Y;r)
               -\max_{j\in\mathcal{A}^c}\widehat{CD}(X_j,Y;r)\le 0\right)\nonumber\\
    \le\;&\cP\!\Big(
         [CD(X_{j_1},Y;r)-\widehat{CD}(X_{j_1},Y;r)]
         +[\widehat{CD}(X_{j_2},Y;r)-CD(X_{j_2},Y;r)]
         \ge c_{2}n^{-\kappa_{2}}
        \Big)\nonumber\\
    \le\;&\cP\!\left(
          \lvert\widehat{CD}(X_{j_1},Y;r)-CD(X_{j_1},Y;r)\rvert
          \ge\tfrac{1}{2}c_{2}n^{-\kappa_{2}}
        \right)
       +\cP\!\left(
          \lvert\widehat{CD}(X_{j_2},Y;r)-CD(X_{j_2},Y;r)\rvert
          \ge\tfrac{1}{2}c_{2}n^{-\kappa_{2}}
        \right)\nonumber\\
    \le\;&2\cP\!\left(
          \max_{j\in\{1,\dots,p\}}
          \lvert\widehat{CD}(X_j,Y;r)-CD(X_j,Y;r)\rvert
          \ge\tfrac{1}{2}c_{2}n^{-\kappa_{2}}
        \right)\nonumber\\
    =\;&2\cP\!\left(\bigcup_{j=1}^p
          \Big\{
             \lvert\widehat{CD}(X_j,Y;r)-CD(X_j,Y;r)\rvert
             \ge\tfrac{1}{2}c_{2}n^{-\kappa_{2}}
          \Big\}
        \right)\nonumber\\
    \le\;&2\sum_{j=1}^p
        \cP\!\left(
          \lvert\widehat{CD}(X_j,Y;r)-CD(X_j,Y;r)\rvert
          \ge\tfrac{1}{2}c_{2}n^{-\kappa_{2}}
        \right).\nonumber
\end{align}

Applying Theorems~2.2 and~2.3, for $n\ge n_2$ we obtain:

\begin{enumerate}
    \item \textbf{$r=1$ case:}
    \[
        \cP\!\left(
            \min_{j\in\mathcal{A}}\widehat{CD}(X_j,Y;1)
            -\max_{j\in\mathcal{A}^c}\widehat{CD}(X_j,Y;1)>0
        \right)
        \ge 1-2C_2p\exp\!\left(
            -\frac{C_2n^{1-2\kappa_2}}{\log^2 n}
        \right).
    \]

    \item \textbf{$r=2$ case (for any $\beta\in(0,1)$):}
    \begin{align}
        &\cP\!\left(
            \min_{j\in\mathcal{A}}\widehat{CD}(X_j,Y;2)
            -\max_{j\in\mathcal{A}^c}\widehat{CD}(X_j,Y;2)>0
         \right)\nonumber\\
        \ge\;&1-2C_2p\Big[
            \exp\!\left(-\frac{C_2n^{1-4\kappa_2}}{\log^2 n}\right)
            +\exp\!\left(-C_2n^{\beta(1-2\kappa_2)}\right)
        \Big].\nonumber
    \end{align}
\end{enumerate}

The remainder follows directly from the proof of Theorem~3 in \cite{liu2022model}.

\qed

\subsection{Proof of Theorem \ref{thm:FDR_control}}

In this proof, we define the inactive feature set 
\[
\widetilde{\mathcal{A}}^c=\left\{j\in\{1,\cdots,p\}:X_j\perp Y\right\},
\]
and denote its cardinality by $\tilde{p}_0=|\widetilde{\mathcal{A}}^c|$. The sparsity assumption implies $\tilde{p}_0/p_0\to 1$. By the definition of FDR, we have
\begin{align}
    \operatorname{FDR}[T_{\alpha}]
    &=\eE\left[\operatorname{FDP}[T_{\alpha}]\right]\nonumber\\
    &=\eE\left[\frac{\#\{j\in \widetilde{\mathcal{A}}^c: \wh{U}_j\geqslant T_{\alpha}\}}{\#\{j:\wh{U}_j\geqslant T_{\alpha}\}}\right]\nonumber\\
    &=\eE\left[\frac{\#\{j\in \widetilde{\mathcal{A}}^c: \wh{U}_j\geqslant T_{\alpha}\}}{1+\#\{j: \wh{U}_j\leqslant -T_{\alpha}\}}\cdot\frac{1+\#\{j: \wh{U}_j\leqslant -T_{\alpha}\}}{\#\{j:\wh{U}_j\geqslant T_{\alpha}\}}\right]\nonumber\\
    &\leqslant\eE\left[\frac{\#\{j\in \widetilde{\mathcal{A}}^c: \wh{U}_j\geqslant T_{\alpha}\}}{1+\#\{j\in \widetilde{\mathcal{A}}^c: \wh{U}_j\leqslant -T_{\alpha}\}}\cdot\frac{1+\#\{j: \wh{U}_j\leqslant -T_{\alpha}\}}{\#\{j:\wh{U}_j\geqslant T_{\alpha}\}}\right]\nonumber\\
    &\leqslant\alpha \eE\left[\frac{\#\{j\in \widetilde{\mathcal{A}}^c: \wh{U}_j\geqslant T_{\alpha}\}}{1+\#\{j\in \widetilde{\mathcal{A}}^c: \wh{U}_j\leqslant -T_{\alpha}\}}\right],\label{eq:FDR results}
\end{align}
where the last inequality follows from \eqref{eq:threshold}. The expectation is taken with respect to $\{\mathcal{W}_{r}(\wh{f}_{\wh{S}_j,\wh{T}}, \phi\otimes \phi)\}_{j=1}^p$ and $\{W_i\}_{i=1}^m$.

Next, we derive an upper bound for
\[
\eE\left[\frac{\#\{j\in \widetilde{\mathcal{A}}^c: \wh{U}_j\geqslant T_{\alpha}\}}{1+\#\{j\in \widetilde{\mathcal{A}}^c: \wh{U}_j\leqslant -T_{\alpha}\}}\right].
\]
Without loss of generality, assume
\[
\left|\wh{U}_1\right| \geqslant\left|\wh{U}_2\right| \geqslant \cdots \geqslant\left|\wh{U}_p\right|>0,
\]
and define $\left|\wh{U}_{p+1}\right|=0$. To determine the threshold $T_{\alpha}$, different values of $t$ are considered in \eqref{eq:threshold}, ranging from the smallest $t=\left|\wh{U}_{p+1}\right|$ to the largest $t=\left|\wh{U}_{1}\right|$. In this procedure, $T_{\alpha}$ is a stopping time. Moreover, since the quantity
\[
\frac{\#\{j\in \widetilde{\mathcal{A}}^c: \wh{U}_j\geqslant t\}}{1+\#\{j\in \widetilde{\mathcal{A}}^c: \wh{U}_j\leqslant -t\}}
\]
does not change for $t=\left|\wh{U}_{j}\right|$ with $j\not\in \widetilde{\mathcal{A}}^c$, it suffices to consider $t=\left|\wh{U}_{j}\right|$ for $j\in \widetilde{\mathcal{A}}^c$ only.

Accordingly, we relabel the indices of $\widetilde{\mathcal{A}}^c$ as $\left\{1, \ldots, \tilde{p}_0\right\}$, and assume
\[
\left|\wh{U}_1\right| \geqslant\left|\wh{U}_2\right| \geqslant \cdots \geqslant\left|\wh{U}_{\tilde{p}_0}\right|>0,\qquad \left|\wh{U}_{\tilde{p}_0+1}\right|=0,
\]
and we again view $T_{\alpha}$ as a stopping time. Equivalently, for $k=\tilde{p}_0+1, \tilde{p}_0, \cdots, 1$, define
\begin{align}
M_k 
&=\frac{\#\left\{j\in \widetilde{\mathcal{A}}^c: \wh{U}_j\geqslant \left|\wh{U}_k\right|\right\}}{1+\#\left\{j\in \widetilde{\mathcal{A}}^c: \wh{U}_j\leqslant -\left|\wh{U}_k\right|\right\}}\nonumber\\
&=\frac{\#\left\{j: j\leqslant k \text{ and } \wh{U}_j\geqslant 0\right\}}{1+\#\left\{j: j\leqslant k \text{ and } \wh{U}_j<0\right\}}.\nonumber
\end{align}

By the definition of $B_j$ in Assumption \ref{asmpt:weak dependence} and setting $\widetilde{S}^c_k\doteq\sum_{j=1}^k B_j$, we obtain
\begin{align}
M_k
&=\frac{\sum_{j=1}^k\left[1-B_j\right]}{1+\sum_{j=1}^kB_j}\nonumber\\
&=\frac{k+1}{1+\widetilde{S}^c_k}-1.\nonumber
\end{align}

Given the sample $\{W_i\}_{i=1}^m$, let $\mathcal{F}_{k}$ denote the $\sigma$-field generated by $\left\{\sum^k_{j=1}B_j, B_{k+1}, \cdots, B_{\tilde{p}_0+1}\right\}$. With respect to the process $\{M_k\}_{k=\tilde{p}_0+1}^1$ and the backward filtration $\mathcal{F}_{\tilde{p}_0+1} \subset \cdots \subset \mathcal{F}_{1}$, $T_{\alpha}$ is a stopping time in reverse time (from $\tilde{p}_0+1$ to $1$). It is also easy to see that $\wh{U}_j$ are identically distributed for $j \in \widetilde{\mathcal{A}}^c$, and hence $\left\{B_1, \cdots, B_k\right\}$ are identically distributed with respect to $\mathcal{F}_{k}$.

By Lemma 8 in \cite{tong2023model}, we have
\[
\cP\left(B_k=1 \mid \mathcal{F}_{k}\right)=\frac{\widetilde{S}^c_k}{k}.
\]
Consequently, on the one hand, if $\widetilde{S}^c_k=0$, then $\widetilde{S}^c_{k-1}=0$, and thus $M_{k-1}=k-1<k=M_k$. On the other hand, if $\widetilde{S}^c_k>0$, then
\begin{align}
\eE\left[M_{k-1} \mid \mathcal{F}_{k}\right]
&= \left[\frac{k}{1+\widetilde{S}^c_k}-1\right] \cP\left(B_k=0 \mid \mathcal{F}_{k}\right)
 +\left[\frac{k}{1+\widetilde{S}^c_k-1}-1\right] \cP\left(B_k=1 \mid \mathcal{F}_{k}\right)\nonumber\\
&=\left[\frac{k}{1+\widetilde{S}^c_k}-1\right] \frac{k-\widetilde{S}^c_k}{k}
 +\left[\frac{k}{\widetilde{S}^c_k}-1\right]\frac{\widetilde{S}^c_k}{k}\nonumber\\
&=\frac{k+1}{1+\widetilde{S}^c_{k}}-1\nonumber\\
&=M_{k}.\nonumber
\end{align}

Thus, $\{M_k\}_{k=\tilde{p}_0+1}^1$ is a super-martingale with respect to $\{\mathcal{F}_{k}\}$. By the Optional Stopping Time Theorem, we obtain
\begin{align}
\eE\left[\frac{\#\{j\in \widetilde{\mathcal{A}}^c: \wh{U}_j\geqslant T_{\alpha}\}}{1+\#\{j\in \widetilde{\mathcal{A}}^c: \wh{U}_j\leqslant -T_{\alpha}\}}\Bigg|\{W_i\}_{i=1}^m\right]
&\leqslant \eE\left[M_{\tilde{p}_0}\mid\{W_i\}_{i=1}^m\right] \nonumber\\
&= \eE\left[\frac{\tilde{p}_0-\widetilde{S}^c}{1+\widetilde{S}^c}\Bigg|\{W_i\}_{i=1}^m\right]\nonumber\\
&= \eE\left[\frac{1-\frac{\widetilde{S}^c}{\tilde{p}_0}}{\frac{1}{\tilde{p}_0}+\frac{\widetilde{S}^c}{\tilde{p}_0}}\Bigg|\{W_i\}_{i=1}^m\right],\nonumber
\end{align}
where $\widetilde{S}^c\doteq\sum_{j=1}^{\tilde{p}_0} B_{j}$.

By taking expectations over $\{W_i\}_{i=1}^m$ again, we have
\begin{equation}
    \eE\left[\frac{\#\{j\in \widetilde{\mathcal{A}}^c: \wh{U}_j\geqslant T_{\alpha}\}}{1+\#\{j\in \widetilde{\mathcal{A}}^c: \wh{U}_j\leqslant -T_{\alpha}\}}\right]
    \leqslant\eE\left[\frac{1-\frac{\widetilde{S}^c}{\tilde{p}_0}}{\frac{1}{\tilde{p}_0}+\frac{\widetilde{S}^c}{\tilde{p}_0}}\right].\label{eq:upper bound needed}
\end{equation}

Next, denote $\Delta S^c=S^c-\widetilde{S}^c$. Then
\begin{align}
    \operatorname{Var}\left[\frac{\widetilde{S}^c}{p_0}\right]
    =&\operatorname{Var}\left[\frac{S^c-\Delta S^c}{p_0}\right]\nonumber\\
    =&\operatorname{Var}\left[\frac{S^c}{p_0}\right]+\operatorname{Var}\left[\frac{\Delta S^c}{p_0}\right]
      -2\eE\left[\frac{S^c}{p_0}\frac{\Delta S^c}{p_0}\right]
      +2\eE\left[\frac{S^c}{p_0}\right]\eE\left[\frac{\Delta S^c}{p_0}\right]\nonumber\\
    \leqslant&\operatorname{Var}\left[\frac{S^c}{p_0}\right]+\operatorname{Var}\left[\frac{\Delta S^c}{p_0}\right]
      +2\eE\left[\frac{S^c}{p_0}\frac{\Delta S^c}{p_0}\right]
      +2\eE\left[\frac{S^c}{p_0}\right]\eE\left[\frac{\Delta S^c}{p_0}\right]\nonumber\\
    \leqslant&\operatorname{Var}\left[\frac{S^c}{p_0}\right]+\operatorname{Var}\left[\frac{\Delta S^c}{p_0}\right]
      +4\frac{p_0-\tilde{p}_0}{p_0}\label{eq:Sc_leq_p_0}\\
    \leqslant&\operatorname{Var}\left[\frac{S^c}{p_0}\right]+\frac{1}{4}\left[\frac{p_0-\tilde{p}_0}{p_0}\right]^2
      +4\frac{p_0-\tilde{p}_0}{p_0}\nonumber\\
    =&o(1),\nonumber
\end{align}
where \eqref{eq:Sc_leq_p_0} follows from $S^c\leqslant p_0$ and $\Delta S^c\leqslant p_0-\tilde{p}_0$, and the last equality follows from Assumption \ref{asmpt:weak dependence}. Hence,
\[
\operatorname{Var}\left[\frac{\widetilde{S}^c}{\tilde{p}_0}\right]
=\left(\frac{p_0}{\tilde{p}_0}\right)^2\operatorname{Var}\left[\frac{\widetilde{S}^c}{p_0}\right]
=o(1).
\]

By Lemma \ref{lemma:B_expectation}, we also have $\eE\left[\frac{\widetilde{S}^c}{\tilde{p}_0}\right]\to\frac{1}{2}$. Therefore, by Markov's inequality,
\[
\frac{\widetilde{S}^c}{\tilde{p}_0}\stackrel{p}{\to}\frac{1}{2},\quad \text{as } n\to\infty.
\]
Since $f(x)=\frac{1-x}{\frac{1}{\tilde{p}_0}+x}$ is bounded and continuous, it follows that
\begin{equation}
    \eE\left[\frac{1-\frac{\widetilde{S}^c}{\tilde{p}_0}}{\frac{1}{\tilde{p}_0}+\frac{\widetilde{S}^c}{\tilde{p}_0}}\right]
    \to \eE\left[\frac{1-\frac{1}{2}}{0+\frac{1}{2}}\right]=1.\label{eq:expectation bound}
\end{equation}

Finally, combining \eqref{eq:FDR results}, \eqref{eq:upper bound needed}, and \eqref{eq:expectation bound}, we obtain
\[
\limsup_{n\rightarrow\infty}\operatorname{FDR}[T_{\alpha}]
\leqslant\alpha\limsup_{n\rightarrow\infty}\eE\left[\frac{1-\frac{\widetilde{S}^c}{\tilde{p}_0}}{\frac{1}{\tilde{p}_0}+\frac{\widetilde{S}^c}{\tilde{p}_0}}\right]
=\alpha.
\]

\qed

\bibliographystyle{agsm}
\bibliography{ref}